\documentclass{sig-alternate2-pods09} 

\input{psfig.sty}
\usepackage{dsfont}
\usepackage{url}
\usepackage{subfigure}
\usepackage{boxedminipage}
\usepackage{amssymb}
\usepackage{amsmath}
\usepackage{stmaryrd}
\usepackage{mathpartir}
\usepackage{pst-node}
\usepackage{pst-tree}
\usepackage{booktabs}
\usepackage{color}
\usepackage{colortbl}

\definecolor{darkgray}{rgb}{0.88,0.88,0.88}
\definecolor{lightgray}{rgb}{0.96,0.96,0.96}

\newcommand{\squishlist}{
   \begin{list}{$\bullet$}
    { \setlength{\itemsep}{0pt}      \setlength{\parsep}{1.75pt}
      \setlength{\topsep}{3pt}       \setlength{\partopsep}{0pt}
      \setlength{\leftmargin}{1.5em} \setlength{\labelwidth}{1em}
      \setlength{\labelsep}{0.5em} } }

\newcommand{\squishend}{
    \end{list}  }

\newtheorem{example}{Example}
\newtheorem{theorem}{Theorem}
\newtheorem{lemma}{Lemma}
\newtheorem{corollary}{Corollary}
\newtheorem{proposition}{Proposition}
\newtheorem{remark}{Remark}
\newtheorem{definition}{Definition}

\DeclareFontFamily{U}{outerjoin}{}
\DeclareFontShape{U}{outerjoin}{m}{n}{ <-> outerjoin10 }{}
\DeclareSymbolFont{outerjoin}{U}{outerjoin}{m}{n}
\DeclareMathSymbol{\louterjoin}{\mathalpha}{outerjoin}{'61}

\def\punto{\hspace*{\fill}\Box}
\def\qed{\hspace*{1pt}\Box}

\newcommand{\evalc}[1]{{\llbracket#1\rrbracket}}

\newcommand{\evalcd}[1]{{\llbracket#1\rrbracket_D}}

\newcommand{\oopt}{{\ \textsc{Opt}\ }}
\newcommand{\oand}{{\ \textsc{And}\ }}
\newcommand{\ounion}{{\ \textsc{Union}\ }}
\newcommand{\ofilter}{{\ \textsc{Filter}\ }}

\newcommand{\set}[2]{\ensuremath{\{$ $#1$ $|$ $#2$ $\}}}
\newcommand{\setone}[1]{\ensuremath{\{#1\}}}

\newcommand{\leftouterjoin}{{\ \louterjoin\ }}

\newcommand{\select}{{\sigma}}
\newcommand{\project}{{\pi}}

\newcommand{\nop}[1]{{}}

\begin{document}

\title{Foundations of SPARQL Query Optimization} 

\author{Michael Schmidt\thanks{The work of this author was funded by Deutsche Forschungsgemeinschaft grant GRK 806/03.}\ \ \ \ \ \ \ \ \ \ \ \ \ \ Michael Meier$^*$\ \ \ \  \ \ \ \ \ \ \ \ \ \ Georg Lausen\\
\\[-0.25cm]
Institut f\"{u}r Informatik\\
Georges-K\"{o}hler-Allee, Geb\"{a}ude 051\\
79110 Freiburg i. Br., Germany\\
\\[-0.25cm]
\{mschmidt, meierm, lausen\}@informatik.uni-freiburg.de
}

\maketitle
\thispagestyle{empty}

\begin{abstract}
The SPARQL query language is a recent W3C standard for processing RDF data,
a format that has been developed to encode information in a machine-readable way.
We investigate the foundations of SPARQL query optimization and (a)~provide
novel complexity results for the SPARQL evaluation problem, showing that the
main source of complexity is operator \textsc{Optional} alone; (b)~propose a
comprehensive set of algebraic query rewriting rules; (c)~present a framework
for constraint-based SPARQL optimization based upon the well-known chase procedure
for Conjunctive Query minimization. In this line, we develop two novel
termination conditions for the chase. They subsume the strongest conditions
known so far and do not increase the complexity of the recognition problem,
thus making a larger class of both Conjunctive and SPARQL queries
amenable to constraint-based optimization. Our results are of immediate
practical interest and might empower any SPARQL query optimizer.
\end{abstract}

%%% sections 

\section{Introduction}
\label{sec:introduction}

The SPARQL Protocol and Query Language is a recent
W3C recommendation that has been developed to extract information from data
encoded using the Resource Description Framework (RDF)~\cite{ghm2004}.
From a technical point of view, RDF databases are collections of
({\it subject},{\it predicate},{\it object}) triples. Each triple encodes the
binary relation {\it predicate} between {\it subject}
and {\it object}, i.e.~represents a single knowledge fact.
Due to their homogeneous structure, RDF databases can be seen 
as labeled directed graphs, where each triple defines an edge from the
{\it subject} to the {\it object} node under label {\it predicate}.
While originally designed to encode knowledge in the Semantic Web in
a machine-readable format, RDF has found its way out of the Semantic
Web community and entered the wider discourse of Computer Science.
Coming along with its application in other areas, such as bio informatics,
data publishing, or data integration, large RDF repositories have been
created (cf.~\cite{swc}). It has repeatedly been observed that the
database community is facing new challenges to cope with the specifics
of the RDF data format~\cite{ammh2007,lgknm2008,shklp2008,wkb2008}.

With SPARQL, the W3C has recommended a declarative query language that allows
to extract data from RDF graphs. SPARQL comes with a powerful graph matching
facility, whose basic construct are so-called triple patterns. During query
evaluation, variables inside these patterns are matched against the RDF input
graph. The solution of the evaluation process is then described by a set of mappings,
where each mapping associates a set of variables with RDF graph components. SPARQL
additionally provides a set of operators (namely \textsc{And},
\textsc{Filter}, \textsc{Optional}, \textsc{Select}, and \textsc{Union}),
which can be used to compose more expressive queries.

One key contribution in this paper is a comprehensive complexity analysis for 
fragments of SPARQL. We follow previous approaches~\cite{pag2006} and use the
complexity of the \textsc{Evaluation} problem as a yardstick: given query $Q$,
data set $D$, and candidate solution $S$ as input, check if $S$ is contained
in the result of evaluating $Q$ on $D$. In~\cite{pag2006} it has been shown that
full SPARQL is \textsc{PSpace}-complete, which is bad news
from a complexity point of view. We show that yet operator
\textsc{Optional} alone makes the \textsc{Evaluation} problem \textsc{PSpace}-hard.
Motivated by this result, we further refine our analysis and prove better
complexity bounds for fragments with restricted nesting depth of
\textsc{Optional} expressions.
%Even more,  we show that the nesting depth of
%\textsc{Optional}-expressions is the source of complexity and present better
%complexity bounds for restricted fragments.

Having established this theoretical background, we turn towards SPARQL query
optimization. The semantics of SPARQL is formally defined on top of a
compact algebra over mapping sets. In the evaluation process, the SPARQL
operators are first translated into algebraic operations, which are then
directly evaluated on the data set.
The SPARQL Algebra (SA) comprises operations such as join, union,
left outer join, difference, projection, and selection, akin to the
operators defined in Relational Algebra (RA). At first glance, there are
many parallels between SA and RA; in fact, the study in~\cite{ag2008}
reveals that SA and RA have exactly the same expressive power.
Though, the technically involved proof in~\cite{ag2008} indicates that a
semantics-preserving SA-to-RA translation is far from being trivial
(cf.~\cite{c2005}). Hence, although both algebras provide similar operators, there
are still very fundamental differences between both. One of the most striking
discrepancies, as also argued in~\cite{pag2006}, is that joins in RA are
rejecting over null-values, but in SA, where the schema is loose
in the sense that mappings may bind an arbitrary set of variables, joins
over unbound variables (essentially the equivalent of RA null-values)
are always accepting.

One direct implication is that not all equivalences that hold in RA also hold
in SA, and vice versa, which calls for a study of SA by its own. In response,
we present an elaborate study of SA in the second part of the paper. 
We survey existing and develop new algebraic equivalences,
covering various SA operators, their interaction, and their relation
to the RA counterparts. When interpreted as rewriting rules, these equivalences
form the theoretical foundations for transferring established RA optimization
techniques, such as filter pushing, into the SPARQL context. Going beyond
the adaption of existing techniques, we also address SPARQL-specific issues,
e.g.~provide rules for simplifying expressions involving (closed-world) negation,
which can be expressed in SPARQL syntax using a combination of
\textsc{Optional} and \textsc{Filter}.

We note that in the past much research effort has been spent in
processing RDF data with traditional systems, such as relational DBMSs
or datalog
engines~\cite{ammh2007,lgknm2008,wkb2008,nw2008,fb2008,shklp2008,p2007},
thus falling back on established optimization strategies. Some of them
(e.g.~\cite{wkb2008,fb2008}) work well in practice, but are limited to
small fragments, such as \textsc{And}-only queries. More complete approaches
(e.g.~\cite{ammh2007}) suffer from performance
bottlenecks for complex queries, often caused by poor optimization results
(cf.~\cite{lgknm2008,shklp2008,shlp2009}).
For instance, \cite{shklp2008} identifies
deficiencies of existing schemes for queries involving negation, a problem
that we tackle in our analysis. This also shows that traditional approaches
are not laid out for the specific challenges that come along with SPARQL
processing and urges the need for a thorough investigation of SA.
%, e.g.~by adapting the SPARQL-specific
%rewriting rules for queries involving negation presented in this paper.

In the final part of the paper we study constraint-based query
optimization in the context of SPARQL, also known as Semantic Query Optimization (SQO). SQO has been applied successfully in other contexts before, such as
Conjunctive Query (CQ) optimization (e.g.,
\cite{bv1984}), relational databases (e.g., \cite{k1981}), and deductive
databases (e.g., \cite{cgm1990}). We demonstrated the prospects of
SQO for SPARQL in~\cite{lms2008}, and in this work we lay the foundations
for a schematic semantic optimization approach. Our SQO scheme builds
upon the Chase \& Backchase (C\&B)
algorithm~\cite{dpt2006}, an extension of the well-known chase
procedure for CQ optimization~\cite{mms1979,bv1984,jk1982}.
One key problem with the chase is that it might not always terminate.
Even worse, it has recently been shown that for an arbitrary set of constraints it
is undecidable if it terminates or not \cite{dnr2008}. There exist, however, sufficient
conditions for the termination of the chase; the best condition known so far is
that of {\it stratified} constraints \cite{dnr2008}. The definition of stratification
uses a former termination condition for the chase, namely {\it weak acyclicity} \cite{fkmp2003}.
In this paper, we present two provably stronger termination conditions, making a larger
class of CQs amenable to the semantic optimization process and generalizing the methods
introduced in \cite{fkmp2003,dnr2008}.

Our first condition, called {\it safety}, strictly subsumes weak acyclicity
and the second, {\it safe restriction}, strictly subsumes stratification.
They do not increase the complexity of the recognition problem, i.e.~safety is
checkable in polynomial time (like weak acyclicity) and safe restriction by a
$\textsc{coNP}$-algorithm (like stratification). We emphasize that our results
immediately carry over to data exchange~\cite{fkmp2003}
and integration~\cite{l2002}, query answering using views~\cite{h2001}, and
the implication problem for constraints. Further, they apply to
the core chase introduced in~\cite{dnr2008} (there it was proven
that the termination of the chase implies termination of the core chase).

In order to optimize SPARQL queries, we translate \textsc{And}-blocks of 
the query into CQs, optimize them using the C\&B-algorithm, and translate the
outcome back into SPARQL. Additionally, we provide optimization rules that go beyond
such simple queries, showing that in some cases \textsc{Optional}- and
\textsc{Filter}-queries can be simplified. With respect to chase termination,
we introduce two alternate SPARQL-to-CQ translation schemes. They differ
w.r.t.~the termination conditions that they exhibit for the subsequent chase,
i.e.~our sufficient chase termination conditions might guarantee termination
for the first but not for the second translation, and vice versa.

Our key contributions can be summarized as follows.

\squishlist
\item We present previously unknown complexity results for fragments of the SPARQL
query language, showing that the main source of complexity
is operator \textsc{Optional} alone. Moreover, we prove there are better
bounds when restricting the nesting depth of \textsc{Optional} expressions.

\item
We summarize existent and establish new equivalences over SPARQL Algebra.
Our extensive study characterizes the algebraic operators and their interaction, and
might empower any SPARQL query optimizer. We also indicate an erratum in~\cite{pag2006}
and discuss its implications. 

\item 
Our novel SQO scheme for SPARQL can be used to optimize \textsc{And}-only queries
under a set of constraints. Further, we provide rules for semantic optimization
of queries involving operators \textsc{Optional} and \textsc{Filter}.

\item We present two novel sufficient termination conditions for the chase,
which strictly generalize previous conditions. This improvement empowers the
practicability  of many important research areas,
like e.g. \cite{fkmp2003,l2002,h2001,dpt2006}.

\squishend

{\bf Structure.} We start with some preliminaries in
Section~\ref{sec:preliminaries} and present the complexity results for
SPARQL fragments in Section~\ref{sec:complexity}. The subsequent discussion
of query optimization divides into algebraic optimization
(Section~\ref{sec:algebra}) and semantic optimization
(Section~\ref{sec:sqo}). The latter discussion is complemented
by the chase termination conditions presented in Section~\ref{sec:chase}.
Finally, Section~\ref{sec:conclusion} contains some closing remarks.

\section{Preliminaries}
\label{sec:preliminaries}

{\bf RDF.} We follow the notation from~\cite{pag2006}. We
consider three disjoint sets $B$ (blank nodes), $I$ (IRIs), and $L$ (literals)
and use the shortcut $BIL$ to denote the union of the sets $B$, $I$, and $L$.
By convention, we indicate literals by quoted strings (e.g.``Joe'', ``30'')
and prefix blank nodes with ``\_:''. An RDF triple 
($v_1$,$v_2$,$v_3$) $\in BI \times\ I\ \times\ BIL$
connects subject $v_1$ through predicate $v_2$ to object~$v_3$. 
An RDF database, also called document, is a finite set of triples.
We refer the interested reader
to~\cite{ghm2004} for an elaborate discussion of RDF. 

{\bf SPARQL Syntax.}
Let $V$ be an infinite set of variables disjoint from $BIL$.
We start with an abstract syntax for SPARQL, where we abbreviate
operator \textsc{Optional} as \textsc{Opt}.

\begin{definition}
\em
We define SPARQL {\it expressions} recursively as follows.
(1)~A triple pattern $t \in BIV \times IV \times BILV$
is an expression. (2)~Let $Q_1$, $Q_2$ be expressions
and $R$ a filter condition. Then $Q_1 \ofilter R$, $Q_1 \ounion Q_2$, $Q_1 \oopt Q_2$, and $Q_1 \oand Q_2$
are expressions.$\punto$
\label{def:expression}
\end{definition}

In the remainder of the paper we restrict our discussion to {\it safe} filter
expressions $Q \ofilter R$, where the variables occurring in $R$ form a subset of
the variables in $Q$. As shown in~\cite{ag2008}, this restriction does not
compromise expressiveness.
%, i.e.~non-safe filters can always be simulated by safe ones.

Next, we define SPARQL {\it queries} on top of expressions.\footnote{We
do not consider the remaining SPARQL query forms \textsc{Ask},
\textsc{Construct}, and \textsc{Describe} in this paper.}

\begin{definition}
\em
Let $Q$ be a SPARQL expression and let $S \subset V$
a finite set of variables. A SPARQL {\it query} is an expression of the form
$\textsc{Select}_S(Q)$.$\punto$
\label{def:query}
\end{definition}

{\bf SPARQL Semantics.} 
A {\em mapping} is a partial function $V \rightarrow BIL$ from a subset of
variables $V$ to RDF terms $BIL$. The domain of a mapping $\mu$, written
$\textit{dom}(\mu)$, is defined as the subset of $V$ for which $\mu$ is defined. 
As a naming convention, we distinguish variables from elements in $BIL$ through
a leading question mark symbol.
Given two mappings $\mu_1$, $\mu_2$, we say $\mu_1$
is {\em compatible} with $\mu_2$ if $\mu_1(?x) = \mu_2(?x)$ for all
$?x \in \textit{dom}(\mu_1) \cap \textit{dom}(\mu_2)$.
We write $\mu_1\sim\mu_2$ if $\mu_1$ and $\mu_2$ are compatible,
and $\mu_1\not\sim\mu_2$ otherwise. Further, we write ${\it vars}(t)$ to denote all
variables in triple pattern $t$ and by $\mu(t)$ we denote the triple pattern
obtained when replacing all variables $?x \in {\it dom}(\mu) \cap {\it vars}(t)$
in $t$ by $\mu(?x)$.

Given variables $?x$, $?y$ and constants $c$, $d$, a filter condition
$R$ is either an atomic filter condition of the form
$\textit{bound}(?x)$ (abbreviated as $\textit{bnd}(?x)$),
$?x=c$, $?x=?y$, or a combination of atomic conditions using connectives
$\neg$, $\land$, $\lor$. Condition $\textit{bnd}(?x)$
applied to a mapping set $\Omega$ returns all mappings in $\Omega$ for
which $?x$ is bound, 
i.e.~$\{ \mu \in \Omega \mid ?x \in \textit{dom}(\mu) \}$. The conditions
$?x=c$ and $?x= ?y$ are equality checks, comparing the values of $?x$ with
$c$ and $?y$, respectively. These checks fail whenever one of the variables
is not bound. We write $\mu \models R$ if
mapping $\mu$ satisfies filter condition $R$ (see Definition~\ref{def:filter}
in Appendix~\ref{app:equivalences56} for a formal definition).
The semantics of SPARQL is then formally defined using a compact
algebra over mapping sets (cf.~\cite{pag2006}). The definition of the
algebraic operators join~$\Join$, union $\cup$, set minus $\setminus$,
left outer join $\leftouterjoin$, projection $\project$, and
selection $\select$ is given below.

\begin{definition}
\em
Let $\Omega$, $\Omega_l$, $\Omega_r$ denote mapping sets, $R$ a filter condition,
and $S \subset V$ a finite set of variables. We define the algebraic operations
$\Join$, $\cup$, $\setminus$, $\leftouterjoin$, $\pi$, and $\sigma$:

\begin{center}
\begin{tabbing}
x \= xxxxxxl \= xxl \= \kill
\>$\Omega_l \Join \Omega_r$\>${:=}$\>$\{ \mu_l \cup \mu_r \mid \mu_l \in \Omega_l, \mu_r \in \Omega_r: \mu_l\sim\mu_r\}$\\
\\[-0.33cm]
\>$\Omega_l \cup \Omega_r$\>${:=}$\>$\{ \mu \mid \mu \in \Omega_l\ \textit{or}\ \mu \in \Omega_r \}$\\
\\[-0.33cm]
\>$\Omega_l \setminus \Omega_r$\>${:=}$\>$\{ \mu_l \in \Omega_l \mid \textit{for all}\ \mu_r \in \Omega_r: \mu_l\not\sim\mu_r\}$\\
\\[-0.33cm]
\>$\Omega_l \leftouterjoin \Omega_r$\>${:=}$\>$(\Omega_l \Join \Omega_r)\ \cup\ (\Omega_l \setminus \Omega_r)$\\
\\[-0.33cm]
\>$\project_S(\Omega)$\>${:=}$\>$\{ \mu_1 \mid \exists \mu_2: \mu_1 \cup \mu_2 \in \Omega \land {\it dom}(\mu_1) \subseteq S$\\
\>\>\>$\ \ \ \ \ \ \ \ \  \land\ {\it dom}(\mu_2) \cap S = \emptyset\}$\\
\\[-0.3cm]
\>$\select_R(\Omega)$\>${:=}$\>$\{ \mu \in \Omega \mid \mu \models R\}$
\end{tabbing}
\vspace{-0.38cm}
$\punto$
\vspace{0.1cm}
\end{center}
\end{definition}

We follow the compositional, set-based semantics proposed in~\cite{pag2006}
and define the result of evaluating SPARQL query $Q$ on document $D$ using operator
$\evalc{.}_D$ defined below.

\begin{definition}
\em
Let $t$ be a triple pattern, $Q_1$, $Q_2$ SPARQL expressions, $R$ a filter
condition, and $S \subset V$ a finite set of variables. The semantics of
SPARQL evaluation over document $D$ is defined as follows.

\begin{center}
\begin{tabbing}
x \= lxxxxxxxxxxxxxxxl \= xxx \= \kill
\>$\evalcd{t}\ {:=}\ \{ \mu \mid \textit{dom}(\mu)=\textit{vars}(t)\ \textit{and}\ \mu(t) \in D\}$\\
\\[-0.33cm]
\>$\evalcd{Q_1 \oand Q_2}$\>${:=}$\>$\evalcd{Q_1} \Join \evalcd{Q_2}$\\
\\[-0.33cm]
\>$\evalcd{Q_1 \oopt Q_2}$\>${:=}$\>$\evalcd{Q_1} \leftouterjoin \evalcd{Q_2}$\\
\\[-0.33cm]
\>$\evalcd{Q_1 \ounion Q_2}$\>${:=}$\>$\evalcd{Q_1} \cup \evalcd{Q_2}$\\
\\[-0.33cm]
\>$\evalcd{Q_1 \ofilter R}$\>${:=}$\>$\select_R(\evalcd{Q_1})$\\
\\[-0.33cm]
\>$\evalcd{\textsc{Select}_S(Q_1)}$\>${:=}$\>$\project_S(\evalcd{Q_1})$
\end{tabbing}
\vspace{-0.34cm}
$\punto$
\end{center}
\end{definition}

Finally, we extend the definition of function {\it vars}. Let $Q$ be a SPARQL expression,
$A$ a SPARQL Algebra expression, and $R$ a filter condition. By ${\it vars}(A)$, ${\it vars}(Q)$,
and ${\it vars}(R)$ we denote the set of variables in $A$, $Q$, and $R$, respectively.
Further, we define function ${\it safeVars}(A)$, which denotes the subset of
variables in ${\it vars}(A)$ 
that are inevitably bound when evaluating $A$ on any document $D$.

\begin{definition}
\em
Let $A$ be a SPARQL Algebra expression, $S \subset V$ a finite set of variables,
and $R$ a filter condition. We define function ${\it safeVars}(A)$ recursively
on the structure of expression $A$ as follows.

\begin{tabbing}
x \= xxxxxxxxxxxxxxxx \= \kill
\>${\it safeVars}(\evalcd{t})$\>${:=}\ {\it vars}(t)$\\
\\[-0.3cm]
\>${\it safeVars}(A_1 \Join A_2)$\>${:=}\ {\it safeVars}(A_1) \cup {\it safeVars}(A_2)$\\
\\[-0.3cm]
\>${\it safeVars}(A_1 \cup A_2)$\>${:=}\ {\it safeVars}(A_1) \cap {\it safeVars}(A_2)$\\
\\[-0.3cm]
\>${\it safeVars}(A_1 \setminus A_2)$\>${:=}\ {\it safeVars}(A_1)$\\
\\[-0.3cm]
\>${\it safeVars}(A_1 \leftouterjoin A_2)$\>${:=}\ {\it safeVars}(A_1)$\\
\\[-0.3cm]
\>${\it safeVars}(\pi_S({A_1}))$\>${:=}\ {\it safeVars(A_1)} \cap S$\\
\\[-0.3cm]
\>${\it safeVars}(\sigma_R({A_1}))$\>${:=}\ {\it safeVars(A_1)}$
\end{tabbing}
\vspace{-0.58cm}
$\punto$
\label{def:safevars}
\end{definition}

%\begin{example} 
%\em
%Consider the following query, which extracts the
%names of all $30$ years-old persons (variable ?n) and, optionally,
%their email address (variable ?e).
%
%\begin{tabbing}
%x \= xxxx \= x \= x \= x \= \kill
%\>Q =\>$\textsc{Select}_{\{?n,?e\}}(\textsc{Filter}_{?a=30}$(\\
%\>\>\>$(((?p,type,Person) \oand (?p,name,?n))$\\
%\>\>\>\>$\oand (?p,age,?a)) \oopt (?p,email,?e)))$
%\end{tabbing}
%
%It is left to the reader to verify that, for database
%
%\begin{tabbing}
%x \= xxxl \= x \= \kill
%\>$D=$\>\{\>$(\_:$$P1,type,Person), (\_:$$P1,name,$``Joe''$),$\\
%\>\>\>$(\_:$$P1,age,$``30''$), (\_:$$P1,email,$``joe@tl.d''$),$\\
%\>\>\>$(\_:$$P2,type,$``Person''$), (\_:$$P2, age, $``31''$)\ \}$,
%\end{tabbing}
%
%we have $\evalcd{Q} = \{\{?n \mapsto $``Joe''$,?e \mapsto $``joe@tl.d''$\}\}$.$\punto$
%\label{ex:translationandsemantics}
%\end{example}

{\bf Relational Databases, Constraints and Chase.} We assume that the reader is familiar
with first-order logic and relational databases.  We denote by $dom(I)$ the domain of
the relational database instance $I$, i.e.~the set of constants and null values that occur
in $I$. The constraints we consider, are tuple-generating dependencies (TGD) and
equality-generating dependencies (EGD). TGDs have the form
$\forall \overline{x}(\varphi(\overline{x}) \rightarrow \exists \overline{y} \psi(\overline{x},\overline{y}))$
and EGDs have the form $\forall \overline{x}(\varphi(\overline{x}) \rightarrow x_i=x_j)$.
A more exact definition of these types of constraints can be found in Appendix~\ref{app:defs-chase}.
In the rest of the paper $\Sigma$ stands for a fixed set of TGDs and EGDs. If an instance $I$ is not a model of some constraint $\alpha$, then we write $I \nvDash \alpha$.

We now introduce the chase as defined in~\cite{dnr2008}. 
A chase step $I \stackrel{\alpha, \overline{a}}{\rightarrow} J$ takes a relational database
instance $I$ such that $I \nvDash \alpha(\overline{a})$ and adds tuples (in case of TGDs)
or collapses some elements (in case of EGDs) such that the resulting relational database
$J$ is a model of $\alpha(\overline{a})$. If $J$ was obtained from $I$ in that kind,
we sometimes also write $I\overline{a} \oplus C_{\alpha}$ instead of $J$. A chase
sequence is a sequence of relational database instances $I_0, I_1,...$ such that
$I_{s+1}$ is obtained from $I_s$ by a chase step. A chase sequence $I_0,...,I_n$ is
terminating if $I_n \models \Sigma$. In this case, we set $I^{\Sigma} := I_n$ as
the result ($I^{\Sigma}$ is defined only unique up to homomorphic
equivalence, but this will suffice). Otherwise, $I^{\Sigma}$ is undefined.
$I^{\Sigma}$ is also undefined in case the chase fails.
More details can be found in Appendix~\ref{app:defs-chase}.
The chase does not always terminate and there has been different
work on sufficient termination conditions. In~\cite{fkmp2003} the
following condition, based on the notion of dependency graph, was introduced.
The dependency graph
$\mbox{dep}(\Sigma):=(V,E)$ of a set of constraints $\Sigma$ is the directed graph
defined as follows. $V$ is the set of positions that occur in $\Sigma$.
There are two kind of edges in $E$. Add them as follows: for every TGD 
$\forall \overline{x} (\varphi(\overline{x}) \rightarrow \exists \overline{y} \psi(\overline{x},\overline{y})) \in \Sigma$
 and for every $x$ in $\overline{x}$ that occurs in $\psi$ and every occurrence of $x$
in $\varphi$ in position $\pi_1$

\squishlist
	\item for every occurrence of $x$ in $\psi$ in position $\pi_2$, add an edge $\pi_1 \rightarrow \pi_2$ (if it does not already exist).
	\item for every existentially quantified variable $y$ and for every occurrence of $y$ in a position $\pi_2$, add a special edge $\pi_1 \stackrel{*}{\rightarrow} \pi_2$ (if it does not already exist).
\squishend
A set $\Sigma$ of TGDs and EGDs is called \textit{weakly acyclic} iff $\mbox{dep}(\Sigma)$ has no cycles  through a special edge. In \cite{dnr2008} weak acyclicity was lifted to stratification. Given two TGDs or EGDs $\alpha = \forall \overline{x_1}\varphi, \beta = \forall \overline{x_2}\psi$, we define $\alpha \prec \beta$ (meaning that firing $\alpha$ may cause $\beta$ to fire) iff there exist relational database instances $I, J$ and $\overline{a} \in dom(I)$, $\overline{b} \in dom(J)$ s.t.

\squishlist
	\item  $I \nvDash \psi(\overline{b})$ (possibly $\overline{b}$ is not in $dom(I)$),
	\item $I \stackrel{\alpha, \overline{a}}{\longrightarrow} J$, and
	\item $J \nvDash \psi(\overline{b})$.
\squishend

The chase graph $G(\Sigma)=(\Sigma,E)$ of a set of constraints $\Sigma$ contains a
directed edge $(\alpha,\beta)$ between two constraints iff $\alpha \prec \beta$. We
call $\Sigma$ stratified iff the set of constraints in every cycle of $G(\Sigma)$ are
weakly acyclic. It is immediate that weak acyclicity implies stratification; further,
it was proven in~\cite{dnr2008} that the chase always terminates for stratified constraint sets.
%This is also reviewed in Appendix \ref{app:defs-chase}.

A Conjunctive Query (CQ) is an expression of the form $ans(\overline{x}) \leftarrow \varphi(\overline{x},\overline{z})$, where $\varphi$ is a CQ of relational atoms and $\overline{x}, \overline{z}$ are tuples of variables and constants. Every variable in $\overline{x}$ must also occur in $\varphi$. The semantics of such a query on a database instance $I$ is $q(I) := \set{\overline{a}}{I \models \exists \overline{z} \varphi(\overline{a},\overline{z})}$.

Let $q, q'$ be CQs and $\Sigma$ be a set of constraints. We write $q \sqsubseteq_{\Sigma} q'$
if for all database instances $I$ such that $I \models \Sigma$ it holds
that $q(I) \subseteq q'(I)$ and say that $q$ and $q'$ are $\Sigma$-equivalent
($q \equiv_\Sigma q'$) if $q \sqsubseteq_{\Sigma} q'$ and $q' \sqsubseteq_{\Sigma} q$.
In \cite{dpt2006} an algorithm was presented that, given $q$ and $\Sigma$,
lists all $\Sigma$-equivalent minimal (with respect to the number of atoms in the body) rewritings (up to isomorphism) of $q$. This algorithm,
called Chase \& Backchase, uses the chase and therefore does not necessarily
terminate. We denote its output by $\textit{cb}_{\Sigma}(q)$ (if it terminates).

{\bf General mathematical notation.} The natural numbers $\mathbb{N}$ do not include $0$;
$\mathbb{N}_0$ is used as a shortcut for $\mathbb{N} \cup \{ 0 \}$. 
For $n \in \mathbb{N}$, we denote by $[n]$ the set $\setone{1,...,n}$.
Further, for a set $M$, we denote by $2^M$ its powerset.

\section{SPARQL Complexity}
\label{sec:complexity}
\vspace{-.2cm}

We introduce operator shortcuts $\cal{A} := \textsc{And}$,
$\cal{F} := \textsc{Filter}$, $\cal{O} := \textsc{Opt}$, $\cal{U} := \textsc{Union}$, 
and denote the class of expressions that can be constructed using a set of operators
by concatenating their shortcuts. Further, by $\cal{E}$ we denote the whole class
of SPARQL expressions, i.e.~$\cal{E} := \cal{AFOU}$. The terms {\it class} and
{\it fragment} are used interchangeably.

We first present a complete complexity study for all possible expression
classes, which complements the study in~\cite{pag2006}. We assume the reader
to be familiar with basics of complexity theory, yet summarize the
background in Appendix \ref{app:complexityback}, to be self-contained. 
We follow~\cite{pag2006} and take the combined complexity
of the \textsc{Evaluation} problem as a yardstick:

\vspace{-0.1cm}
\begin{tabbing}
;\= xxxxxxxxxxxx \= \kill
\>\textsc{Evaluation}: given a mapping $\mu$, a document $D$, and an\\
\>\>expression/query $Q$ as input: is $\mu \in \evalcd{Q}$?
\end{tabbing}
\vspace{-0.1cm}

The theorem below summarizes previous results from~\cite{pag2006}.

\begin{theorem} 
\em
~\cite{pag2006} The \textsc{Evaluation} problem is

\squishlist
\item in \textsc{PTime} for class $\cal{AF}$; membership in \textsc{PTime}
for classes $\cal{A}$ and $\cal{F}$ follows immediately,
\item \textsc{NP}-complete for class $\cal{AFU}$, and
\item \textsc{PSpace}-complete for classes $\cal{AOU}$ and $\cal{E}$.
\squishend

\vspace{-0.58cm}
$\punto$
\label{th:pag}
\end{theorem}

Our first goal is to establish a more precise characterization of the \textsc{Union}
operator. As also noted in~\cite{pag2006}, its design was subject to
controversial discussions in the SPARQL working
group\footnote{See the discussion of disjunction in Section~6.1 in \url{http://www.w3.org/TR/2005/WD-rdf-sparql-query-20050217/}.}, and we pursue the goal to improve 
the understanding of the operator and its relation to others, beyond the known
\textsc{NP}-completeness result for class $\cal{AFU}$. The following theorem gives the
results for all missing \textsc{Opt}-free fragments.

\begin{theorem}
\em
The \textsc{Evaluation} problem is

\squishlist
\item in \textsc{PTime} for classes $\cal{U}$ and $\cal{FU}$, and
\item \textsc{NP}-complete for class $\cal{AU}$.
\squishend

\vspace{-0.6cm}
$\punto$
\label{th:andunion}
\end{theorem}

The hardness part of the \textsc{NP}-completeness proof for fragment $\cal{AU}$
is a reduction from \textsc{SetCover}. The interested reader will find details
and other technical results of this section in Appendix~\ref{app:complexity}.
Theorems~\ref{th:pag} and~\ref{th:andunion} clarify that the source of complexity
in \textsc{Opt}-free fragments is the
combination of \textsc{And} and \textsc{Union}. In particular, adding or removing
\textsc{Filter}-expressions in no case affects the complexity.

We now turn towards an investigation of the complexity of operator
\textsc{Opt} and its interaction with other operators. 
The \textsc{PSpace}-completeness results for classes 
$\cal{AOU}$ and $\cal{AFOU}$ stated in~Theorem~\ref{th:pag} give only
partial answers to the questions. One of the main results in this
section is the following.

\begin{theorem}
\em
\textsc{Evaluation} is \textsc{PSpace}-complete for $\cal{O}$.$\punto$
\label{th:opt}
\end{theorem}

\vspace{-.5cm}
This result shows that already operator \textsc{Opt}
alone makes the \textsc{Evaluation} problem really hard. Even more, it upgrades 
the claim in~\cite{pag2006} that ``the main source of complexity
in SPARQL comes from the combination of \textsc{Union} and \textsc{Opt} operators'',
by showing that \textsc{Union} (and \textsc{And}) are not necessary to obtain
\textsc{PSpace}-hardness. The intuition of this result is that the algebra operator
$\leftouterjoin$ (which is the algebraic counterpart of operator \textsc{Opt})
is defined using operators $\Join$, $\cup$, and $\setminus$; the mix of these
algebraic operations compensates for missing \textsc{And} and \textsc{Union} operators
at syntax level.
%To prove Theorem~\ref{th:opt} we first show that the (\textsc{PSpace}-complete)
%\textsc{QBF} problem can be encoded using \textsc{And} and \textsc{Opt}, and
%then present a rewriting rule that allows us to rewrite \textsc{And}- through
%\textsc{Opt}-expression.
The corollary below follows from Theorems~\ref{th:pag} and~\ref{th:opt}
and makes the complexity study of the expression fragments complete.

\begin{corollary} 
\em
The \textsc{Evaluation} problem for any expression fragment involving
\textsc{Opt} is \textsc{PSpace}-complete.$\punto$
\label{cor:optpspace}
\end{corollary}

\vspace{-.1cm}
Due to the high complexity of \textsc{Opt}, an interesting question is
whether we can find natural syntactic conditions that lower the complexity
of fragments involving \textsc{Opt}. 
%We observe that all hardness proofs for
%such fragments rely on a nesting of \textsc{Opt} expressions and propose a
In fact, a restriction of the nesting depth of \textsc{Opt} expressions
constitutes such a condition.
We define the \textsc{Opt}-rank {\it r} of an expression as its deepest
nesting of \textsc{Opt} expressions: for triple pattern $t$,
expressions $Q$, and condition $R$, we define ${\it r}(Q)$ recursively on the
structure of $Q$ as {\it r}$(t):=0$,
{\it r}$(Q_1 \ofilter R):={\it r}(Q_1)$,
{\it r}$(Q_1 \oand Q_2)$=${\it r}(Q_1 \ounion Q_2):={\it max}({\it r}(Q_1),{\it r}(Q_2))$, and
{\it r}$(Q_1 \oopt Q_2):= {\it max}({\it r}(Q_1),{\it r}(Q_2)) + 1$.

By $\cal{E}$$_{\leq n}$ we denote the class of expressions $Q \in \cal{E}$
with $r(Q) \leq n$. The following theorem shows that, when restricting the
\textsc{Opt}-rank of expressions, the \textsc{Evaluation} problem falls into
a class in the polynomial hierarchy.

\begin{theorem}
\em
For any $n \in \mathbb{N}_0$, the \textsc{Evaluation} problem is $\Sigma^P_{n+1}$-complete
for the SPARQL fragment $\cal{E}$$_{\leq n}$.$\punto$
\label{th:ph}
\end{theorem}

\vspace{-.1cm}
Observe that \textsc{Evaluation} for class $\cal{E}$$_{\leq 0}$ is complete for
$\Sigma^P_1$=\textsc{NP}, thus obtaining the result for \textsc{Opt}-free
expressions (cf.~Theorem~\ref{th:pag}). With increasing
nesting-depth of \textsc{Opt} expressions we climb up the polynomial hierarchy (PH).
This is reminiscent of the \textsc{Validity}-problem for quantified
boolean formulae, where the number of
quantifier alternations fixes the complexity class in the PH. In fact, the
hardness proof (see~Appendix~\ref{app:complexityopt}) makes these
similarities explicit.

We finally extend our study to SPARQL queries, i.e.~fragments involving top-level
projection in the form of a \textsc{Select}-operator (see Def.~\ref{def:query}).
We extend the notation for classes as follows.
Let $F$ be an expression fragment. We denote by $F_+$ the class
of queries of the form $\textsc{Select}_S(Q)$, where $S \subset V$ is a finite set
of variables and $Q \in F$ is an expression. The next theorem shows that
we obtain (top-level) projection for free in fragments that are at least \textsc{NP}-complete.

\begin{theorem}
\em
Let $C$ be a complexity class and $F$ a class of expressions. If
\textsc{Evaluation} is \textsc{C}-complete for $F$ and
$\textsc{C} \supseteq \textsc{NP}$ then \textsc{Evaluation} is
also $\textsc{C}$-complete for $F_+$.$\punto$
\label{th:select1}
\end{theorem}

\vspace{-.55cm}
In combination with Corollary~\ref{cor:optpspace} we immediately obtain
\textsc{PSpace}-completeness for query classes involving operator \textsc{Opt}.
Similarly, all \textsc{Opt}-free query fragments involving both
\textsc{And} and \textsc{Union} are \textsc{NP}-complete.
We conclude our complexity analysis with the following theorem, which shows
that top-level projection makes the \textsc{Evaluation} problem for
\textsc{And}-only expressions considerably harder.

\begin{theorem}
\textsc{Evaluation} is \textsc{NP}-complete for $\cal{A}$$_+$.$\punto$
\label{th:select2}
\end{theorem}

\section{SPARQL Algebra}
\label{sec:algebra}

\begin{figure*}

\begin{boxedminipage}{17.8cm}

\begin{tabular}{lll}

{\small
\begin{minipage}{6cm}
\verb!I. Idempotence and Inverse!
\vspace{-.15cm}
\begin{tabbing}
\= xxxxxxxlll \= xl \= xxxxxxxxxxxxxxxxxxxxxxxxxxxx \= \kill
\>$A \cup A$\>$\equiv$\>$A$\>{\small \it (UIdem)}\\
\>$A^- \Join A^-$\>$\equiv$\>$A^-$\>{\small \it (JIdem)}\\
\>$A^- \leftouterjoin A^-$\>$\equiv$\>$A^-$\>{\small \it (LIdem)}\\
\>$A \setminus A$\>$\equiv$\>$\emptyset$\>{\small \it (Inv)}
\end{tabbing}

\vspace{-.05cm}
\verb!II. Associativity!
\vspace{-.15cm}
\begin{tabbing}
\= xxxxxxxxxxxxlll \= xl \= xxxxxxxxxxxxxxxxxxxxxxx \= \kill
\>$(A_1 \cup A_2) \cup A_3$\>$\equiv$\>$A_1 \cup (A_2 \cup A_3)$\>{\small \it (UAss)}\\
\>$(A_1 \Join A_2) \Join A_3$\>$\equiv$\>$A_1 \Join (A_2 \Join A_3)$\>{\small \it (JAss)}
\end{tabbing}

\vspace{-.05cm}
\verb!III. Commutativity!
\vspace{-.15cm}
\begin{tabbing}
\= xxxxxxxl \= xl \= llxxxxxxxxxxxxxxxxxxxxxxxxxxxx \= \kill
\>$A_1 \cup A_2$\>$\equiv$\>$A_2 \cup A_1$\>{\small \it (UComm)}\\
\>$A_1 \Join A_2$\>$\equiv$\>$A_2 \Join A_1$\>{\small \it (JComm)}
\end{tabbing}

\vspace{-.05cm}
\verb!IV. Distributivity!
\vspace{-.15cm}
\begin{tabbing}
\= xxxxxxxxxxxxxl \= xl \= llxxxxxxxxxxxxxxxxxxxxxx \= \kill
\>$(A_1 \cup A_2) \Join A_3$\>$\equiv$\>$(A_1 \Join A_3) \cup (A_2 \Join A_3)$\>{\small \it (JUDistR)}\\
\>$A_1 \Join (A_2 \cup A_3)$\>$\equiv$\>$(A_1 \Join A_2) \cup (A_1 \Join A_3)$\>{\small \it (JUDistL)}\\
\>$(A_1 \cup A_2) \setminus A_3$\>$\equiv$\>$(A_1 \setminus A_3) \cup (A_2 \setminus A_3)$\>{\small \it (MUDistR)}\\
\>$(A_1 \cup A_2) \leftouterjoin A_3$\>$\equiv$\>$(A_1 \leftouterjoin A_3) \cup (A_2 \leftouterjoin A_3)$\>{\small \it (LUDistR)}
\end{tabbing}
\end{minipage}
}

&
\
&

{\small
\begin{minipage}{8cm}
\verb!V. Filter Decomposition and Elimination!
\vspace{-.15cm}
\begin{tabbing}
x \= xxxxxxxxxxxlll \= xl \= llxxxxxxxxxxxxxxxxxxxxxxxxl \= \kill
\>$\select_R(A_1 \cup A_2)$\>$\equiv$\>$\select_R(A_1) \cup \select_R(A_2)$\>{\small \it (SUPush)}\\
\>$\select_{R_1 \land R_2}(A)$\>$\equiv$\>$\select_{R_1}(\select_{R_2}(A))$\>{\small \it (SDecompI)}\\
\>$\select_{R_1 \lor R_2}(A)$\>$\equiv$\>$\select_{R_1}(A) \cup \select_{R_2}(A)$\>{\small \it (SDecompII)}\\
\>$\select_{R_1}(\select_{R_2}(A))$\>$\equiv$\>$\select_{R_2}(\select_{R_1}(A))$\>{\small \it (SReord)}\\
\\[-0.2cm]
\>$\select_{{\it bnd}(?x)}(A_1)$\>$\equiv$\>$A_1$, if $?x \in {\it safeVars}(A_1)$\>{\small \it (BndI)}\\
\>$\select_{{\it bnd}(?x)}(A_1)$\>$\equiv$\>$\emptyset$, if $?x \not \in {\it vars}(A_1)$\>{\small \it (BndII)}\\
\>$\select_{\neg{\it bnd}(?x)}(A_1)$\>$\equiv$\>$\emptyset$, if $?x \in {\it safeVars}(A_1)$\>{\small \it (BndIII)}\\
\>$\select_{\neg{\it bnd}(?x)}(A_1)$\>$\equiv$\>$A_1$, if $?x \not \in {\it vars}(A_1)$\>{\small \it (BndIV)}\\
\>\\[-0.15cm]
If $?x \in {\it safeVars}(A_2) \setminus {\it vars}(A_1)$, then\\
\>$\select_{{\it bnd}(?x)}(A_1 \leftouterjoin A_2) \equiv A_1 \Join A_2$\>\>\>{\small \it (BndV)}
\end{tabbing}

\vspace{.1cm}
\verb!VI. Filter Pushing!

\noindent
The following rules hold if ${\it vars}(R) \subseteq {\it safeVars(A_1)}$.
\vspace{-0.1cm}
\begin{tabbing}
x \= xxxxxxxxxxxlll \= xl \= llxxxxxxxxxxxxxxxxxxxxxxxxxl \= \kill
\>$\select_R(A_1 \Join A_2)$\>$\equiv$\>$\select_R(A_1) \Join A_2$\>{\small \it (SJPush)}\\
\>$\select_R(A_1 \setminus A_2)$\>$\equiv$\>$\select_R(A_1) \setminus A_2$\>{\small \it (SMPush)}\\

\>$\select_R(A_1 \leftouterjoin A_2)$\>$\equiv$\>$\select_R(A_1) \leftouterjoin A_2$\>{\small \it (SLPush)}
\end{tabbing}
\end{minipage}
}

\end{tabular}
\end{boxedminipage}

\vspace{-.2cm}
\caption{SA equivalences for $\mathds{A}$-expr.~$A$, $A_1$, $A_2$, $A_3$; $\mathds{A}^-$-expr. $A^-$; filter condition $R$; variable $?x$.}
\label{fig:rewritings}
\vspace{-.15cm}
\end{figure*}

We next present a rich set of algebraic equivalences for SPARQL Algebra.
In the interest of a complete survey we include equivalences
that have been stated before in~\cite{pag2006}.\footnote{Most 
equivalences in~\cite{pag2006} were established at the syntactic level.
In summary, rule {\it (MJ)} in Proposition~\ref{prop:minus} and about
half of the equivalences in Figure~\ref{fig:rewritings} are borrowed
from~\cite{pag2006}. We indicate these rules in the proofs in
Appendix~\ref{app:rewritings}.}
Our main contributions in
this section are (a)~a systematic extension of previous rewriting rules,
(b)~a correction of an erratum in~\cite{pag2006}, and (c)~the development
and discussion of rewriting rules for SPARQL expressions involving negation.

We focus on two fragments of SPARQL algebra, namely the full class of algebra
expressions $\mathds{A}$ (i.e., algebra expressions build using operators
$\cup$, $\Join$, $\setminus$, $\leftouterjoin$, $\pi$, and $\sigma$)
and the union- and projection-free expressions $\mathds{A}^-$ 
(build using only operator $\Join$, $\setminus$, $\leftouterjoin$, and $\sigma$).
We start with a property that separates $\mathds{A}^-$ from
$\mathds{A}$, called {\it incompatibility property}.\footnote{Lemma~2
in~\cite{pag2006} also builds on this observation.}

\begin{proposition}
\em
Let $\Omega$ be the mapping set obtained from evaluating an $\mathds{A}^-$-expression
on any document $D$. All pairs of distinct mappings in $\Omega$ are incompatible.$\punto$
\label{prop:incomp}
\end{proposition}

Figure~\ref{fig:rewritings}(\verb$I-IV$) surveys rewriting rules that hold with respect
to common algebraic laws (we write $A \equiv B$ if $A$ is equivalent to $B$ on
any document $D$). Group \verb!I! contains results obtained when combining an expression
with itself
using the different operators. It is interesting to see that
{\it (JIdem)} and {\it (LIdem)} hold only for fragment $\mathds{A}^-$;
in fact, it is the incompatibility property that makes the equivalences valid.
The associativity and commutativity rules were introduced in~\cite{pag2006}
and we list them for completeness. Most interesting is distributivity.
We observe that $\Join$, $\setminus$, $\leftouterjoin$ are right-distributive
over $\cup$, and $\Join$ is also left-distributive over $\cup$.
The listing in Figure~\ref{fig:rewritings} is complete in the following sense:

\begin{lemma}
\em
Let $O_1 := \{\ \Join, \setminus, \leftouterjoin\}$ and $O_2 := O_1 \cup \{ \cup \}$. 

\squishlist
\item The two equivalences {\it (JIdem)} and {\it (LIdem)} in general do not hold for fragments
larger than $\mathds{A}$$^-$.
\item Associativity and Commutativity do not hold for operators $\setminus$ and
$\leftouterjoin$.
\item Neither $\setminus$ nor $\leftouterjoin$ are left-distributive
over $\cup$.
\item Let $o_1 \in O_1$, $o_2 \in O_2$, and $o_1 \not = o_2$. Then~$o_2$ is
neither left- nor right-commutative over $o_1$.
\squishend

\vspace{-0.48cm}
$\punto$
\label{lemma:completeness}
\end{lemma}

Cases (3) and (4) rule out distributivity for all operator combinations
different from those listed in Figure~\ref{fig:rewritings}. This result
implies that Proposition 1(3) in~\cite{pag2006} is wrong:

\begin{example} 
\em
We show that the SPARQL equivalence

{\small
\begin{center}
$A_1 \oopt (A_2 \ounion A_3) \equiv (A_1 \oopt A_2) \ounion (A_1 \oopt A_3)$
\end{center}
}

stated in Proposition~1(3) in~\cite{pag2006} does not hold in the general case.
We choose database $D$=$\{(0,c,1)\}$ and set $A_1$=$(0,c,?a)$,
$A_2$=$(?a,c,1)$, and $A_3$=$(0,c,?b)$. Then 
$\evalcd{A_1 \oopt (A_2 \ounion A_3)} = \{ \{ ?a \mapsto 1, ?b \mapsto 1\} \}$, but
$\evalcd{(A_1 \oopt A_2) \ounion (A_1 \oopt A_3)}$ evaluates to
$\{ \{ ?a \mapsto 1 \}, \{ ?a \mapsto 1, ?b \mapsto 1 \} \}$. The results differ.$\punto$
\end{example}

\begin{remark}
\em
This erratum calls the existence of the union normal form stated in
Proposition~$1$ in~\cite{pag2006} into question, as it builds upon the invalid
equivalence. We actually do not see how to fix or compensate for this rule, so it
remains an open question if such a union normal form exists or not.
The non-existence would put different results into perspective, since --
based on the claim that \textsc{Union} can always be pulled to the top --
the authors restrict the subsequent discussion to \textsc{Union}-free expressions.
For instance, results on well-defined patterns, normalization, and equivalence
between compositional and operational semantics are applicable only to
queries that can be brought into union normal form. Arguably, this
class may comprise most of the SPARQL queries that arise in practice (queries without
union or with union only at the top-level also constitute very
frequent patterns in other query languages, such as SQL).
Still, a careful reinvestigation would be necessary to extend the results to
queries beyond that class.$\punto$
\end{remark}

Figure~\ref{fig:rewritings}(\verb!V-VI!) presents rules for decomposing, eliminating,
and rearranging (parts of) filter conditions.  In combination with rewriting rules
\verb!I-IV! they provide a powerful framework for manipulating filter expressions in
the style of RA filter rewriting and pushing.
Most interesting is the use of {\it safeVars} as a sufficient
precondition for {\it (SJPush)}, {\it (SMPush)}, and {\it (SLPush)}.\footnote{A
variant of rule {\it (SJPush)}, restricted to \textsc{And}-only
queries, has been stated (at syntax level) in Lemma~1(2) in~\cite{pag2006}.}
The need for this precondition arises from the fact that joins over mappings
are accepting for unbound variables. In RA,
where joins over null values are rejecting, the situation is less complicated.
For instance, given two RA relations $A_1$, $A_2$ and a (relational) filter $R$, 
{\it (SJPush)} is applicable whenever the schema of $A_1$ contains all attributes
in $R$. We conclude this discussion with the remark that,
for smaller fragments of SPARQL conditions, weaker preconditions for the rules in
group \verb!VI! exist. For instance, if $R = e_1 \land \dots \land e_n$ is a
conjunction of atomic equalities $e_1, \dots, e_n$, then the
equivalences in group \verb!VI! follow from the (weaker) condition
${\it vars}(R) \subseteq {\it vars}(A) \land {\it vars}(B) \cap {\it vars}(R) \subseteq {\it safeVars}(A)$.

We pass on a detailed discussion of operator $\pi$, also 
because -- when translating SPARQL queries into algebra expression -- this operator
appears only at the top-level. Still, we emphasize that also for this operator
rewriting rules exist, e.g.~allowing to project away unneeded variables at an
early stage. Instead, in the remainder of this section we will present a thorough
discussion of operator~$\setminus$. The latter, in contrast to the other algebraic
operations, has no direct counterpart at the syntactic level. This complicates
the encoding of queries involving negation and,
as we will see, poses specific challenges to the optimization scheme. We start with
the remark that, as shown in~\cite{ag2008}, operator $\setminus$ can always
be encoded at the syntactic level through a combination of operators \textsc{Opt},
\textsc{Filter}, and {\it bnd}. We illustrate the idea by example.

\begin{example}
\em
The following SPARQL expression $Q_1$ and the corresponding algebra expression $A_1$
select all persons for which {\bf no} name is specified in the data set. 

{\small 
\begin{tabbing}
x \= xxxxxxxxxxxxxx \= xxxxxxxxxxxxxxxx \= \kill
\>$Q_1 = \textsc{Filter}_{\neg {\it bnd(?n)}}((?p,type,Person) \oopt$\\
\>\>$((?p,type,Person) \oand (?p,name,?n)))$\\
\>$A_1 = \sigma_{\neg {\it bnd(?n)}}(\evalc{(?p,type,Person)} \leftouterjoin$\\
\>\>$(\evalc{(?p,type,Person)} \Join \evalc{(?p,name,?n)}))$
\end{tabbing}
}
\vspace{-0.5cm}
$\punto$
\label{ex:minus}
\end{example}

From an optimization point of view it would be desirable to have a clean translation
of this constellation using operator $\setminus$, but the semantics maps $Q_1$
into $A_1$, which contains operators $\sigma$, $\leftouterjoin$,
$\Join$, and predicate {\it bnd}, rather than $\setminus$. In fact, a better
translation (based on $\setminus$) exists for a class of practical queries
and we will provide rewriting rules for such a transformation.

\begin{proposition}
\em 
Let $A_1$, $A_2$ be $\mathds{A}$-expressions and $A_1^-$, $A_1^-$ be
$\mathds{A}$$^-$-expressions. The following equivalences hold.

{\small
\begin{tabbing}
x \= xxxxxxxxxxxlll \= xl \= llxxxxxxxxxxxxxxxxxx \= \kill
\>$(A_1 \setminus A_2) \setminus A_3$\>$\equiv$\>$(A_1 \setminus A_3) \setminus A_2$\>{\small \it (MReord)}\\
\>$(A_1 \setminus A_2) \setminus A_3$\>$\equiv$\>$A_1 \setminus (A_2 \cup A_3)$\>{\small \it (MMUCorr)}\\
\>$A_1 \setminus A_2$\>$\equiv$\>$A_1 \setminus (A_1 \Join A_2)$\>{\small \it (MJ)}\\
\>$A_1^- \leftouterjoin A_2^-$\>$\equiv$\>$A_1^- \leftouterjoin (A_1^- \Join A_2^-)$\>{\small \it (LJ)}
\end{tabbing}
}
\label{prop:minus}
\vspace{-0.6cm}
$\punto$
\end{proposition}

Rules {\it (MReord)} and {\it (MMUCorr)} are general-purpose rewriting rules, listed
for completeness. Most important in our context is rule {\it (LJ)}.
It allows to eliminate redundant subexpressions in
the right side of $\leftouterjoin$-expressions (for $\mathds{A}^-$ expressions),
e.g.~the application of {\it (LJ)} simplifies $A_1$ to
$A_1'$$=\sigma_{\neg {\it bnd(?n)}}(\evalc{(?p,type,Person)} \leftouterjoin \evalc{(?p,name,?n)})$. The following lemma allows for further simplification.

\begin{lemma}
\em
Let $A_1^-$, $A_2^-$ be $\mathds{A}^-$-expressions, $R$ a filter condition,
and $?x \in {\it safeVars}(A_2) \setminus {\it vars}(A_1)$ a variable. Then
$\sigma_{\neg {\it bnd}(?x)}(A_1^- \leftouterjoin A_2^-) \equiv A_1^- \setminus A_2^-$
holds.$\punto$
\label{lemma:minus}
\end{lemma}

The application of the lemma to query $A_1'$ yields the
expression $A_1''=\evalc{(?p,type,Person)} \setminus \evalc{(?p,name,?n)}$.
Combined with rule {\it (LJ)}, we have established a powerful mechanism that
often allows to make simulated negation explicit.

\section{Semantic SPARQL Query Optimization}
\label{sec:sqo}
\vspace{-.1cm}

This chapter complements the discussion of algebraic optimization
with constraint-based, semantic query optimization (SQO).
The key idea of SQO is to find semantically equivalent queries over a
database that satisfies a set of integrity constraints. These
constraints might have been specified by the user,
extracted from the underlying
database, or hold implicitly when SPARQL is evaluated on RDFS data
coupled with an RDFS inference system.\footnote{Note that
the SPARQL semantics disregards RDFS inference, but assumes that
it is realized in a separate layer.}
More precisely, given a query $Q$ and a set of constraints $\Sigma$
over an RDF database $D$ s.t.~$D \models \Sigma$, we want to enumerate (all)
queries $Q'$ that compute the same result on $D$. We write $Q \equiv_\Sigma Q'$
if $Q$ is equivalent to $Q'$ on each database $D$ s.t.~$D \models \Sigma$.
Following previous approaches~\cite{dpt2006}, we focus on TGDs and EGDs,
which cover a broad range of practical constraints over RDF, such as functional
and inclusion dependencies. When talking about constraints
in the following we always mean TGDs or EGDs. We refer the interested reader
to~\cite{lms2008} for motivating examples and a study of constraints for RDF.
We represent each constraint $\alpha \in \Sigma$ by a first-order
logic formula over a ternary relation $T_D(s,p,o)$ that stores
all triples contained in RDF database $D$ and use $T$ as the 
corresponding relation symbol. For instance, the
constraint $\forall x_1, x_2 ( T(x_1,p_1,x_2) \rightarrow \exists y_1 T(x_1,p_2,y_1))$
states that each RDF resource with property $p_1$ also has property $p_2$. Like in the case of conjunctive queries we call a $\cal{A}_+$ query minimal if there is no equivalent $\cal{A}_+$ query with fewer triple patterns.

Our approach relies on the Chase \& Backchase (C\&B) algorithm
for semantic optimization of CQs proposed in~\cite{dpt2006}.
Given a CQ $q$ and a set $\Sigma$ of constraints as input, the algorithm
outputs all semantically equivalent and minimal $q' \equiv_\Sigma q$ whenever
the underlying chase algorithm terminates. We defer the discussion of chase 
termination to the subsequent section and use the C\&B algorithm as a
black box with the above properties. Our basic idea is as follows. First, we
translate \textsc{And}-only blocks (or queries), so-called basic graph
patterns (BGPs), into CQs and then apply C\&B to optimize them. We
introduce two alternate translation schemes below.

\begin{definition}
\em
Let $S \subset V$ be a finite set of variables and $Q \in \cal{A}$$_+$ be a
SPARQL query defined as 

\begin{tabbing}
x \= \kill
\>$Q = \textsc{Select}_S((s_1,p_1,o_1) \oand \dots \oand (s_n,p_n,o_n))$.
\end{tabbing}

We define the translation $C_1(Q) :=  q$, where

\begin{tabbing}
x \= \kill
\>$q: {\it ans}(\overline{s}) \leftarrow T(s_1,p_1,o_1), \dots, T(s_n,p_n,o_n)$,
\end{tabbing}

and $\overline{s}$ is a vector of variables containing exactly the
variables in $S$. We define $C_1^{-1}(q)$ as follows. It takes a CQ in
the form of $q$ as input and returns $Q$ if it is a valid SPARQL query,
i.e.~if $s_i \in BIV$, $p_i \in IV$, $o_i \in BILV$ for all $i \in [n]$;
otherwise, $C_1^{-1}(q)$ is undefined.$\punto$
\label{def:c1}
\end{definition}

\begin{definition} \em
Let $\Sigma$ be a set of RDF constraints, $D$ an RDF database, and
$\alpha = \forall \overline{x} (\phi(\overline{x}) \rightarrow \exists \overline{y} \psi(\overline{x} ,\overline{y})) \in \Sigma$. 
%Further set $R(D) := \set{a_2(a_1,a_3)}{(a_1,a_2,a_3) \in D}$.
We use $h(T(a_1,a_2,a_3)) := a_2(a_1,a_3)$ if $a_2$ is not a variable,
otherwise we set it to the empty string.
For a conjunction $\bigwedge_{i=1}^{n} T(\overline{a}_i)$ of atoms, we set
$h(\bigwedge_{i=1}^{n} T(\overline{a}_i)) := \bigwedge_{i=1}^{n} h(T(\overline{a}_i))$. 
Then, we define the constraint $\alpha'$ as $\forall \overline{x} (h(\phi(\overline{x})) \rightarrow \exists \overline{y} h(\psi(\overline{x} ,\overline{y})))$.
We set $\Sigma' := \{\alpha'\ |\ \alpha \in \Sigma \}$ if all $\alpha'$ are constraints, otherwise $\Sigma' := \emptyset$.\\

Let $S \subset V$ be a set of variables, $Q \in \mathcal{A}_+$ defined as 

\begin{tabbing}
x \= \kill
\>$Q = \textsc{Select}_S((s_1,p_1,o_1) \oand ... \oand (s_n,p_n,o_n))$,
\end{tabbing}

and assume that $p_i$ is never a variable. We define the translation
$C_2(Q) := ans(\overline{s}) \leftarrow p_1(s_1,o_1), ... ,p_n(s_n,o_n)$,
where vector $\overline{s}$ contains exactly the variables in $S$.
For a CQ $q$$: ans(\overline{s}) \leftarrow R_1(x_{11},x_{12}),...,R_n(x_{n1},x_{n2})$,
we denote by $C_2^{-1}(q)$ the expression

\begin{tabbing}
x \= \kill
\>$\textsc{Select}_{S}((x_{11},R_1,x_{12}) \oand ... \oand (x_{n1},R_n,x_{n2}))$
\end{tabbing}

if it is a SPARQL query, else $C_2^{-1}(q)$ is undefined. $\punto$
\label{def:c2}
\end{definition} 

$C_1(Q)$ and $C_1^{-1}(Q)$ constitute straightforward translations from SPARQL
\textsc{And}-only queries to CQs and back. The definition of $C_2(Q)$ and
$C_2^{-1}(Q)$ was inspired by the work in \cite{skct2005} and is
motivated by the observation that in many real-world SPARQL queries variables
do not occur in predicate position; it is applicable only in this context.
Given that the second translation scheme is not always applicable, 
the reader may wonder why we introduced it. The reason is that the translation
schemes are different w.r.t.~the termination conditions for the subsequent
chase that they exhibit. We will come back to this issue when discussing
termination conditions for the chase in the next section
(see~Proposition~\ref{schemes-vgl}).

The translation schemes, although defined for $\cal{A}_+$ queries,
directly carry over to $\cal{A}$-expressions, i.e.~each expression
$Q \in \cal{A}$ can be rewritten into the equivalent
$\cal{A}_+$-expression $\textsc{Select}_{{\it vars}(Q)}(Q)$.
Coupled with the C\&B algorithm, they provide a sound approach to
semantic query optimization for \textsc{And}-only queries whenever 
the underlying chase algorithm terminates, as stated by the following
lemma. 

\begin{lemma} \label{direct} \em Let $Q$ an $\cal{A}_+$-expression, $D$
a database, and $\Sigma$ a set of EGDs and TGDs. 
\squishlist
	\item If $cb_{\Sigma}(C_1(Q))$ terminates then $\forall Q' \in \mathcal{A}_+$:\\$Q' \in C_1^{-1}(cb_{\Sigma}(C_1(Q))) \Rightarrow  Q' \equiv_\Sigma Q$ and $Q'$ minimal.

	\item If $C_2(Q)$ is defined, $|\Sigma'|=|\Sigma|$ and $cb_{\Sigma'}(C_2(Q))$ terminates then so does $cb_{\Sigma}(C_1(Q))$.

\item If $C_2(Q)$ is defined, $|\Sigma'|=|\Sigma|$ and $cb_{\Sigma'}(C_2(Q))$ terminates then $\forall Q' \in \mathcal{A}_+$:\\$Q' \in C_1^{-1}(cb_{\Sigma}(C_1(Q))) \Rightarrow  Q' \equiv_\Sigma Q$.$\punto$ % and $Q'$ minimal.$\punto$
\squishend
\end{lemma}

The converse direction in bullets one and three does not hold in general,
i.e.~the scheme is not complete. Before we address this issue, we illustrate
the problem by example.

\begin{example}
\em
Let the two expressions $Q_1 := (?x,b,'l')$, $Q_2 := (?x,b,'l') \oand (?x,a,c)$,
and let the constraint
$\forall x_1, x_2, x_3 (T(x_1,x_2,x_3) \rightarrow T(x_3,x_2,x_1))$
be given. By definition, there are no RDF databases that contain a literal 
in a predicate position because, according to the constraint, such a literal would also occur in the subject position, which is not allowed. Therefore, the answer to both
expressions $Q_1$ and $Q_2$ is always the empty set, which implies $Q_1 \equiv_{\Sigma} Q_2$. But 
it is easy to verify that $C_1(Q_1) \equiv_{\Sigma} C_1(Q_2)$ does not hold.
The reason for this discrepancy is that the universal plan~\cite{dpt2006}
of the queries is not a valid SPARQL query.$\punto$
\label{ex:noncompleteness}
\end{example}

We formalize this observation in the next lemma, i.e.~provide a
precondition that guarantees completeness.\footnote{We might expect
that situations as the one sketched in Example~\ref{ex:noncompleteness}
occur rarely in practice, so the condition in Lemma~\ref{direct2}
may guarantee completeness in most practical scenarios.}
For a CQ $q$, we denote by $U(q)$ its universal
plan~\cite{dpt2006}, namely the conjunctive query $q'$ obtained from
$q$ by chasing its body.

\begin{lemma} \label{direct2} \em Let $D$ be a database and $Q$ an
$\cal{A}_+$-expression such that $C_1^{-1}(U(C_1(Q))) \in \cal{A}_+$.
\squishlist
	\item If $cb_{\Sigma}(C_1(Q))$ terminates then $\forall Q' \in \mathcal{A}_+$ such that $C_1^{-1}(U(C_1(Q'))) \in \cal{A}_+$:\\$Q' \in C_1^{-1}(cb_{\Sigma}(C_1(Q))) \Leftrightarrow  Q' \equiv_\Sigma Q$ and $Q'$ minimal.

\item If $C_2(Q)$ is defined, $|\Sigma'|=|\Sigma|$ and $cb_{\Sigma'}(C_2(Q))$ terminates then $\forall Q'$$\in$$\mathcal{A}_+$ s.t.~$C_1^{-1}(U(C_1(Q'))) \in \cal{A}_+$: $Q' \in C_1^{-1}(cb_{\Sigma}(C_1(Q))) \Leftrightarrow  Q' \equiv_\Sigma Q$ and $Q'$ minimal.$\punto$
\squishend
\end{lemma}

%\begin{proposition} \label{scheme-prop} \em Let $\Sigma$ be a set of constraints and $D$ a RDF dataset. If the chase with $(\Sigma(\Sigma),R(D))$ terminates and $|\Sigma(\Sigma)| = |\Sigma|$, so does  the chase on $(\Sigma,D)$. $\punto$
%\end{proposition}
%The proof is immediate and therefore omitted.

%\begin{proposition} \em
%Let $Q \in \mathcal{A}_+$ and $\Sigma$ a set of constraints such that $|\Sigma(\Sigma)| = |\Sigma|$. If $cb_{\Sigma(\Sigma)}(C_2(Q))$ terminates, then it holds for all $Q' \in \mathcal{A}_+$ (up to isomorphism) that $Q' \in C_2^{-1}(cb_{\Sigma(\Sigma)}(C_2(Q))) \Leftrightarrow Q' \equiv_{\Sigma} Q$. $\punto$
%\end{proposition}

By now we have established a mechanism that allows us to enumerate equivalent
queries of SPARQL \textsc{And}-only queries, or BGPs inside queries. Next,
we provide extensions that go beyond \textsc{And}-only queries.
The first rule in the following lemma shows that sometimes \textsc{Opt} can 
be replaced by \textsc{And}; informally spoken, it applies when the expression
in the \textsc{Opt} clause is implied by the constraints. The second rule can be
used to eliminate redundant BGPs in \textsc{Opt}-subexpressions. 

\begin{lemma} \label{elim-opt}
\em
Let $Q_1, Q_2, Q_3 \in \mathcal{A}$ and $S \subset V$ a finite set of variables.

\squishlist
\item If $Q_1 \equiv_\Sigma \textsc{Select}_{{\it vars}(Q_1)} (Q_1 \oand Q_2)$ then\\
 $Q_1 \oopt Q_2 \equiv_\Sigma Q_1 \oand Q_2$.

\item If $Q_1 \equiv_\Sigma Q_1 \oand Q_2$ then\\
$(Q_1 \oopt (Q_2 \oand Q_3)) \equiv_\Sigma Q_1 \oopt Q_3$. $\punto$

\squishend
\end{lemma}

Note that the preconditions are always expressed in terms of \textsc{And}-only
queries and projection, thus can be checked using our translation schemes and
the C\&B algorithm.
%by translating the expressions into CQs,
%chasing them to the universal plan, and checking for the existence of a
%containment mapping.
We conclude our discussion of SQO with a lemma that gives rules for the
elimination of redundant filter expressions.

\begin{lemma} \label{elim-filter}
\em
Let $Q_1, Q_2 \in \mathcal{A}$, $S \subset V \backslash \setone{?y}$ a set of variables, $?x, ?y \in {\it vars}(Q_2)$, $\Sigma$ a set of constraints, and $D$ a documents s.t.~$D \models \Sigma$. Further let $Q_2 \frac{?x}{?y}$ be obtained from $Q_2$ by replacing each occurrence of $?y$ by $?x$.

\squishlist
\item If $Q_1 \equiv_\Sigma \textsc{Select}_{{\it vars}(Q_1)}(Q_1 \oand Q_2)$ then\\
$\evalcd{\textsc{Filter}_{\neg {\it bnd}(?x)}(Q_1 \oopt Q_2)} = \emptyset$.

\item If $\textsc{Select}_{S}(Q_2) \equiv_\Sigma \textsc{Select}_{S}(Q_2 \frac{?x}{?y})$ then\\
$\textsc{Select}_{S}(\textsc{Filter}_{?x=?y}(Q_2)) \equiv_\Sigma \textsc{Select}_{S}(Q_2 \frac{?x}{?y})$.

\item If $\textsc{Select}_{S}(Q_2) \equiv_\Sigma \textsc{Select}_{S}(Q_2 \frac{?x}{?y})$ then\\
 $\evalcd{\textsc{Filter}_{\neg ?x=?y}(Q_2)} = \emptyset$. $\punto$
\squishend

\end{lemma}

We conclude this section with some final remarks. First, we note that semantic
optimization strategies are basically orthogonal to algebraic optimizations,
hence both approaches can be coupled with each other. For instance, we might get
better optimization results when combining the rules for filter decomposition and
pushing in Figure~\ref{fig:rewritings}(\verb!V-VI!) with the semantic rewriting rules for
filter expressions in the lemma above. Second, as discussed in~\cite{dpt2006},
the C\&B algorithm can be enhanced by a cost function, which makes it easy to
factor in cost-based query optimization approaches for SPARQL, e.g.~in the
style of~\cite{ssbkr2008}. This flexibility strengthens the prospectives and
practicability of our semantic optimization scheme. The
study of rewriting heuristics and the integration of a cost function, though,
is beyond the scope of this paper.

\section{Chase Termination}
\label{sec:chase}
\vspace{-0.1cm}

The applicability of the C\&B algorithm, and hence of our SQO scheme presented
in the previous section, depends on the termination of the underlying chase
algorithm. Given an arbitrary set of constraints it is in general undecidable
if the chase terminates for every database instance~\cite{dnr2008}; still,
in the past several sufficient termination conditions have been
postulated~\cite{jk1982,dt2005,dpt2006,fkmp2003,dnr2008}.  The strongest
sufficient conditions known so far are \textit{weak acyclicity}~\cite{fkmp2003},
which was strictly generalized to \textit{stratification} in~\cite{dnr2008},
raising the recognition problem from \textsc{P} to \textsc{coNP}. 
Our SQO approach on top of the C\&B algorithm motivated a reinvestigation of
these termination conditions, and as a key result we present two novel
chase termination conditions for the classical framework of relational databases,
which empower virtually all applications that rely on the chase. 
%, thus broadening the applicability of virtually all applications that rely on the chase.
Whenever we mention
a database or a database instance in this section, we mean a relational database.
We will start our discussion with a small example run of the chase algorithm. The basic
idea is simple: given a database and a set of constraints as input, it fixes
constraint violations in the database instance. Consider for example database 
$\setone{R(a,b)}$ and constraint
$\forall x_1, x_2 (R(x_1,x_2) \rightarrow \exists y R(x_2,y))$.
The chase first adds $R(b,y_1)$ to the instance, where $y_1$ is a fresh null value.
The constraint is still violated, because $y_1$ does not occur in the first
position of an $R$-tuple. So, the chase will add $R(y_1,y_2)$, $R(y_2,y_3)$, $R(y_3,y_4)$,
$\dots$ in subsequent steps, where $y_2, y_3, y_4,$ \ldots are  fresh null values. 
Obviously, the chase algorithm will never terminate in this toy example.
% Let e.g. $\setone{R(a,b)}$ a simple database and 
%$\forall x_1, x_2 (R(x_1,x_2) \rightarrow \exists y S(x_2,y))$ a
%constraint\footnote{In contrast to the SPARQL conventions, we stick to the
%habits in database theory and do not write question marks in front of variables.}.
%In this case the constraint is not satisfied because there is no tuple in the relation
%$S$ with the value $b$ in its first position, in fact the relation $S$ is empty. So, the
%chase algorithm adds a tuple $S(b,y_1)$ to the database, where $y_1$ is a null value.
%Now the constraint is satisfied and the chase terminates and returns
%$\setone{R(a,b), S(b,y_1)}$. Unfortunately, the situation is not always so easy. 

\begin{figure}[t]
\begin{center}
\psfig{figure=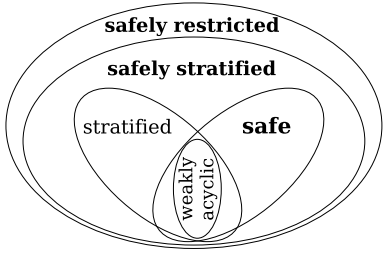,width=5cm}
\end{center}
\vspace{-0.5cm}
\caption{Chase termination conditions.}
\vspace{-0.2cm}
\label{fig:survey}
\end{figure}

Figure~\ref{fig:survey} summarizes the results of this section
and puts them into context.  First, we will introduce the novel
class of {\it safe} constraints, which guarantees the termination of the
chase. It strictly subsumes weak acyclicity, but is different from
stratification. Building upon the definition of safety, we then present
{\it safely restricted} constraints as a consequent advancement of our
ideas. The latter class strictly subsumes all remaining termination
conditions known so far. Finally, we will show that, based on our framework,
we can easily define another class called {\it safely stratified constraints},
which is strictly contained in the class of safely restricted constraints, but
also subsumes weak acyclicity and safeness.

\begin{figure}[t]
\begin{center}
	\begin{tabular}[t]{cc|cc} 
\psfig{figure=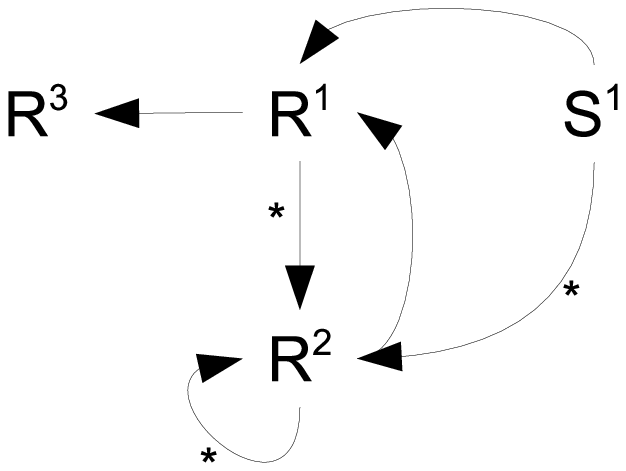,width=3cm}
		& \hspace{0.2cm} & \hspace{0.2cm} &
	\raisebox{1cm}{\psfig{figure=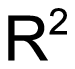,width=0.5cm}}\\
	\end{tabular}
\end{center}
\vspace{-0.3cm}
\caption{Left: Dependency graph. Right: Corresponding propagation graph (it has no edges).}
\vspace{-0.2cm}
\label{tab:notweak--safe}
\end{figure}

{\bf Safe Constraints.}
The basic idea of the first termination condition is to keep track of
positions a
newly introduced labeled null may be copied to. Consider for instance
constraint $R(x_1,x_2,x_3), S(x_2) \rightarrow \exists y R(x_2,y,x_1)$,
which is not weakly
acyclic. Its dependency graph is depicted in Figure \ref{tab:notweak--safe}
(left). As illustrated in the toy example in the beginning of this section,
a cascading of labeled nulls (i.e.~new labeled null values that are created
over and over again) may cause a non-terminating chase sequence. However,
we can observe that for the constraint above
such a cascading of fresh labeled nulls cannot occur, i.e.~no
fresh labeled null can repeatedly create new labeled nulls in position
$R^2$ while copying itself to position $R^1$. The reason is that the constraint
cannot be violated with a fresh labeled null in $R^2$,
i.e.~if $R(a_1,a_2,a_3)$ and 
$S(a_2)$ hold, but $\exists y R(a_2,y,a_1)$ does not, then $a_2$ is
never a newly created labeled null. This is due to the fact that $a_2$
must also occur in relation $S$,
which is not modified when chasing only with this single constraint.
Consequently, the chase sequence always terminates. 
We will later see that this is not a mere coincidence: the constraint is safe.

To formally define safety, we first introduce the notion of affected positions.
Intuitively, a position is affected if, during the application of the chase, a newly
introduced labeled null can be copied or created in it. Thus, the set of affected
positions is an overestimation of the positions in which a null value that
was introduced during the chase may occur.

\begin{definition} \cite{cgk2008} \em
Let $\Sigma$ be a set of TGDs. The set of affected positions $\mbox{aff}(\Sigma)$ of $\Sigma$ is defined inductively as follows. Let $\pi$ be a position in the head of an $\alpha \in$~$\Sigma$. 
\squishlist
	\item If an existentially quantified variable appears in $\pi$, then $\pi \in \mbox{aff}(\Sigma)$.
	\item If the same universally quantified variable $X$ appears both in position $\pi$, and only in affected positions in the body of $\alpha$, then $\pi \in \mbox{aff}(\Sigma)$. $\punto$
\squishend
\end{definition}

Although we borrow this definition from \cite{cgk2008} our focus is different. We extend known classes of constraints for which the chase terminates. The focus in \cite{cgk2008} is on query answering in cases the chase may not terminate. Our work neither subsumes \cite{cgk2008} nor the other way around.
Like in the case of weak acyclicity, we define the safety condition with the help of the absence of  cycles containing special edges in some graph. We call this graph propagation graph.

\begin{definition} \em
Given a set of TGDs $\Sigma$, the propagation graph $\mbox{prop}(\Sigma):=(\mbox{aff}(\Sigma),E)$ is the directed graph defined as follows. There are two kinds of edges in $E$. Add them as follows: for every TGD 
$\forall \overline{x}(\phi(\overline{x}) \rightarrow \exists \overline{y} \psi(\overline{x},\overline{y})) \in \Sigma$
 and for every $x$ in $\overline{x}$ that occurs in $\psi$ and every occurrence of $x$ in $\phi$ in position $\pi_1$
\squishlist
	\item if $x$ occurs only in affected positions in $\phi$ then, for every occurrence of $x$ in $\psi$ in position $\pi_2$, add an edge $\pi_1 \rightarrow \pi_2$ (if it does not already exist).
	\item if $x$ occurs only in affected positions in $\phi$ then, for every existentially quantified variable $y$ and for every occurrence of $y$ in a position $\pi_2$, add a special edge $\pi_1 \stackrel{*}{\rightarrow} \pi_2$ (if it does not already exist).$\punto$
\squishend
\end{definition}

\begin{definition} \em
A set $\Sigma$ of constraints is called \textit{safe} iff $\mbox{prop}(\Sigma)$ has no cycles going through a special edge. $\punto$
\end{definition}

The intuition of these definitions is that we forbid an unrestricted cascading of null values, i.e.~with the help of the propagation graph we impose a partial order on the affected positions such that any newly introduced null value can only be created in a position that has a higher rank in that partial order in comparison to null values that may occur in the body of a TGD. To state this more precisely, assume a TGD of the form $\forall \overline{x}(\phi(\overline{x}) \rightarrow \exists \overline{y} \psi(\overline{x},\overline{y}))$ is violated. Then, $I \models \phi(\overline{a})$ and $I \nvDash \exists \overline{y} \psi(\overline{a},\overline{y}))$ must hold. The safety condition ensures that any position in the body that has a newly created labeled null from $\overline{a}$ in itself and also occurs in the head of the TGD has a strictly lower rank in our partial order than any position in which some element from $\overline{y}$ occurs. The main difference in comparison to weak acyclicity is that we look in a refined way (see affected positions) on where a labeled null can be propagated to. We note that given a set of constraints it can be decided in polynomial time whether it is safe.

\begin{example} \em
Consider the TGD $R(x_1,x_2,x_3), S(x_2) \rightarrow \exists y R(x_2,y,x_1)$ from
before. The dependency graph is depicted in Figure \ref{tab:notweak--safe} on the
left side and its propagation graph on the right side. The only affected position
is $R^2$. From the respective definitions it follows that this constraint is safe,
but not weakly acyclic. $\punto$
\end{example}

Note that if $\Sigma$ is safe, then every subset of $\Sigma$ is safe, too. We will now compare safety to other termination conditions. In the example, the propagation graph is a subgraph of the dependency graph. This is not a mere coincidence.

\begin{theorem} \label{rel-safety} \em
Let $\Sigma$ be a set of constraints.

\squishlist
	\item Then, $\mbox{prop}(\Sigma)$ is a subgraph of $\mbox{dep}(\Sigma)$. It holds that if $\Sigma$ is weakly acyclic, then it is also safe.
	\item There is some $\Sigma$ that is safe, but not stratified.
	\item There is some $\Sigma$ that is stratified, but not safe. $\punto$
\squishend
\end{theorem}

The next result shows that safety guarantees termination while retaining polynomial time data complexity.

\begin{theorem} \label{safeterm} \em
Let $\Sigma$ be a fixed set of safe constraints. Then, there exists a polynomial $Q \in \mathbb{N}[X]$ such that for any database instance $I$, the length of every chase sequence is bounded by $Q(||I||)$, where $||I||$ is the number of distinct values in $I$. $\punto$
\end{theorem}

{\bf Safely Restricted Constraints.}
In this section we generalize the method of stratification from~\cite{dnr2008}
to a condition which we call {\it safe restriction}. The chase graph from~\cite{dnr2008}
will be a special case of our new notion. We then define the notion of safe restriction
and show that the chase always terminates for constraints obeying it.

Let $\alpha := S(x_2,x_3), R(x_1,x_2,x_3) \rightarrow \exists y R(x_2,y,x_1)$
and $\beta := R(x_1,x_2,x_3) \rightarrow S(x_1,x_3)$. It can be seen that $\alpha \prec \beta$
and $\beta \prec \alpha$. Further, $\setone{\alpha, \beta}$ is not weakly
acyclic, so it follows that $\setone{\alpha, \beta}$ is not stratified. Still, the chase will
always terminate: A firing of $\alpha$ may cause a null value to appear in position $R^2$,
but a firing of $\beta$ will never introduce null values in the head of $\beta$ although
$\beta \prec \alpha$ holds. This is the key observation for the upcoming definitions.
First, we will refine the relation $\prec$ from~\cite{dnr2008}. This refinement
helps us to detect if during the chase null values might be copied to the head of
some constraint. Let $pos(\Sigma)$ denote the set of positions that occur in the body
of some constraint in $\Sigma$.

\begin{definition} \label{def-verf} \em
Let $\Sigma$ a set of constraints and $P \subseteq pos(\Sigma)$. For all $\alpha, \beta \in \Sigma$, we define $\alpha \prec_{P} \beta$ iff there are tuples $\overline{a}, \overline{b}$ and a database instance $I$ s.t.
\squishlist
	\item $I \nvDash \alpha(\overline{a})$,
	\item $\beta$ is not applicable on $\overline{b}$ and $I$,
	\item $I\overline{a}\oplus C_{\alpha} \nvDash \beta(\overline{b})$,
	\item null values in $I$ occur only in positions from $P$, and 
		\item the firing of $\beta$ in the case of bullet three copies some null value from $I\overline{a}\oplus C_{\alpha}$ to the head of $\beta$. $\punto$
\squishend\end{definition}

We next introduce a notion for affected positions
relative to a constraint and a set of positions.

\begin{definition} \em
For any set of positions $P$ and tgd $\alpha$ let $\mbox{aff-cl}(\alpha,P)$ be the set of positions $\pi$ from the head of $\alpha$ such that either

\squishlist
	\item the variable in $\pi$ occurs in the body of $\alpha$ only in positions from $P$ or
	\item $\pi$ contains an existentially quantified variable. $\punto$
\squishend
\end{definition}

The latter definition and the refinement of $\prec$ will help us to define the notion of a
restriction system,
which is a strict generalization of the chase graph introduced
in~\cite{dnr2008}. 

\begin{definition} \label{rest} \em
A {\it restriction system} is a pair $(G'(\Sigma),f)$, where
$G'(\Sigma) := (\Sigma,E)$ is a directed graph and
$f: \Sigma \rightarrow 2^{pos(\Sigma)}$ is a function such that
\squishlist
        \item forall TGDs $\alpha$ and forall $(\alpha,\beta) \in E$:\\$\mbox{aff-cl}(\alpha,f(\alpha))  \cap pos(\setone{\beta}) \subseteq f(\beta)$,
                \item forall EGDs $\alpha$ and forall $(\alpha,\beta) \in E$:\\$f(\alpha) \cap pos(\setone{\beta}) \subseteq f(\beta)$, and
        \item forall $\alpha, \beta \in \Sigma$: $\alpha \prec_{f(\alpha)} \beta \implies (\alpha,\beta) \in E$.$\punto$
\squishend
\end{definition}

We illustrate this definition by an example. It also shows that restriction systems always exist.

\begin{example} \label{alle} \em
Let $\Sigma$ a set of constraints. Then, $(G(\Sigma),f)$, where $f(\alpha):=pos(\setone{\alpha})$ for all $\alpha \in \Sigma$ is a restriction system for $\Sigma$. $\punto$
\end{example}

Based on the novel technical notion of restriction systems we can easily define
a new class of constraints.

\begin{definition} \label{def-rest} \em
$\Sigma$ is called safely restricted if and only if there is a restriction system $(G'(\Sigma),f)$ for $\Sigma$ such that every strongly connected component in $G'(\Sigma)$ is safe. $\punto$
\end{definition}

The next theorem shows that safe restriction strictly extends the notion of stratification and safety.

\begin{theorem} \label{rest-safe-weak} \em If $\Sigma$ is stratified or safe, then it is also safely restricted. There is some $\Sigma$ that is safely restricted but neither safe nor stratified. $\punto$
\end{theorem}

Definition \ref{def-rest} implies that safely restricted constraints can be recognized by a $\Sigma_2^P$-algorithm. However, with the help of a canonical restriction system, we can show that safe restriction can be decided in $\textsc{coNP}$ (like stratification).

\begin{theorem} \label{rest-comp} \em
Given constraint set $\Sigma$ it can be checked by a $\textsc{coNP}$-algorithm whether $\Sigma$ is safely restricted. $\punto$
\end{theorem}

The next theorem is the main contribution of this section. It states that the chase will always terminate in polynomial time data complexity for safely restricted constraints. 

\begin{theorem} \label{chase-main} \em
Let $\Sigma$ be a fixed set of safely restricted constraints. Then, there exists a polynomial $Q \in \mathbb{N}[X]$ such that for any database instance $I$, the length of every chase sequence is bounded by $Q(||I||)$, where $||I||$ is the number of distinct values in $I$. $\punto$
\end{theorem}

To the best of our knowledge safe restriction is the most general sufficient
termination condition for TGDs and EGDs.
We finally compare the chase graph to restriction systems. The reader might
wonder what happens if we substitute weak acyclicity with safety in the
definition of stratification (in the preliminaries). 

\begin{definition} \em
We call $\Sigma$ \textit{safely stratified} iff the constraints in every cycle of $G(\Sigma)$ are safe. $\punto$
\end{definition}

We obtain the
following result, showing that with the help of restriction systems, we
strictly extended the method of the chase graph from \cite{dnr2008}.

\begin{theorem} \label{safestrat} \em Let $\Sigma$ be a set of constraints.
 \squishlist
	\item If $\Sigma$ is weakly acyclic or safe, then it is safely stratified.
	\item If $\Sigma$ is  safely stratified, then it is safely restricted.
	\item There is some set of constraints that is safely restricted, but not safely stratified. $\punto$
\squishend
\end{theorem}

Note that we used safety instead of safe stratification in the definition
of safe restrictedness although safely stratified constraints are the provably
larger class. This is due to the fact that safety is easily checkable and would 
not change the class of constraints. The next proposition clarifies this issue.

\begin{proposition} \label{falsch} \em
$\Sigma$ is safely restricted iff there is a restriction system $(G'(\Sigma),f)$ for $\Sigma$ such that every strongly connected component in $G'(\Sigma)$ is safely stratified. $\punto$
\end{proposition}

In the previous section we proposed two SPARQL translation schemes and it is left
to explain why we introduced two alternative schemes. The next proposition
states that the two schemes behave differently with respect to safe restriction.

\begin{proposition} \label{schemes-vgl} \em Let $\Sigma$ be a non-empty set of constraint set over a ternary relation symbol $T$.
\squishlist
	\item There is some $\Sigma$ that is safely restricted, but $\Sigma' = \emptyset$, i.e.~the second translation scheme is not applicable.
	
	\item There is some $\Sigma$ such that $|\Sigma|=|\Sigma'|$ and $\Sigma'$ is safely restricted, but $\Sigma$ is not. $\punto$
\squishend
\end{proposition}

Referring back to Lemma \ref{direct}, this means we might check both $\Sigma$
or $\Sigma'$ for safe restrictedness, and can guarantee termination of the chase
if at least one of them is safely restricted.

\section{Conclusion}
\label{sec:conclusion}
\vspace{-.15cm}

We have discussed several facets of the SPARQL query language. Our
complexity analysis extends prior investigations~\cite{pag2006}
and (a)~shows that the combination of \textsc{And} and \textsc{Union}
is the main source of complexity in \textsc{Opt}-free SPARQL fragments
and (b)~clarifies that yet operator \textsc{Opt} alone makes SPARQL
evaluation \textsc{PSpace}-complete. We also show that, when restricting
the nesting depth of \textsc{Opt}-expressions, we obtain better complexity
bounds.

The subsequent
study of SPARQL Algebra lays the foundations for transferring established
Relational Algebra optimization techniques into the context of SPARQL.
Additionally, we considered specifics of the SPARQL query language, such as
rewriting of SPARQL queries involving negation. The algebraic optimization
approach is complemented by a powerful framework for semantic query
optimization. We argue that a combination of both algebraic and semantic
optimization will push the limits of existing SPARQL implementations and
leave the study of a schematic rewriting approach and good rewriting
heuristics as future work.

Finally, our results on
chase termination empower the practicability of SPARQL query optimization
in the presence of constraints and directly carry over to other
applications that rely on the chase, such as~\cite{fkmp2003,l2002,h2001,dpt2006}.

%%% bibliography
{
\small
\bibliographystyle{abbrv}
\bibliography{main}
}

%%% appendix
\newpage
\appendix
% one file for each section
\section{Proofs of the Complexity Results}
\label{app:complexity}

This section contains the complexity proofs of the \textsc{Evaluation} problem
for the fragments studied in Section~\ref{sec:complexity}. We refer the interested
reader to~\cite{pag2006} for the proof of Theorem~\ref{th:pag}.
We start with some basics from complexity theory.

\subsection{Background from Complexity Theory} \label{app:complexityback}

{\bf Complexity Classes.} As usual, we denote by \textsc{PTime} (or \textsc{P},
for short) the complexity class comprising all problems that can be decided
by a deterministic Turing Machine (TM) in polynomial time, by \textsc{NP} the
set of problems that can be decided by a non-deterministic TM in polynomial
time, and by \textsc{PSpace} the class of problems that can be decided by a
deterministic TM within polynomial space bounds.

\subsubsection*{The Polynomial Hierarchy}

Given a complexity class $\textsc{C}$ we denote by $\textsc{coC}$ the set of decision
problems whose complement can be decided by a TM in class $\textsc{C}$.
Given complexity classes $\textsc{C}_1$ and $\textsc{C}_2$, the class $\textsc{C}_1^{\textsc{C}_2}$
captures all problems that can be decided by a TM $M_1$ in class $\textsc{C}_1$
enhanced by an oracle TM $M_2$ for solving problems in class $\textsc{C}_2$. Informally,
engine $M_1$ can use $M_2$ to obtain a {\it yes}/{\it no}-answer for a problem in
$\textsc{C}_2$ in a single step. We refer the interested reader to \cite{ab2007} for a more formal
discussion  of oracle machines. Finally, we define the classes $\Sigma^P_i$ and $\Pi^P_i$
recursively as

\begin{tabbing}
x \= \kill
\>$\Sigma^P_0 = \Pi^P_0 {:=} \textsc{P}$ and $\Sigma^P_{n+1} {:=} \textsc{NP}^{\Sigma^P_n}$, and put\\
\>$\Pi^P_{n+1} {:=} \textsc{coNP}^{\Sigma^P_n}$.
\end{tabbing}

The polynomial hierarchy \textsc{PH}~\cite{s1976} is then defined as

\begin{tabbing}
x \= \kill
\>$\textsc{PH} = \bigcup_{i \in \mathbb{N}_0} \Sigma^P_i$
\end{tabbing}

It is folklore that $\Sigma^P_i = co\Pi^P_i$,
and that  $\Sigma^P_i \subseteq \Pi^P_{i+1}$ and $\Pi^P_i \subseteq \Sigma^P_{i+1}$
holds. Moreover, the following inclusion hierarchies for $\Sigma^P_i$ and
$\Pi^P_i$ are known.

\begin{tabbing}
x \= \kill
\>$\textsc{P} = \Sigma^P_0 \subseteq \textsc{NP} = \Sigma^P_1 \subseteq \Sigma^P_2 \subseteq \dots \subseteq \textsc{PSPACE}$, and\\
\>$\textsc{P} = \Pi^P_0 \subseteq \textsc{coNP} = \Pi^P_1 \subseteq \Pi^P_2 \subseteq \dots \subseteq \textsc{PSPACE}$.
\end{tabbing}

\subsubsection*{Complete Problems}

We consider completeness only with respect to polynomial-time many-one
reductions. QBF, the tautology test for quantified boolean formulas, is known
to be \textsc{PSpace}-complete \cite{ab2007}. Forms of QBF with restricted
quantifier alternation are complete for classes $\Pi^P_i$ or $\Sigma^P_i$ depending
on the question if the first quantified of the formula is $\forall$ or $\exists$.
A more thorough introduction to complete problems in the polynomial hierarchy can
be found in \cite{ab2007}. Finally, the $\textsc{NP}$-completeness of the
$\textsc{SetCover}$-problem and the \textsc{3Sat}-problem is folklore.

\subsection{OPT-free Fragments (Theorem~\ref{th:andunion})}
\label{app:complexitynoopt}

\subsubsection*{Fragment $\cal{U}$: UNION (Theorem~\ref{th:andunion}(1))}

For  a \textsc{Union}-only expression $P$ and data set $D$ it suffices to check if
$\mu \in \evalcd{t}$  for any triple pattern $t$ in $P$. This can easily be
achieved in polynomial time.$\qed$

\subsubsection*{Fragment $\cal{FU}$: FILTER + UNION (Theorem~\ref{th:andunion}(1))}

We present a \textsc{PTime}-algorithm that solves the \textsc{Evaluation} problem
for this fragment. It is defined on the structure of the input expression
$P$ and returns {\it true} if $\mu \in \evalcd{P}$, {\it false} otherwise. We
distinguish three cases. (a) If $P = t$ is a triple pattern, we return {\it true} if
and only if $\mu \in \evalcd{t}$. (b) If
$P = P_1 \ounion P_2$ we (recursively) check if
$\mu \in \evalcd{P_1} \lor \mu \in \evalcd{P_2}$ holds. (c) If $P = P_1 \ofilter R$
for any filter condition $R$ we return {\it true} if and only if
$\mu \in \evalcd{P_1} \land R \models \mu$. It is easy to see that the above
algorithm runs in polynomial time. Its correctness
follows from the definition of the algebraic operators $\cup$ and $\select$.$\qed$

\subsubsection*{Fragment $\cal{AU}$: AND + UNION (Theorem~\ref{th:andunion}(2))}

In order to show that \textsc{Evaluation} for this fragment is \textsc{NP}-complete
we have to show membership and hardness.

{\it Membership in \textsc{NP}.}
Let $P$ be a SPARQL expression composed of operators \textsc{And} and \textsc{Union},
$D$ a document, and $\mu$ a mapping. We provide an \textsc{NP}-algorithm that
returns {\it true} if $\mu \in \evalcd{P}$, and {\it false} otherwise. Our algorithm is
defined on the structure of $P$. (a) If $P = t$ return {\it true} if $\mu \in \evalcd{t}$,
{\it false} otherwise. (b) If $P = P_1 \ounion P_2$, we return the truth value of
$\mu \in \evalcd{P_1} \lor \mu \in \evalcd{P_2}$. (c) If $P = P_1 \oand P_2$, we guess
a decomposition $\mu = \mu_1 \cup \mu_2$ and return the truth value of
$\mu_1 \in \evalcd{P_1} \land \mu_2 \in \evalcd{P_2}$. Correctness of the algorithm
follows from the definition of the algebraic operators $\Join$ and $\cup$.
It  can easily be realized by a non-deterministic TM that runs in polynomial time,
which proves membership in \textsc{NP}.

{\it \textsc{NP}-Hardness.}
We reduce the \textsc{SetCover} problem to the \textsc{Evaluation} problem for SPARQL (in
polynomial time). \textsc{SetCover} is known to be \textsc{NP}-complete, so the
reduction gives us the desired hardness result.

The decision version of \textsc{SetCover} is defined as follows. Let $U = \{ u_1, \dots, u_k \}$
be a universe, $S_1, \dots S_n \subseteq U$ be sets over $U$, and let $k$ be positive integer. Is there
a set $I \subseteq \{ 1, \dots, n \}$ of size $\mid I \mid \leq k$ s.t.~$\bigcup_{i \in I} S_i = U$?

We use the fixed database $D := \{ (a,b,1) \}$ for our encoding and represent each set
$S_i = \{ x_1, x_2, \dots, x_m \}$ by a SPARQL expression of the form

\begin{tabbing}
x \= \kill
\>$P_{S_i} := (a,b,?X_1) \oand \dots \oand (a,b,?X_m)$.
\end{tabbing}

The set $S = \{ S_1, \dots, S_n \}$ of all $S_i$ is then encoded as

\begin{tabbing}
x \= \kill
\>$P_S := P_{S_1} \ounion \dots \ounion P_{S_n}$.
\end{tabbing}

Finally we define the SPARQL expression

\begin{tabbing}
x \= xxxxx \= \kill
\>$P := P_S \oand \dots \oand P_S$,\\
\>\>where $P_S$ appears exactly $k$ times.
\end{tabbing}

It is straightforward to show that \textsc{SetCover} 
is true if and only if $\mu = \{ ?U_1 \mapsto 1, \dots, ?U_k \mapsto 1 \} \in \evalcd{P}$.
The intuition of the encoding is as follows. $P_S$ encodes all
subsets $S_i$. A set element, say $x$, is represented in SPARQL by a mapping from 
variable $?X$ to value $1$. The encoding of $P$ allows us to merge
(at most) $k$ arbitrary sets $S_i$. We finally check if the universe $U$ can
be constructed this way.$\qed$

\begin{remark}
\em
The proof above relies on the fact that mapping $\mu$ is part of the input of the
\textsc{Evaluation} problem. In fact, when fixing $\mu$ the resulting (modified)
version of the \textsc{Evaluation} problem can be solved in $\textsc{PTime}$.$\punto$
\end{remark}

\subsection{Fragments Including Operator OPT}
\label{app:complexityopt}

We now discuss several fragments including $\textsc{Opt}$.
One goal here is to show that fragment $\cal{O}$ is \textsc{PSpace}-complete
(c.f.~Theorem~\ref{th:opt}); \textsc{PSpace}-completeness for all fragments involving
\textsc{Opt} then follows (cf.~Corollary~\ref{cor:optpspace}).
Given the \textsc{PSpace}-completeness results for fragment
$\cal{E} = \cal{AFOU}$, it suffices to prove {\it hardness} for all smaller fragments;
membership is implicit. Our road map is as follows.

\begin{enumerate}
\item We first show \textsc{PSpace}-hardness for fragment $\cal{AFO}$.
\item We then show \textsc{PSpace}-hardness for fragment $\cal{AO}$.
\item Next, a rewriting of operator \textsc{And} by \textsc{Opt} is presented,
which can be used to eliminate all \textsc{And} operators in the proof of (2).
\textsc{PSpace}-completeness for $\cal{O}$ then is shown using this
rewriting rule.
\item Finally, we prove Theorem~\ref{th:ph}, i.e.~show that fragment
$\cal{E}$$_{\leq n}$ is $\Sigma^P_{n+1}$-complete, making use
of part (1).
\end{enumerate}

\subsubsection*{Fragment $\cal{AFO}$: AND + FILTER + OPT}

We present a (polynomial-time) reduction from \textsc{QBF} to \textsc{Evaluation} for
fragment $\cal{AFO}$. The \textsc{QBF} problem is known to be
\textsc{PSpace}-complete, so this reduction gives us the desired \textsc{PSpace}-hardness
result. Membership in \textsc{PSpace}, and hence \textsc{PSpace}-completeness, then
follows from Theorem~\ref{th:pag}(3). \textsc{QBF} is defined as follows.

\begin{tabbing}
x \= xxxxx \= xxx \= \kill
\>\textsc{QBF}: given a quantified boolean formula of the form\\
\\[-0.2cm]
\>\>\>$\varphi = \forall x_1 \exists y_1 \forall x_2 \exists y_2 \dots \forall x_m \exists y_m \psi$,\\
\\[-0.2cm]
\>\>where $\psi$ is a quantifier-free boolean formula,\\
\>\>as input: is $\varphi$ valid?
\end{tabbing}

The following proof was inspired by the proof of Theorem~3 in~\cite{pag2006}: 
we encode the inner formula $\psi$ using \textsc{And} and \textsc{Filter},
and then adopt the translation scheme for the quantifier sequence
$\forall \exists \forall \exists \dots$ proposed in~\cite{pag2006}.

First note that, according to the problem statement, $\psi$ is a quantifier-free
boolean formula. We assume w.l.o.g.~that $\psi$ is composed of $\land$, $\lor$ and
$\neg$.\footnote{In~\cite{pag2006} $\psi$ was additionally restricted to be in CNF.
We relax this restriction here.} We use the fixed database

\begin{tabbing}
x \= \kill
\>$D := \{ (a,\textit{tv},0), (a,\textit{tv},1), (a,\textit{false},0), (a,\textit{true},1) \}$
\end{tabbing}

and denote by $V=\{ v_1, \dots v_l \}$ the set of variables appearing in $\psi$.
Formula $\psi$ then is encoded as

\begin{tabbing}
x \= xxxxxxxxxx \= \kill
\>$P_\psi {:=} ((a,tv,?V_1) \oand (a,tv,?V_2) \oand \dots$\\
\>\>$\oand (a,tv,?V_l)) \ofilter {\it f}(\psi)$,
\end{tabbing}

where ${\it f}(\psi)$ is a function that generates a SPARQL condition that
mirrors the boolean formula $\psi$. More precisely, ${\it f}$ is defined
recursively on the structure of $\psi$ as

\begin{tabbing}
x \= xxxxxxxxx \= xxx \= \kill
\>{\it f}($v_i$)\>${:=}$\>$?V_i$ = 1\\
\>{\it f}($\psi_1 \land \psi_2$)\>${:=}$\>{\it f}($\psi_1$) $\land$ {\it f}($\psi_2$)\\
\>{\it f}($\psi_1 \lor\psi_2$)\>${:=}$\>{\it f}($\psi_1$) $\lor$ {\it f}($\psi_2$)\\
\>{\it f}($\neg \psi_1$)\>${:=}$\>$\neg$ {\it f}($\psi_1$)
\end{tabbing}

In our encoding $P_\psi$, the $\textsc{And}$-block generates all possible valuations for the
variables, while the \textsc{Filter}-expression retains exactly those valuations that
satisfy formula $\psi$. It is straightforward to show that $\psi$ is satisfiable
if and only if there exists a mapping $\mu \in \evalcd{P_\psi}$ and, moreover, for each
mapping $\mu \in \evalcd{P_\psi}$ there is a truth assignment $\rho_\mu$ defined as
$\rho_\mu(x) = \mu(?X)$ for all variables $?X_i, ?Y_i \in {\it dom}(\mu)$ such that
$\mu \in \evalcd{P_\psi}$ if and only if $\rho_\mu$ satisfies $\psi$. Given
$P_\psi$, we can encode the quantifier-sequence using a series of nested
\textsc{Opt} statements as shown in~\cite{pag2006}. To make the proof
self-contained, we shortly summarize this construction.

SPARQL variables $?X_1, \dots, ?X_m$ and $?Y_1, \dots Y_m$ are used to represent
variables $x_1, \dots x_m$ and $y_1, \dots, y_m$, respectively. In addition to
these variables, we use fresh variables $?A_0, \dots ?A_m$, $?B_0, \dots ?B_m$,
and operators \textsc{And} and \textsc{Opt} to encode the quantifier sequence
$\forall x_1 \exists y_1 \dots \forall x_m \exists y_m$. For each $i \in [m]$
we define two expressions $P_i$ and $Q_i$

\begin{tabbing}
x \= xx \= xx \= \kill
\>$P_i$\>${:=}$\>$((a,tv,?X_1) \oand \dots \oand (a,tv,?X_i) \oand$\\
\>\>\>$\ \ (a,tv,?Y_1) \oand \dots \oand (a,tv,?Y_{i-1}) \oand$\\
\>\>\>$\ \ (a,{\it false},?A_{i-1}) \oand (a,{\it true},?A_i))$,\\
\>$Q_i$\>${:=}$\>$((a,tv,?X_1) \oand \dots \oand (a,tv,?X_i) \oand$\\
\>\>\>$\ \ (a,tv,?Y_1) \oand \dots \oand (a,tv,?Y_i) \oand$\\
\>\>\>$\ \ (a,{\it false},?B_{i-1}) \oand (a,{\it true},?B_i))$,
\end{tabbing}

and encode $P_\varphi$ as

\begin{tabbing}
x \= xxxxx \= x \= x \= x \= x \= \kill
\>$P_\varphi {:=}$\>$((a,{\it true},?B_0)$\\
\>\>\>$\oopt (P_1 \oopt (Q_1$\\
\>\>\>\>$\oopt (P_2 \oopt (Q_2$\\
\>\>\>\>\>$\dots$\\
\>\>\>\>\>$\oopt (P_m \oopt (Q_m \oand P_\psi)) \dots )))))$\\
\end{tabbing}

It can be shown that $\mu = \{ ?B_0 \mapsto 1 \} \in \evalcd{P_\varphi}$ iff
$\varphi$ is valid, which completes the reduction. We refer the reader
to the proof of Theorem~3 in~\cite{pag2006} for this part of the proof.$\qed$

\begin{remark}
\em
The proof for this fragment ($\cal{AFO}$) is subsumed by the subsequent
proof, which shows \textsc{PSpace}-hardness for a smaller fragment. It was included
to illustrate how to encode quantifier-free boolean formulas that are not in CNF.
Some of the following proofs build upon this construction.$\punto$
\end{remark}

\subsubsection*{Fragment $\cal{AO}$: AND + OPT}

We reduce the \textsc{QBF} problem to \textsc{Evaluation} for class $\cal{AO}$.
We encode a quantified boolean formula of the form

\begin{tabbing}
x \= \kill
\>$\varphi = \forall x_1 \exists y_1 \forall x_2 \exists y_2 \dots \forall x_m \exists y_m \psi$,
\end{tabbing}

where $\psi$ is a quantifier-free formula in conjunctive normal form (CNF), i.e.
$\psi$ is a conjunction of clauses

\begin{tabbing}
x \= \kill
\>$\psi=C_1 \land \dots \land C_n$,
\end{tabbing}

where the $C_i$, $1 \leq i \leq n$,
are disjunctions of literals.\footnote{In the previous proof (for
fragment $\cal{AFO}$) there was no such restriction for formula $\psi$.
Still, it is known that \textsc{QBF} is also $\textsc{PSpace}$-complete
when restricting to formulae in CNF.} By $V$
we denote the variables in $\psi$ and by $V_{C_i}$ the variables appearing in
clause $C_i$ (either as positive of negative literals). We use
the following database, which is polynomial in the size of the query.

\begin{tabbing}
x \= xxxx \= \kill
\>$D :=$\>$\{ (a,\textit{tv},0), (a,\textit{tv},1), (a,\textit{false},0), (a,\textit{true},1) \}\ \cup$\\
\>\>$\{ (a,var_i,v) \mid v \in V_{C_i} \}\ \cup\ \{ (a,v,v) \mid v \in V \}$
\end{tabbing}

For each $C_i = v_1 \lor \dots \lor v_j \lor \neg v_{j+1} \lor \dots \lor \neg v_k$,
where the $v_1 \dots v_j$ are positive and the $v_{j+1} \dots v_k$ are negated
variables (contained in $V_{C_i}$), we define a separate SPARQL expression

\begin{tabbing}
x \= l \= l \= l \= l \= l \= l \= l \= l \= l \= l \= l \= \kill
\>$P_{C_i} {:=} (\dots((\dots($\\
\>\>$(a,var_i,?var_i)$\\
\>\>\>$\ \textsc{Opt}\ ((a,v_1,?var_i) \oand (a,true,?V_1)))$\\
\>\>\>\>$\ \ \ldots$\\
\>\>\>\>\>$\ \ \ \textsc{Opt}\ ((a,v_j,?var_i) \oand (a,true,?V_j)))$\\
\>\>\>\>\>\>$\ \ \ \ \textsc{Opt}\ ((a,v_{j+1},?var_i) \oand (a,false,?V_{j+1})))$\\
\>\>\>\>\>\>\>$\ \ \ \ \ \ldots$\\
\>\>\>\>\>\>\>\>$\ \ \ \ \ \ \textsc{Opt}\ ((a,v_k,?var_i) \oand (a,false,?V_k)))$
\end{tabbing}

and encode formula $\psi$ as 

\begin{tabbing}
x \= \kill
\>$P_\psi := P_{C_1} \oand \dots \oand P_{C_n}$.
\end{tabbing}

It is straightforward to verify that $\psi$ is satisfiable if and only if there is
a mapping $\mu \in \evalcd{P_\psi}$ and, moreover, for each $\mu \in \evalcd{P_\psi}$
there is a truth assignment $\rho_\mu$ defined as $\rho_\mu(x) = \mu(?X)$ for
all variables $?X_i, ?Y_i \in {\it dom}(\mu)$ such that $\mu \in \evalcd{P_\psi}$ if and
only if $\rho_\mu$ satisfies $\psi$. Now, given $P_\psi$, we encode the
quantifier-sequence using only operators \textsc{Opt} and \textsc{And}, as shown in
the previous proof for fragment $\cal{AFO}$. For the resulting encoding
$P_\varphi$, it analogously holds that
$\mu = \{ ?B_0 \mapsto 1 \} \in \evalcd{P_\varphi}$ iff
$\varphi$ is valid.$\qed$

We provide a small example that illustrates the translation scheme for QBF
presented in in the proof above.

\begin{example}
\em
We show how to encode the QBF

\begin{tabbing}
x \= xl \= \kill
\>$\varphi$\>$= \forall x_1 \exists y_1 (x_1 \Leftrightarrow y_1)$\\
\>\>$= \forall x_1 \exists y_1 ((x_1 \lor \neg y_1) \land (\neg x_1 \lor y_1))$,
\end{tabbing}

where $\psi = ((x_1 \lor \neg y_1) \land (\neg x_1 \lor y_1))$ is in CNF.
It is easy to see that the QBF formula $\varphi$ is a tautology.
The variables in $\psi$ are $V = \{ x_1, y_1\}$; further, we have $C_1 = x_1 \lor \neg y_1$,
$C_2 = \neg x_1 \lor y_1$, and $V_{C_1} = V_{C_2} = V = \{ x_1, y_1 \}$. Following
the construction in the proof we set up the database 

\begin{tabbing}
x \= xxxxx \= \kill
\>$D :=\ $\{\>$(a,\textit{tv},0), (a,\textit{tv},1), (a,\textit{false},0), (a,\textit{true},1),$\\
\>\>$(a,var_1,x_1), (a,var_1,y_1), (a,var_2,x_1),$\\
\>\>$(a,var_2,y_1), (a,x_1,x_1), (a,y_1,y_1) \}$
\end{tabbing}

and define expression $P_\psi = P_{C_1} \oand P_{C_2}$, where

\begin{tabbing}
x \= xxxxxxxx \= x \= \kill
\>$P_{C_1} {:=} ((a,var_1,?var_1)$\\
\>\>$\oopt ((a,x_1,?var_1) \oand (a,true,?X_1)))$\\
\>\>\>$\oopt ((a,y_1,?var_1) \oand (a,false,?Y_1))$\\
\>$P_{C_2} {:=} ((a,var_2,?var_2)$\\
\>\>$\oopt ((a,y_1,?var_2) \oand (a,true,?Y_1)))$\\
\>\>\>$\oopt ((a,x_1,?var_2) \oand (a,false,?X_1))$.
\end{tabbing}

When evaluating these expressions we get: 

\begin{tabbing}
x \= xxxxxx \= xx \= xx \= xx \= \kill
\>$\evalcd{P_{C_1}}$\>$=(\{\{?var_1 \mapsto x_1\},\{?var_1 \mapsto y_1\}\}$\\
\>\>\>$\leftouterjoin \{\{?var_1 \mapsto x_1,?X_1 \mapsto 1\}\})$\\
\>\>\>\>$\leftouterjoin \{\{?var_1 \mapsto y_1,?Y_1 \mapsto 0\}\}$\\
\>$= \{\{?var_1 \mapsto x_1,?X_1 \mapsto 1\},\{?var_1 \mapsto y_1, ?Y_1 \mapsto 0\}\}$\\
\\[-0.2cm]
\>$\evalcd{P_{C_2}}$\>$=\{\{?var_2 \mapsto x_1\},\{?var_2 \mapsto y_1\}\}$\\
\>\>\>$\leftouterjoin \{\{?var_2 \mapsto y_1,?Y_1 \mapsto 1\}\}$\\
\>\>\>\>$\leftouterjoin \{\{?var_2 \mapsto x_1,?X_1 \mapsto 0\}\}$\\
\>$= \{\{?var_2 \mapsto x_1,?X_1 \mapsto 0\},\{?var_2 \mapsto y_2, ?Y_2 \mapsto 1\}\}$\\
\\[-0.2cm]
\>$\evalcd{P_\psi}$\>$= \evalcd{P_{C_1} \oand P_{C_2}}$\\
\>$= \{\{?var_1 \mapsto x_1,?var_2 \mapsto y_1,?X_1 \mapsto 1,?Y_1 \mapsto 1\},$\\
\>$\ \ \ \ \ \ \ \{?var_1 \mapsto y_1,?var_2 \mapsto x_1,?X_1 \mapsto 0,?Y_1 \mapsto 0\}\}$
\end{tabbing}

Finally, we set up the expressions $P_1$ and $Q_1$, as described in the proof
for fragment $\cal{AOF}$

\begin{tabbing}
x \= xx \= xx \= \kill
\>$P_1$\>${:=}$\>$((a,tv,?X_1) \oand (a,{\it false},?A_0)$\\
\>\>\>$\oand (a,{\it true},?A_1))$\\
\>$Q_1$\>${:=}$\>$((a,tv,?X_1) \oand (a,tv,?Y_1)$\\
\>\>\>$\oand (a,{\it false},?B_0) \oand (a,{\it true},?B_1))$
\end{tabbing}

and encode the quantified boolean formula $\varphi$ as

\begin{tabbing}
x \= \kill
\>$P_\varphi := (a,{\it true},?B_0) \oopt (P_1 \oopt (Q_1 \oand P_\psi))$
\end{tabbing}

We leave it as an exercise to verify that the mapping
$\mu = \{ ?B_0 \mapsto 1\}$ is contained in $\evalcd{P_\varphi}$. This
result confirms that the original formula $\psi$ is valid.$\punto$
\end{example}

\subsubsection*{Fragment $\cal{O}$: OPT-only (Theorem~\ref{th:opt})}

We start with a transformation rule for operator \textsc{And}; it essentially 
expresses the key idea of the subsequent proof.

\begin{lemma}
\em
Let

\begin{itemize}
\item $Q, Q_1, Q_2, \dots, Q_n$ ($n \geq 2$) be SPARQL expressions,
\item $S = \it{vars}(Q) \cup \it{vars}(Q_1) \cup {\it vars}(Q_2) \cup \dots \cup \it{vars}(Q_n)$,
denote the set of variables in $Q, Q_1, Q_2, \dots, Q_n$
\item $D = \{ (a,true, 1), (a,false,0), (a,tv,0), (a,tv,1) \}$ be a fixed database,
\item $?V_2, ?V_3, \dots, ?V_n$ be a set of $n-1$ fresh variables,
i.e.~$S \cap \{ ?V_2, ?V_3, \dots, ?V_n \} = \emptyset$ holds.
\end{itemize}

Further, we define

\begin{tabbing}
x \= xx \= xxxxxxxxxxl \= xx \= \kill
\>$Q' {:=} ((\dots ((Q \oopt V_2) \oopt V_3) \dots ) \oopt V_n)$,\\
\>$Q'' {:=} ((\dots((Q_1 \oopt (Q_2 \oopt V_2))$\\
\>\>\>$\oopt (Q_3 \oopt V_3))$\\
\>\>\>\>$\dots$\\
\>\>\>\>$\oopt (Q_n \oopt V_n))),$\\
\>$V_i {:=} (a,true,?V_i)$, and\\
\>$\overline{V}_i {:=} (a,false,?V_i)$.
\end{tabbing}

The following claims hold. 

\begin{tabbing}
x \= l \= xxlxxxxxxxxxxxxxxx \= xx \= \kill
\>(1) $\evalcd{Q'} = \{ \mu \cup \{ ?V_2 \mapsto 1, \dots, ?V_n \mapsto 1\} \mid \mu \in \evalcd{Q} \}$,\\
\>(2) $\evalcd{Q' \oopt (Q_1 \oand Q_2 \oand \dots \oand Q_n)}$\\
\>\>$= \llbracket Q' \oopt (\dots((Q'' \oopt \overline{V}_2)$\\
\>\>\>$\oopt \overline{V}_3)$\\
\>\>\>$\ \ \dots$\\
\>\>\>\ \ $\oopt \overline{V}_n)\rrbracket_D$
\end{tabbing}
\label{lemma:andrewriting}
\vspace{-0.6cm}
$\punto$
\end{lemma}

The second part of the lemma provides a way to rewrite an \textsc{And}-only
expression that is encapsulated in the right side of an \textsc{Opt}-expression
by means of an \textsc{Opt}-only expression. Before proving the lemma, we
illustrate the construction by means of a small example.

\begin{example}
\em
Let $D$ be the database given in the previous lemma and consider the
expressions

\begin{tabbing}
x \= xx \= xxxxxxxxxxxxl \= xxxxxxxx \= \kill
\>$Q$\>$:= (a,tv,?a)$\>,i.e.~$\evalcd{Q}$\>$=\{ \{ ?a \mapsto 0 \}, \{ ?a \mapsto 1 \} \}$\\
\>$Q_1$\>$:= (a,true,?a)$\>,i.e.~$\evalcd{Q_1}$\>$=\{ \{ ?a \mapsto 1 \} \}$\\
\>$Q_2$\>$:= (a,false,?b)$\>,i.e.~$\evalcd{Q_2}$\>$=\{ \{ ?b \mapsto 0 \} \}$
\end{tabbing}

As for the part~(1)~of the lemma we observe that

\begin{tabbing}
x \= xxxxl \= \kill
\>$\evalcd{Q'}$\>$= \evalcd{Q \oopt V_2}$\\
\>\>$= \evalcd{Q \oopt (a,true,?V_2)}$\\
\>\>$= \{ \{ ?a \mapsto 0, ?V_2 \mapsto 1 \}, \{ ?a \mapsto 1, ?V_2 \mapsto 1 \} \}$.
\end{tabbing}

Concerning part~(2) it holds that the left side

\begin{tabbing}
x \= \kill
\>$\evalcd{Q' \oopt (Q_1 \oand Q_2)}$\\
\>$= \evalcd{Q'} \leftouterjoin \{ \{ ?a \mapsto 1, ?b \mapsto 0 \} \}$\\
\>$= \{ \{ ?a \mapsto 0, ?V_2 \mapsto 1 \}, \{ ?a \mapsto 1, ?b \mapsto 0, ?V_2 \mapsto 1\} \}$
\end{tabbing}

is equal to the right side

\begin{tabbing}
x \= \kill
\>$\evalcd{Q' \oopt ((Q_1 \oopt (Q_2 \oopt V_2)) \oopt \overline{V_2})}$\\
\>$= \evalcd{Q'} \leftouterjoin (\{\{?a \mapsto 1, ?b \mapsto 0, ?V_2 \mapsto 1\}\} \leftouterjoin \evalcd{\overline{V_2}})$\\
\>$= \evalcd{Q'} \leftouterjoin \{\{?a \mapsto 1, ?b \mapsto 0, ?V_2 \mapsto 1\}\}$\\
\>$= \{\{ ?a \mapsto 0, ?V_2 \mapsto 1 \}, \{ ?a \mapsto 1, ?b \mapsto 0, ?V_2 \mapsto 1\}\}$.$\punto$
\end{tabbing}

\end{example}

{\it Proof of Lemma~\ref{lemma:andrewriting}}.
We omit some technical details, but instead give the intuition of the
encoding.
(1)~The first claim follows trivially from the definition of $Q'$, the observations that
each $V_i$ evaluates to $\{\{?V_i \mapsto 1\}\}$, and the fact that all $?V_i$ are unbound in $Q'$
(recall that, by assumption, the $?V_i$ are fresh variables). To prove~(2), we consider
the evaluation of the right side expression, in order to show that it yields the same result as
the left side. First consider
subexpression $Q''$ and observe that the result of evaluating $Q_i \oopt V_i$ is exactly
the result of evaluating $Q_i$ extended by the binding $?V_i \mapsto 1$. In the sequel,
we use $Q^V_i$ as an abbreviation for $Q_i \oopt V_i$, i.e.~we denote 
$Q''$ as $((\dots((Q_1 \oopt Q^V_2) \oopt Q^V_3) \oopt \dots ) \oopt Q^V_n)$. Applying semantics,
we can rewrite $\evalcd{Q''}$ into the form 

\begin{tabbing}
x \= xl \= \kill
\>$\evalcd{Q''}$\\
\>$=$\>$\evalcd{((\dots((Q_1 \oopt Q^V_2) \oopt Q^V_3) \oopt \dots ) \oopt Q^V_n)}$\\ 
\>$=$\>$\evalcd{(Q_1 \oand Q^V_2 \oand Q^V_3 \oand \dots \oand Q^V_n)} \cup P_D$,
\end{tabbing}

where we call the left subexpression of the union {\it join part}, and $P_D$ at the right side
is an algebra expression (over database $D$) with the following property:
for each mapping $\mu \in P_D$ there is at least one $?V_i$ ($2 \leq i \leq n$)
s.t.~$?V_i \not \in {\it dom}(\mu)$. We observe that, in contrast, for each
mapping in the join part
${\it dom}(\mu) \supseteq \{ ?V_2, \dots, ?V_n \}$ holds and, even more,
$\mu(?V_i) = 1$, for $2 \leq i \leq n$. Hence, the mappings in the result
of the join part are identified
by the property that $?V_2, ?V_3, \dots, ?V_n$ are all bound to 1.

Let us next consider the evaluation of the larger expression (on the right
side of the original equation)

\begin{tabbing}
l \= xx \= xxxxxxxxxxxxxx \= xx \= \kill
\>$R := ((\dots((Q'' \oopt \overline{V}_2) \oopt \overline{V}_3) \oopt \dots ) \oopt \overline{V}_n))$.
\end{tabbing}

When evaluating $R$, we obtain exactly the mappings
from $\evalcd{Q''}$, but each mapping $\mu \in \evalcd{Q''}$ is extended by bindings
$?V_i \mapsto 0$ for all $?V_i \not \in {\it dom}(\mu)$ (cf.~the argumentation in for
claim~(1)). As argued before, all mappings in the join part of $Q''$ are complete
in the sense that all $?V_i$ are bound, so these mappings will not be affected.
The remaining mappings (i.e.~those originating from $P_D$) will be extended by bindings
$?V_i \mapsto 0$ for at least one $?V_i$. The resulting situation can be summarized
as follows: all mappings resulting from the join part of $Q''$ bind all variables $?V_i$ to 1;
all mappings in $P_D$ bind all $?V_i$, but at least one of them is bound to 0.

From part (1)~we know that each mapping in $\evalcd{Q'}$ maps all $?V_i$ to 1. 
Hence, when computing
$\evalcd{Q' \oopt R} = \evalcd{Q'} \leftouterjoin \evalcd{R}$, the bindings $?V_i \mapsto 1$
for all $\mu \in \evalcd{Q'}$ serves as a filter that removes the mappings in 
$\evalcd{R}$ originating from $P_D$. This means

\begin{tabbing}
x \= \kill
\>$\evalcd{Q' \oopt R}$\\ 
\>$= \evalcd{Q'} \leftouterjoin \evalcd{R}$\\ 
\>$= \evalcd{Q'} \leftouterjoin \evalcd{(Q_1 \oand Q^V_2 \oand Q^V_3 \oand \dots \oand Q^V_n)}$\\
\>$= \evalcd{Q' \oopt (Q_1 \oand Q^V_2 \oand Q^V_3 \oand \dots \oand Q^V_n)}$.
\end{tabbing}

Even more, we observe that all $?V_i$ are already bound in $Q'$ (all of them to $1$),
so the following rewriting is valid.

\begin{tabbing}
x \= \kill
\>$\evalcd{Q' \oopt R}$\\ 
\>$= \evalcd{Q' \oopt (Q_1 \oand Q^V_2 \oand Q^V_3 \oand \dots \oand Q^V_n)}$\\
\>$= \evalcd{Q' \oopt (Q_1 \oand Q_2 \oand Q_3 \oand \dots \oand Q_n)}$
\end{tabbing}

Thus, we have shown that the equivalence holds. This completes the proof. $\qed$

Given Lemma~\ref{lemma:andrewriting} we are now in the position to prove
\textsc{PSpace}-completeness for fragment~$\cal{O}$. As in previous
proofs it suffices to show hardness; membership follows as before from
the \textsc{PSpace}-completeness of fragment $\cal{E}$.

The proof idea is the following. We show that, in the previous reduction
from $\textsc{QBF}$ to \textsc{Evaluation} for fragment $\cal{AO}$,
each \textsc{And} expression can be rewritten using only \textsc{Opt} operators.
We start with a QBF of the form

\begin{tabbing}
x \= \kill
\>$\varphi = \forall x_1 \exists y_1 \forall x_2 \exists y_2 \dots \forall x_m \exists y_m \psi$,
\end{tabbing}

where $\psi$ is a quantifier-free formula in conjunctive normal form (CNF), i.e.
$\psi$ is a conjunction of clauses

\begin{tabbing}
x \= \kill
\>$\psi=C_1 \land \dots \land C_n$,
\end{tabbing}

where the $C_i$, $1 \leq i \leq n$, are disjunctions of literals.
By $V$ we denote the set of variables
inside $\psi$ and by $V_{C_i}$ the variables appearing in clause
$C_i$ (either in positive of negative form) and use the same
database as in the proof for fragment $\cal{AO}$, namely

\begin{tabbing}
x \= xxxx \= \kill
\>$D :=$\>$\{ (a,\textit{tv},0), (a,\textit{tv},1), (a,\textit{false},0), (a,\textit{true},1) \}\ \cup$\\
\>\>$\{ (a,var_i,v) \mid v \in V_{C_i} \}\ \cup\ \{ (a,v,v) \mid v \in V \}$
\end{tabbing}

The first modification of the proof for class $\cal{AO}$ concerns the
encoding of clauses
$C_i = v_1 \lor \dots \lor v_j \lor \neg v_{j+1} \lor \dots \lor \neg v_k$,
where the $v_1 \dots v_j$ are positive and $v_{j+1} \dots v_k$ are negated 
variables.
In the prior encoding we used both \textsc{And} and \textsc{Opt} operators
to encode them. It is easy to see that we can simply replace each \textsc{And}
operator there through \textsc{Opt} without changing semantics. The reason is that,
for all subexpressions $A \oopt B$ in the encoding, it holds that
${\it vars}(A) \cap {\it vars}(B) = \emptyset$ and $\evalcd{B} \not = \emptyset$;
hence, all mappings in $A$ are compatible with all mappings in $B$ and there
is at least one mapping in $B$. When applying
this modification, we obtain the following $\cal{O}$-encoding for clauses $C_i$.

\begin{tabbing}
x \= l \= l \= l \= l \= l \= l \= l \= l \= l \= l \= l \= \kill
\>$P_{C_i} {:=} (\dots((\dots($\\
\>\>$(a,var_i,?var_i)$\\
\>\>\>$\ \textsc{Opt}\ ((a,v_1,?var_i) \oopt (a,true,?V_1)))$\\
\>\>\>\>$\ \ \ldots$\\
\>\>\>\>\>$\ \ \ \textsc{Opt}\ ((a,v_j,?var_i) \oopt (a,true,?V_j)))$\\
\>\>\>\>\>\>$\ \ \ \ \textsc{Opt}\ ((a,v_{j+1},?var_i) \oopt (a,false,?V_{j+1})))$\\
\>\>\>\>\>\>\>$\ \ \ \ \ \ldots$\\
\>\>\>\>\>\>\>\>$\ \ \ \ \ \ \textsc{Opt}\ ((a,v_k,?var_i) \oopt (a,false,?V_k)))$,
\end{tabbing}

Let us next consider the $P_i$ and $Q_i$ used for simulating the quantifier alternations.
The original definition of these expression was given in the proof for fragment
$\cal{AFO}$. With a similar argumentation as before we can replace each occurrence
of operator \textsc{And} through \textsc{Opt} without changing the semantics of the whole
expression. This results in the following $\cal{O}$ encodings for $P_i$ and $Q_i$,
$i \in [m]$.

\begin{tabbing}
x \= xx \= xx \= \kill
\>$P_i$\>${:=}$\>$((a,tv,?X_1) \oopt \dots \oopt (a,tv,?X_i) \oopt$\\
\>\>\>$\ \ (a,tv,?Y_1) \oopt \dots \oopt (a,tv,?Y_{i-1}) \oopt$\\
\>\>\>$\ \ (a,{\it false},?A_{i-1}) \oopt (a,{\it true},?A_i))$\\
\>$Q_i$\>${:=}$\>$((a,tv,?X_1) \oopt \dots \oopt (a,tv,?X_i) \oopt$\\
\>\>\>$\ \ (a,tv,?Y_1) \oopt \dots \oopt (a,tv,?Y_i) \oopt$\\
\>\>\>$\ \ (a,{\it false},?B_{i-1}) \oopt (a,{\it true},?B_i))$,\\
\end{tabbing}

In the underlying proof for $\cal{AO}$, the conjunction $\psi$
is encoded as $P_{C_1} \oand \dots \oand P_{C_n}$, thus we have not yet eliminated
all \textsc{And}-operators. We shortly summarize what we have achieved so far:

\begin{tabbing}
x \= xxxxl \= x \= x \= x \= x \= \kill
\>$P_\varphi {:=}$\>$((a,{\it true},?B_0)$\\
\>\>\>$\oopt (P_1 \oopt (Q_1$\\
\>\>\>\>$\dots$\\
\>\>\>\>$\oopt (P_{m-1} \oopt (Q_{m-1}$\\
\>\>\>\>\>$\oopt (P'))) \dots ))),$ where\\
\\[-0.2cm]
\>$P' = P_m \oopt (Q_m \oand (P_\psi))$\\ 
\>$\ \ \ \ \ \ = P_m \oopt (Q_m \oand P_{C_1} \oand \dots \oand P_{C_n})$   
\end{tabbing}

Note that $P'$ is the only expression that still contains \textsc{And} operators
(where $Q_m$, $P_{C_1}$, $\dots$, $P_{C_n}$ are already \textsc{And}-free). 
We now exploit the rewriting given in Lemma~\ref{lemma:andrewriting}.
In particular, we replace $P'$ in $P_\varphi$ by the expression $P'_*$ defined as

\begin{tabbing}
x \= xx \= xxxxxxxxxxxx \= x \= x \= x \= \kill
\>$P'_* := Q' \oopt$\\
\>\>$((\dots((Q'' \oopt \overline{V}_2) \oopt \overline{V}_3) \oopt \dots ) \oopt \overline{V}_{n+1}))$,\\
\\[-0.2cm]
\>where\\
\>$Q' {:=} ((\dots ((P_m \oopt V_2) \oopt V_3) \dots ) \oopt V_{n+1})$,\\
\>$Q'' {:=} ((\dots((Q_m \oopt (P_{C_1} \oopt V_2))$\\
\>\>\>$\oopt (P_{C_2} \oopt V_3))$\\
\>\>\>\>$\dots$\\
\>\>\>\>$\oopt (P_{C_n} \oopt V_{n+1}))),$\\
\>$V_i {:=} (a,true,?V_i)$,\\
\>$\overline{V}_i {:=} (a,false,?V_i)$,\\
\>and the $?V_i$ ($i \in \{2, \dots, n+1\}$) are fresh variables.
\end{tabbing}

The resulting $P_\varphi$ is now an $\cal{O}$-expression.
From Lemma~\ref{lemma:andrewriting} it follows that the result of evaluating
$P'_*$ is obtained from the result of $P'$ by extending each mapping in $P'$
by additional variables, more precisely by
$\{ ?V_2 \mapsto 1, ?V_3 \mapsto 1, \dots, ?V_{n+1} \mapsto 1 \}$, i.e.~the results
are identical modulo this extension. It is straightforward to verify
that these additional bindings do not harm the construction, i.e.~it holds
that
$\{ ?B_0 \mapsto 1 \} \in \evalcd{P_\varphi}$ iff $\varphi$ is valid.$\qed$

\subsubsection*{$\Sigma^P_{n+1}$-completeness of Fragment $\cal{E}$$_{\leq n}$
(Theorem~\ref{th:ph})}

We start with two lemmas that will be used in the proof. 

\begin{lemma}
\em
Let

\begin{tabbing}
x \= \kill
\>$D = \{ (a,tv,0), (a,tv,1), (a,true,1), (a,false,0) \}$
\end{tabbing}

be an RDF database and $F = \forall x_1 \exists y_1 \dots \forall x_m \exists y_m \psi$
($m \geq 1$) be a QBF, where $\psi$ is a quantifier-free boolean formula.
There is an $\cal{E}$$_{\leq 2m}$ encoding ${\it enc}(F)$ of $F$ s.t.

\begin{enumerate}
\item $F$ is valid exactly if $\{ ?B_0 \mapsto 1\} \in \evalcd{{\it enc}(F)}$
\item $F$ is invalid exactly if all mappings $\mu' \in \evalcd{{\it enc}(F)}$
are of the form $\mu' = \mu_1' \cup \mu_2'$, where $\mu_1'\sim\mu_2'$ and
$\mu_1' = \{ ?B_0 \mapsto 1, ?A_1 \mapsto 1 \}$.$\punto$
\end{enumerate}
\label{lemma:pimembership}
\vspace{-0.4cm}
\end{lemma}

{\it Proof:} The lemma follows from the
\textsc{PSpace}-hardness proof for fragment $\cal{AFO}$, where
we have shown how to encode \textsc{QBF} for a (possibly non-CNF) inner formula $\psi$.$\qed$

\begin{lemma}
\em
Let $A$ and $B$ SPARQL expressions for which the evaluation
problem is in $\Sigma^P_i$, $i \geq 1$, and let $R$ a \textsc{Filter} condition.
The following claims hold.

\begin{enumerate}
\item The \textsc{Evaluation} problem for the SPARQL expression $A \ounion B$ is in $\Sigma^P_i$.
\item The \textsc{Evaluation} problem for the SPARQL expression $A \oand B$ is in $\Sigma^P_i$.
\item The \textsc{Evaluation} problem for the SPARQL expression $A \ofilter R$ is in $\Sigma^P_i$.$\punto$
\end{enumerate}
\label{lemma:lin}
\end{lemma}

{\it Proof:}
1. According to the SPARQL semantics we have that $\mu \in \evalc{A \ounion B}$
if and only if $\mu \in \evalc{A}$ or $\mu \in \evalc{B}$. By assumption, both
conditions can be checked individually in $\Sigma^P_i$, and so can both checks in
sequence.

2. It is easy to see that $\mu \in \evalc{A \oand B}$ iff $\mu$ can be
decomposed into two compatible mappings $\mu_1$ and $\mu_2$
s.t.~$\mu = \mu_1 \cup \mu_2$ and $\mu_1 \in \evalc{A}$ and $\mu_2 \in \evalc{B}$.
By assumption, testing $\mu_1 \in \evalc{A}$ ($\mu_2 \in \evalc{B}$) is in
$\Sigma^P_i$. Since $i \geq 1$, this complexity class is at least
$\Sigma^P_1 = \textsc{NP}$. So we can guess a decomposition
$\mu = \mu_1 \cup \mu_2$ and test for the two conditions
one after the other. Hence, the whole procedure is in $\Sigma^P_i$.

3. $\mu \in \evalc{A \ofilter R}$ holds iff $\mu \in \evalc{A}$, which can be tested in
$\Sigma^P_1$ by assumption, and $R$ satisfies $\mu$, which can be tested in polynomial
time. Since $\Sigma^P_i \supseteq \textsc{NP} \supseteq P$ for $i \geq 1$, the whole
procedure is still in $\Sigma^P_i$.$\qed$

We are now ready to prove Theorem~\ref{th:ph}. The proof divides into
two parts, i.e. hardness and membership. The hardness proof is a reduction from
\textsc{QBF} with a fixed number of quantifier alternations. Second,
we prove by induction on the \textsc{Opt}-rank that there exists a
$\Sigma^P_{n+1}$-algorithm to solve the \textsc{Evaluation} problem for
$\cal{E}$$_{\leq n}$ expressions.

\medskip

{\it Hardness.} We consider a QBF of the form

\begin{tabbing}
x \= \kill
\>$\varphi = \exists x_0 \forall x_1 \exists x_2 \dots Q x_n \ \psi$,
\end{tabbing}

where $n \geq 1$, $Q = \exists$ if $n$ is even, $Q = \forall$
if $n$ is odd, and $\psi$ is a quantifier-free boolean formula.
It is known that the \textsc{Validity} problem for such formulae is
$\Sigma^P_{n+1}$-complete. We now present a (polynomial-time) reduction
from the \textsc{Validity} problem for these quantified boolean formulae
to the \textsc{Evaluation} problem for the $\cal{E}$$_{\leq n}$
fragment, to prove $\Sigma^P_{n+1}$-hardness. We distinguish two cases.

{\it Case 1:} Let $Q = \exists$, so the formula is of the form

\begin{tabbing}
x \= \kill
\>$F = \exists y_0 \forall x_1 \exists y_1 \dots \forall x_m \exists y_m \psi$.
\end{tabbing}

The formula $F$ has $2m+1$ quantifier alternations, so we need to find an
$\cal{E}$$_{\leq 2m}$ encoding for this expressions. We rewrite $F$ into
an equivalent formula $F = F_1 \lor F_2$, where

\begin{tabbing}
x \= xxxx \= \kill
\>$F_1 =$\>$\forall x_1 \exists y_1 \dots \forall x_m \exists y_m (\psi \land y_0)$, and\\
\>$F_2 =$\>$\forall x_1 \exists y_1 \dots \forall x_m \exists y_m (\psi \land \neg y_0)$.
\end{tabbing}

According to Lemma~\ref{lemma:pimembership} there is a fixed document $D$ and
$\cal{E}$$_{\leq 2m}$ encodings ${\it enc}(F_1)$ and ${\it enc}(F_2)$ (for $F_1$ and
$F_2$, respectively)
s.t.~$\evalcd{{\it enc}(F_1)}$ ($\evalcd{{\it enc}(F_2)}$) contains the mapping
$\mu = \{ ?B_0 \mapsto 1 \}$ if and only if $F_1$ ($F_2$) is valid. Then the expression
${\it enc}(F) = {\it enc}(F_1) \ounion {\it enc}(F_2)$ contains $\mu$
if and only if $F_1$ or $F_2$ is valid, i.e.~iff $F$ is valid. Clearly, ${\it enc}(F)$
is an $\cal{E}$$_{\leq 2m}$ expression, so ${\it enc}(F)$ constitutes the desired
$\cal{E}$$_{\leq 2m}$ encoding of the \textsc{Evaluation} problem.

{\it Case 2:} Let $Q = \forall$, so the formula is of the form

\begin{tabbing}
x \= \kill
\>$F = \exists x_0 \forall y_0 \exists x_1 \forall x_1 \dots \exists x_m \forall y_m \psi$.
\end{tabbing}

$F$ has $2m+2$ quantifier alternations, so we need to provide a reduction into
the $\cal{E}$$_{\leq 2m+1}$ fragment. We eliminate the outer $\exists$-quantifier
by rewriting $F$ as $F = F_1 \lor F_2$, where

\begin{tabbing}
x \= xxxx \= \kill
\>$F_1 =$\>$\forall y_0 \exists x_1 \forall y_1 \dots \exists x_m \forall y_m (\psi \land y_0)$, and\\
\>$F_2 =$\>$\forall y_0 \exists x_1 \forall y_1 \dots \exists x_m \forall y_m (\psi \land \neg 
y_0)$.
\end{tabbing}

Abstracting from the details of the inner formula, both $F_1$ and $F_2$ are of the form

\begin{tabbing}
x \= \kill
\>$F' = \forall y_0 \exists x_1 \forall y_1 \dots \exists x_m \forall y_m \psi'$,
\end{tabbing}

where $\psi'$ is a quantifier-free boolean formula. We now show (*) that we can encode
$F'$ by $\cal{E}$$_{\leq 2m+1}$ expressions ${\it enc}(F')$ that, evaluated on a
fixed document $D$, yields a fixed mapping $\mu$ exactly if $F'$ is valid.
This is sufficient, because then the expression
${\it enc}(F_1) \ounion {\it enc}(F_2)$ is an $\cal{E}$$_{\leq 2m+1}$
that contains $\mu$ exactly if the original formula $F = F_1 \lor F_2$ is valid.
We again start by rewriting $F'$:

\begin{tabbing}
x \= xxx \= xx \= \kill
\>$F'$\>$=$\>$\forall y_0 \exists x_1 \forall y_1 \dots \exists x_m \forall y_m \psi'$\\
\>\>$=$\>$\neg \exists y_0 \forall x_1 \exists y_1 \dots \forall x_m \exists y_m \neg \psi'$\\
\>\>$=$\>$\neg (F_1' \lor F_2')$, where\\
\\[-0.3cm]
\>$F_1'$\>$=$\>$\forall x_1 \exists y_1 \dots \forall x_m \exists y_m (\neg \psi' \land y_0)$, and\\
\>$F_2'$\>$=$\>$\forall x_1 \exists y_1 \dots \forall x_m \exists y_m (\neg \psi' \land \neg y_0)$.
\end{tabbing}

According to Lemma~\ref{lemma:pimembership}, each $F_i'$ can be encoded by an
$\cal{E}$$_{\leq 2m}$ expressions ${\it enc}(F_i')$ s.t., on the fixed database
$D$ given there, (1)~$\mu = \{ ?B_0 \mapsto 1 \} \in \evalcd{F_i'}$ iff $F_i'$ is
valid and (2)~if $F_i'$ is not valid, then all mappings $\evalcd{{\it enc}(F_i')}$ bind
both variables $?A_1$ and $?B_0$ to $1$. Then the same conditions (1)~and (2)~hold
for ${\it enc}(F_1') \ounion {\it enc}(F_2')$ exactly if $F_1 \lor F_2$
is valid. Now consider the expression
$Q = ((a,{\it false},?A_1) \oopt ({\it enc}(F_1) \ounion {\it enc}(F_2))$. This expression
contains $\mu' = \{ ?A_1 \mapsto 0 \}$ (when evaluated on the database given in
Lemma~\ref{lemma:pimembership}) if and only if $F_1' \lor F_2'$ is not valid
(since otherwise, there is a compatible mapping for $\mu'$, namely $\{ ?B_0 \mapsto 1 \}$
in $\evalcd{{\it enc}(F_1) \ounion {\it enc}(F_2)}$).
In summary, this means $\mu' \in \evalcd{Q}$ if and only if $\neg (F_1' \lor F_2') = F'$
holds. Since both ${\it enc}(F_1)$ and ${\it enc}(F_2)$ are $\cal{E}$$_{\leq 2m}$
expressions, $Q$ is contained in $\cal{E}$$_{\leq 2m+1}$, so (*) holds.

\medskip

\begin{figure*}[t]
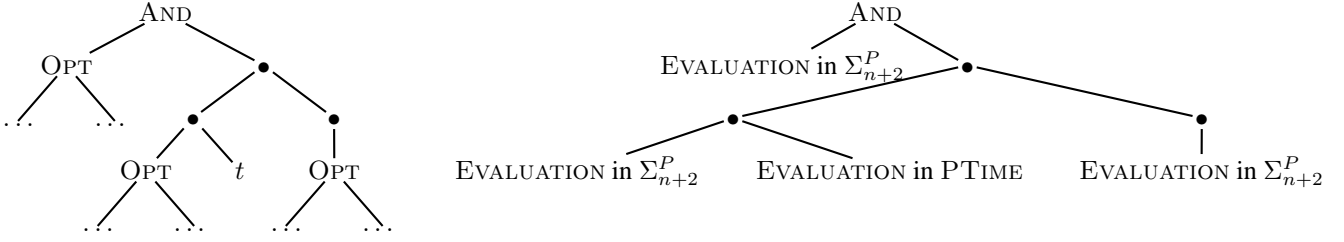

\hspace{-1.5cm}
\begin{center}
\begin{tabular}{ccc}
\pstree[treefit=tight,edge=\ncline,levelsep=0.7cm,nodesep=1pt]{\TR{$\oand$}}{
	\pstree{\TR{$\oopt$}}{
		\TR{$\dots$}
		\TR{$\dots$}
	}
	\pstree{\TR{$\bullet$}}{
		\pstree{\TR{$\bullet$}}{
			\pstree{\TR{$\oopt$}}{
				\TR{$\dots$}
				\TR{$\dots$}
			}
			\TR{$t$}
		}
		\pstree{\TR{$\bullet$}}{
			\pstree{\TR{\oopt}}{
				\TR{$\dots$}
				\TR{$\dots$}
			}
		}
	}
}
&
\ 
&
\pstree[treefit=tight,edge=\ncline,levelsep=0.7cm,nodesep=1pt]{\TR{$\oand$}}{
	\TR{\textsc{Evaluation} in $\Sigma^P_{n+2}$}
	\pstree{\TR{$\bullet$}}{
		\pstree{\TR{$\bullet$}}{
			\TR{\textsc{Evaluation} in $\Sigma^P_{n+2}$}
			\TR{\textsc{Evaluation} in \textsc{PTime}}
		}
		\pstree{\TR{$\bullet$}}{
			\TR{\textsc{Evaluation} in $\Sigma^P_{n+2}$}
		}
	}
}
\end{tabular}
\end{center}
\caption{(a) AND-expression with increased OPT-rank; (b) The OPT-expressions and leaf nodes
have been replaced by the complexity class of their EVALUATION problem.}
\label{fig:samplefig}
\end{figure*}

{\it Membership.} We next prove membership of $\cal{E}$$_{\leq n}$ expressions in
$\Sigma^P_{n+1}$ by induction on the \textsc{Opt}-rank. Let us assume
that for each $\cal{E}$$_{\leq n}$ expression
($n \in \mathbb{N}_0$) \textsc{Evaluation} is in $\Sigma^P_{n+1}$. As stated
in Theorem~\ref{th:pag}(2), \textsc{Evaluation} is $\Sigma^P_1$ = \textsc{NP}-complete
for \textsc{Opt}-free expressions, so the hypothesis holds for the basic case.
In the induction step we increase the \textsc{Opt}-rank from $n$ to $n+1$.

We distinguish several cases, depending on the structure of the expression,
say $A$, with \textsc{Opt}-rank $n+1$.

\medskip

{\it Case 1: Checking if $\mu \in \evalc{A \oopt B}$.} First note that
$A \oopt B$ is in $\cal{E}$$_{\leq n+1}$, and from the definition of \textsc{Opt}-rank
$r$ it follows immediately that both $A$ and $B$ are in $\cal{E}$$_{\leq n}$.
Hence, by induction hypothesis, both $A$ and $B$ can be evaluated in $\Sigma^P_{n+1}$.
By semantics, we have that
$\evalc{A \oopt B} = \evalc{A \oand B} \cup (\evalc{A} \setminus \evalc{B})$,
so $\mu$ is  in $\evalc{A \oopt B}$ iff it is generated by the~(a) left or~(b)
right side of the union. Following Lemma~\ref{lemma:lin}, part~(a) can be checked in
$\Sigma^P_{n+1}$. (b) The more interesting part is to check if
$\mu \in \evalc{A} \setminus \evalc{B}$. According to the semantics of operator
$\setminus$, this check can be formulated as $C = C_1 \land C_2$, where
$C_1 = \mu \in \evalc{A}$ and
$C_2=\neg \exists \mu' \in \evalc{B}: \mu\ {\it and}\ \mu'\ {\it are\ compatible}$.
By induction hypothesis, $C_1$ can be checked in $\Sigma^P_{n+1}$. We argue that also
$\neg C_2 = \exists \mu' \in \evalc{B}: \mu\ {\it and}\ \mu'\ {\it are\ compatible}$
can be evaluated in $\Sigma^P_{n+1}$: we can guess a mapping $\mu'$ (in \textsc{NP}),
then check if $\mu \in \evalc{B}$ (in $\Sigma^P_{n+1}$), and finally check if
$\mu$ and $\mu'$ are compatible (in polynomial time). Since
$P \subseteq NP \subseteq \Sigma^P_{n+1}$, all these checks in sequence can be done
in $\Sigma^P_{n+1}$. Checking if the inverse problem, i.e.~$C_2$, holds is then
possible in $co\Sigma^P_{n+1} = \Pi^P_{n+1}$.
Summarizing cases (a) and (b) we observe that (a)~$\Sigma^P_{n+1}$ and (b)~$\Pi^P_{n+1}$
are contained in $\Sigma^P_{n+2}$, so both checks in sequence can be executed in
$\Sigma^P_{n+2}$, which completes case 1.

\medskip

{\it Case 2: Checking if $\mu \in \evalc{A \oand B}.$}
Figure~\ref{fig:samplefig}(a) shows the structure of a sample \textsc{And}
expression, where the $\bullet$ symbols represent non-\textsc{Opt} operators
(i.e.~\textsc{And}, \textsc{Union}, or \textsc{Filter}), and $t$ stands for
triple patterns. There is an arbitrary number of \textsc{Opt}-subexpression
(which might, of course, contain \textsc{Opt} subexpression themselves).
Each of these subexpressions has \textsc{Opt}-rank $\leq n+1$. Using the
same argumentation as in case (1), the evaluation problem for all of them
is in $\Sigma^P_{n+2}$. Moreover, there might be triple leaf nodes; the
evaluation problem for such patterns is in \textsc{PTime}, and clearly
$P \supseteq \Sigma^P_{n+2}$. Figure~\ref{fig:samplefig}(b) illustrates the
situation when all \textsc{Opt}-expressions and triple patterns have been
replaced by the complexity of their \textsc{Evaluation} problem.

We then proceed as follows. We apply Lemma~\ref{lemma:lin} repeatedly,
folding the remaining \textsc{And}, \textsc{Union}, and \textsc{Filter}
subexpressions bottom up. The lemma guarantees that these folding operations
do not increase the complexity class, and it is easy to prove that
\textsc{Evaluation} remains in $\Sigma^P_{n+2}$ for the whole expression.

\medskip

{\it Case 3: Checking if $\mu \in \evalc{A \ounion B}$ and
Case 4: Checking if $\mu \in \evalc{A \ofilter R}.$} Similar to case 2.

\medskip

It is worth mentioning that the structural induction is polynomially bounded
by the size of the expression when the nesting depth of \textsc{Opt}-operators
is fixed (which holds by assumption), i.e.~comprises only polynomially many
steps. In all cases except case 1 the recursive calls concern subexpressions
of the original expression. In case 1, $\mu \in \evalc{A \oopt B}$ generates
two checks, namely
$\mu \in \evalc{A \oand B}$ and $\mu \in \evalc{A} \setminus \evalc{B}$.
These two checks might trigger recursive checks again. But since the nesting
depth of \textsc{Opt}-expressions is restricted, it is easy to see that
there is a polynomial that bounds the number of recursive call.
$\qed$

\subsection{Queries: Fragments Including SELECT}

\subsection*{Proof of Theorem~\ref{th:select1}}

Let $F$ be a fragment for which the \textsc{Evaluation}
problem is $\textsc{C}$-complete, where \textsc{C} is a complexity class
s.t.~$\textsc{NP} \subseteq \textsc{C}$. We show that, for a query
$Q \in F_+$, document $D$, and mapping $\mu$, testing
if $\mu \in \evalcd{Q}$ is contained in \textsc{C} (\textsc{C}-hardness
follows trivially from \textsc{C}-hardness of fragment $F$).
By definition, each query in $F_+$ is of the form $Q = \textsc{Select}_S(Q')$,
where $S \subset V$ is a finite set of variables and $Q'$ is an $F$-expression.
By definition of operator \textsc{Select}, $\mu \in \evalcd{Q}$ holds if and
only if there exists a mapping $\mu' \in \evalcd{Q'}$ s.t. $\mu' = \mu \cup \mu''$,
for any mapping $\mu''\sim\mu$. We observe that the domain of candidate mappings $\mu''$
is bounded by the set of variables in $Q'$, i.e.~${\it dom}(\mu'') \subseteq {\it vars}(Q')$.
Hence, we can guess a mapping $\mu''$ (this is possible since we are at least in \textsc{NP})
and subsequently check if $\mu' = \mu \cup \mu'' \in \evalcd{Q'}$, which is also possible
in \textsc{C}. The whole algorithm is in \textsc{C}.$\qed$

\subsection*{Proof of Theorem~\ref{th:select2}}

First, we show that \textsc{Evaluation} for $\cal{A}$$_+$-queries is contained in
\textsc{NP}. By definition, each query in $\cal{A}$$_+$ is of the form
$Q = \textsc{Select}_S(Q')$, where $S \subset V$ is a finite set of variables and
$Q'$ is an \textsc{And}-only expression. Further let $D$ a document and $\mu$ a mapping.
To prove membership, we follow the approach taken in the previous proof
(of Theorem~\ref{th:select1}) and eliminate the \textsc{Select}-clause. More precisely,
we guess a mapping $\mu''\sim\mu$ and check if~$\mu'' \cup \mu \in \evalcd{Q'}$
(we refer the reader to the proof of Theorem~\ref{th:select1} before for more details).
Again, the size of the mapping to be guessed is bounded, and it is easy to see that
the resulting algorithm is in \textsc{NP}.

To prove hardness we reduce $\textsc{3Sat}$, a prototypical \textsc{NP}-complete problem,
to our problem. The proof was inspired
by the reduction of \textsc{3Sat} to the evaluation problem for conjunctive queries
in~\cite{beefgllpw2007}. The $\textsc{3Sat}$ problem is defined as follows.
Let $\psi$ a boolean formula

\begin{tabbing}
x \= \kill
\>$\psi = C_1 \land \dots \land C_n$
\end{tabbing}

in CNF, where each clause $C_i$ is of the form

\begin{tabbing}
x \= \kill
\>$C_i = l_{i1} \lor l_{i2} \lor l_{i3}$,
\end{tabbing}

i.e.~contains exactly three, possibly negated, literals: is $\psi$ satisfiable?
For our encoding
we use the fixed database

\begin{tabbing}
x \= xx \= xxxl \= \kill
\>$D$\>$:= \{ (1,1,1), (1,1,0), (1,0,1), (1,0,0), (0,1,1),$\\
\>\>\>$(0,1,0), (0,0,1), (0,0,0), (0,c,1), (1,c,0) \}$,
\end{tabbing}

where we assume that $0, 1 \in I$ are any IRIs. Further let $V = \{ x_1, \dots x_m \}$
denote the set of variables occurring in formula $\psi$. We set up the SPARQL core expression

\begin{tabbing}
x \= xx \= xx \= \kill
\>$P'$\>${:=}$\>$(L^*_{11}, L^*_{12}, L^*_{13}) \oand \dots \oand (L^*_{n1},L^*_{n2},L^*_{n3})$\\
\>\>\>$\oand (?X_1,c,?\overline{X}_1) \oand \dots \oand (?X_m,c,?\overline{X}_m)$\\
\>\>\>$\oand (0,c,?A)$, where\\
\\[-0.2cm]
\>$L^*_{ij} {:=} ?X_k$ if $l_{ij} = x_k$, and $L^*_{ij} {:=} ?\overline{X}_k$ if $l_{ij} = \neg x_k$.
\end{tabbing}

Finally, set $P {:=} \textsc{Select}_{?A}(P')$. It is straightforward to
verify that $\mu = \{ ?A \mapsto 1 \} \in \evalcd{P}$ iff $\psi$ is satisfiable.$\qed$
  % ms
\section{Algebraic Results}
\label{app:rewritings}

\subsection{Proofs of the Equivalences in Figure~\ref{fig:rewritings}(I-IV)}
\label{app:equivalences14}

\subsubsection*{I. Idempotence and Inverse}

The two equivalences {\it (UIdem)} and {\it (Inv)} follow directly from the
definition of operators $\cup$ and $\setminus$, respectively.

\medskip

\noindent
{\it (JIdem)}. Let $A^-$ be an $\mathds{A}^-$-expression. We show that both
directions of the equivalence hold. $\Rightarrow$: Consider a mapping
$\mu \in A^- \Join A^-$. Then $\mu = \mu_1 \cup \mu_2$ where $\mu_1, \mu_2 \in A^-$
and $\mu_1\sim\mu_2$. Each pair of distinct mappings in $A^-$ is incompatible, so
$\mu_1 = \mu_2$ and, consequently, $\mu_1 \cup \mu_2 = \mu_1$. By assumption,
$\mu_1 \in A^-$ holds and we are done. $\Leftarrow$: Consider
a mapping $\mu \in A^-$. Choose $\mu$ for both the left and right expression
in $A^- \Join A^-$. By assumption, $\mu = \mu \cup \mu$ is contained in the
left side expression of the equation.$\qed$

\medskip

\noindent
{\it (LIdem)}. Let $A^-$ be an $\mathds{A}$$^-$ expression. Then

\begin{tabbing}
x \= xxxxxxxx \= xx \= xxxxxxxxxxxxxxxxxxxxxx \= \kill
\>$A^- \leftouterjoin A^-$\>$=$\>$(A^- \Join A^-) \cup (A^- \setminus A^-)$\>[semantics]\\
\>\>$=$\>$(A^- \Join A^-) \cup \emptyset$\>[{\it (Inv)}]\\
\>\>$=$\>$A^- \Join A^-$\\
\>\>$=$\>$A^-$\>[{\it (JIdem)}],
\end{tabbing}

which proves the equivalence.$\qed$

\subsubsection*{II. Associativity}

{\it (UAss)} and {\it (JAss)} are trivial (cf.~\cite{pag2006}).

\subsubsection*{III. Commutativity}

{\it (UComm)} and {\it (JComm)} are trivial (cf.~\cite{pag2006}).

\subsection*{IV. Distributivity}

\noindent
{\it (JUDistR)}.  We show that both directions of the equivalence hold.
$\Rightarrow$: First assume that $\mu \in (A_1 \cup A_2) \Join A_3$. Then
(according to the definition of $\Join$) $\mu$ is of the form $\mu_{12} \cup \mu_3$
where $\mu_{12} \in A_1 \cup A_2$, $\mu_3 \in A_3$, and $\mu_{12}\sim\mu_3$.
More precisely, $\mu_{12} \in A_1 \cup A_2$ means $\mu_{12}$ in $A_1$ or in
$A_2$, so we distinguish two cases. If (a) $\mu_{12} \in A_1$ then the
subexpression $A_1 \Join A_3$ on the right side generates $\mu = \mu_{12} \cup \mu_3$
(choose $\mu_{12}$ from $A_1$ and $\mu_3$ from $A_3$); similarly, if
(b)~$\mu_{12} \in  A_2$, then the expression $A_2 \Join A_3$ on the right
side generates $\mu$. $\Leftarrow$: Consider a mapping
$\mu \in (A_1 \Join A_3) \cup (A_2 \Join A_3)$. Then $\mu$ is
of the form (a) $\mu = \mu_1 \cup \mu_3$ or of the form (b)
$\mu = \mu_2 \cup \mu_3$ with $\mu_1 \in A_1$
$\mu_2 \in A_2$, $\mu_3 \in A_3$ (where (a)~$\mu_1\sim\mu_3$ or
(b)~$\mu_2\sim\mu_3$ holds, respectively). Case (a): $\mu_1$ is contained
in $A_1$, so it is also contained in $A_1 \cup A_2$. Hence, on the left-hand
side we choose $\mu_1$ from $A_1 \cup A_2$ and $\mu_3$ from $A_3$. By
assumption they are compatible and generate $\mu=\mu_1 \cup \mu_3$.
Case (b) is symmetrical. $\qed$

\medskip

\noindent
{\it (JUDistL)}. The equivalence follows from {\it (JComm)} and
{\it (JUDistR)} (cf.~\cite{pag2006}).$\qed$

\medskip

\noindent
{\it (MUDistR)}. We show that both directions of the equation hold. $\Rightarrow$:
Consider a mapping $\mu \in (A_1 \cup A_2) \setminus A_3$. Hence, $\mu$ is contained in
$A_1$ or in $A_2$ and there is no compatible mapping in $A_3$. If $\mu \in A_1$ then
the right side subexpression $A_1 \setminus A_3$ generates $\mu$, in the other case
$A_2 \setminus A_3$ generates does. $\Leftarrow$: Consider
a mapping $\mu$ in $(A_1 \setminus A_3) \cup (A_2 \setminus A_3)$. Then
$\mu \in (A_1 \setminus A_3)$ or $\mu \in (A_2 \cup A_3$). In the
first case, $\mu$ is contained in $A_1$ and there is no compatible mapping in $A_3$.
Clearly, $\mu$ is then also contained in $A_1 \cup A_2$ and
$(A_1 \cup A_2) \setminus A_3$. The second case is symmetrical.$\qed$

\medskip

\noindent
{\it (LUDistR)}. The following rewriting proves the equivalence.
 
\begin{tabbing}
x \= xxxxxxxxxxxxxxxxxxxxxxxxxxxxxxxxxxx \= \kill
\>$(A_1 \cup A_2) \leftouterjoin  A_3$\\
\>$= ((A_1 \cup A_2) \Join A_3) \cup ((A_1 \cup A_2) \setminus A_3)$\\
\>$\stackrel{(1)}{=} ((A_1 \Join A_3) \cup (A_2 \Join A_3)) \cup ((A_1 \setminus A_3) \cup (A_2 \setminus A_3))$\\
\>$\stackrel{(2)}{=} ((A_1 \Join A_3) \cup (A_1 \setminus A_3)) \cup ((A_2 \Join A_3) \cup (A_2 \setminus A_3))$\\
\>$= (A_1 \leftouterjoin A_3) \cup (A_2 \leftouterjoin A_3)$
\end{tabbing}

Step (1) is an application of {\it (JUDistR)} and {\it (MUDistR)};
in step (2) we applied {\it (UAss)} and {\it (UComm)}.$\qed$

\subsection{Proofs of the Equivalences in Figure~\ref{fig:rewritings}(V-VI)}
\label{app:equivalences56}

In the paper satisfaction was defined informally; to
be self-contained, we repeat the formal definition from~\cite{pag2006}.

\begin{definition}
Given a mapping $\mu$, filter conditions $R$, $R_1$, $R_2$, variables $?x$, $?y$,
and constant $c$, we say that $\mu$ satisfies $R$ (denoted as $\mu \models R$), if

\begin{enumerate}
\item $R$ is $\it{bnd}(?x)$ and $?x \in {\it dom}(\mu)$,\vspace{-0.05cm}
\item $R$ is $?x=c$, $?x \in {\it dom}(\mu)$, and $\mu(?x) = c$,\vspace{-0.05cm}
\item $R$ is $?x=?y$, $\{ ?x, ?y \} \subseteq {\it dom}(\mu)$, and $\mu(?x)$$=$$\mu(?y)$,\vspace{-0.05cm}
\item $R$ is $\neg R_1$ and it is not the case that $\mu \models R_1$,\vspace{-0.05cm}
\item $R$ is $R_1 \lor R_2$ and $\mu \models R_1$ or $\mu \models R_2$,\vspace{-0.05cm}
\item $R$ is $R_1 \land R_2$ and $\mu \models R_1$ and $\mu \models R_2$.
\end{enumerate}
\label{def:filter}
\end{definition}

Recall that, given a set of mappings $\Omega$ and filter condition $R$,
$\select_R(\Omega)$ is defined as the subset of mappings in $\Omega$
that satisfy condition $R$,
i.e.~$\select_R(\Omega) = \{ \mu \in \Omega \mid \mu \models R \}$.

The following proposition states that function ${\it safeVars}(A)$ returns only
variables that are bound in each mapping when evaluating $A$ on any document $D$.
It will be required in some of the subsequent proofs.

\begin{proposition}
\em
Let $A$ be a SPARQL Algebra expression and let $\Omega_A$ denote the mapping
set obtained from evaluating $A$ on any document. Then
$?x \in {\it safeVars}(A) \implies \forall \mu \in \Omega_A: ?x \in {\it dom}(\mu)$.$\punto$
\label{prop:safety}
\end{proposition}

{\it Proof.} The proof is by induction on the structure of $A$ and application
of Definition~\ref{def:safevars}. We omit the details.$\qed$

\subsubsection*{Proofs of Equivalences in Figure~\ref{fig:rewritings}(V-VI)}

\noindent
{\it (SUPush)}. Follows directly from Proposition~1(5) in~\cite{pag2006}.$\qed$

\medskip

\noindent
{\it (SDecompI)}. Follows directly from Lemma~1(1) in~\cite{pag2006}.$\qed$

\medskip

\noindent
{\it (SDecompII)}. Follows directly from Lemma~1(2) in~\cite{pag2006}.$\qed$

\medskip

\noindent
{\it (SReord)}. Follows from the application of {\it (SDecompI)} and the
commutativity of the boolean operator $\land$.$\qed$

\medskip

\noindent
{\it (BndI)}, {\it (BndII)}, {\it (BndIII)}, and {\it (BndIV)} are trivial.

\medskip
\noindent
{\it (BndV).}
Recall that by assumption $?x \not \in {\it vars}(A_1)$.
The following rewriting proves the equivalence.

\begin{tabbing}
x \= xxxxxxxxxxxxxxxxxxxxxxxxxxxxxxxxxxxx \= \kill
\>$\sigma_{{\it bnd}(?x)}(A_1 \leftouterjoin A_2)$\\
\>$= \sigma_{{\it bnd}(?x)}(A_1 \Join A_2) \cup (A_1 \setminus A_2))$\>[semantics]\\
\>$= \sigma_{{\it bnd}(?x)}(A_1 \Join A_2) \cup \sigma_{{\it bnd}(?x)}(A_1 \setminus A_2)$\>[{\it (SUPush)}]\\
\>$= \sigma_{{\it bnd}(?x)}(A_1 \Join A_2)$\>[$*_1$]\\
\>$= A_1 \Join A_2$\>[$*_2$]
\end{tabbing}

$*_1$ follows immediately from assumption $?x \not \in {\it vars}(A_1)$; $*_2$
follows from the observation that $?x \in {\it safeVars}(A_2)$.$\qed$

\medskip

\noindent
{\it (SJPush)}. $\Rightarrow$: Let $\mu \in \sigma_R(A_1 \Join A_2)$.
By semantics, $\mu \models R$. Furthermore, $\mu$ is of the form $\mu_1 \cup \mu_2$, where
$\mu_1 \in A_1$, $\mu_2 \in A_2$, and $\mu_1\sim\mu_2$. Recall that by assumption
${\it vars}(R) \subseteq {\it safeVars}(A_1)$, so we always have that
${\it dom}(\mu_1) \subseteq {\it vars}(R)$ (cf.~Proposition~\ref{prop:safety}), i.e.~each
variable that occurs in $R$ is bound in mapping $\mu_1$. It is easy to verify that
$R \models \mu$ implies $R \models \mu_1$, since both mappings coincide in the variables
that are relevant for evaluating $R$. Consequently $\sigma_R(A_1)$ on the right side generates
$\mu_1$, and clearly $\sigma_R(A_1) \Join A_2$ generates $\mu_1 \cup \mu_2 = \mu$.
$\Leftarrow$: Consider a mapping $\mu \in \sigma_R(A_1) \Join A_2$. Then $\mu$ is
of the form $\mu = \mu_1 \cup \mu_2$, $\mu_1 \in A_1$, $\mu_2 \in A_2$,
$\mu_1\sim\mu_2$, and $\mu_1 \models R$. It is easy to see that then also 
$\mu_1 \cup \mu_2 \models R$, because ${\it dom}(\mu_1) \subseteq {\it vars}(R)$
(as argued in case $\Rightarrow$)
and $\mu_1 \cup \mu_2$ coincides with $\mu_1$ on all variables that are relevant
for evaluating $R$. Hence, $\mu = \mu_1 \cup \mu_2$
is generated by the left side of the equation.$\qed$ 

\medskip

\noindent
{\it (SMPush)}. $\Rightarrow$: Let $\mu \in \sigma_R(A_1 \setminus A_2)$.
By semantics, $\mu \in A_1$ and there is no $\mu_2 \in A_2$ compatible with $\mu_1$
and $\mu \models R$. From these preconditions it follows immediately that
$\mu \in \sigma_R(A_1) \setminus A_2$. $\Leftarrow$: Let
$\mu \in \sigma_R(A_1) \setminus A_2$. Then $\mu \in A_1$, $\mu \models R$, and
there is no compatible mapping in $A_2$. Clearly, then also $\mu \in A_1 \setminus A_2$
and $\mu \in \sigma_R(A_1 \setminus A_2)$.$\qed$
\medskip

\noindent
{\it (SLPush)}. The following rewriting proves the equivalence.

\begin{tabbing}
x \= xxxxxxxxxxxxxxxxxxxxxxxxxxxxxxxxx \= \kill
\>$\sigma_R(A_1 \leftouterjoin A_2)$\\
\>$=\sigma_R((A_1 \Join A_2) \cup (A_1 \setminus A_2))$\>[semantics]\\
\>$=\sigma_R(A_1 \Join A_2) \cup \sigma_R(A_1 \setminus A_2)$\>[{\it (SUPush)}]\\
\>$=(\sigma_R(A_1) \Join A_2) \cup (\sigma_R(A_1) \setminus A_2)$\>[*]\\
\>$=\sigma_R(A_1) \leftouterjoin A_2$\>[semantics] 
\end{tabbing}

* denotes application of {\it (SJPush)} and {\it (SMPush)}.$\qed$

\subsection{Proofs of the Remaining Technical Results}
\label{app:miscalg}

\subsubsection*{Proof of Proposition~\ref{prop:incomp}}

{\it Proof.} Let $A^-$ be an $\mathds{A}^-$ expression. The proof is by induction
on the structure of $A^-$ The basic case is $A^- = \evalc{t}$. By semantics, all mappings
in $A^-$ bind exactly the same set of variables, and consequently the values
of each two distinct mappings must differ in at least one variable, which makes them
incompatible.\footnote{Recall that we assume set semantics.} (Case 1) We assume that
the hypothesis holds and consider an expression
$A^- = A_1^- \Join A_2^-$. Then each mapping $\mu \in A^-$ is of
the form $\mu = \mu_1 \cup \mu_2$ with $\mu_1 \in A_1^-$, $\mu_2 \in A_2^-$, and
$\mu_1\sim\mu_2$. We fix $\mu$ and show that each mapping $\mu' \in A^-$
different from $\mu$ is incompatible. Any mapping in $\mu' \in A^-$ that is different from
$\mu$ is of the form $\mu_1' \cup \mu_2'$ with $\mu_1' \in A_1^-$, $\mu_2' \in A_2^-$
and $\mu_1'$ different from $\mu_1$ or $\mu_2'$ different from $\mu_2$.
Let us w.l.o.g. assume that $\mu_1'$ is different from $\mu_1$. By induction hypothesis,
$\mu_1$ is incompatible with $\mu_1'$. It is easy to verify that then
$\mu = \mu_1 \cup \mu_2$ is incompatible with $\mu'=\mu_1' \cup \mu_2'$, since $\mu_1$
and $\mu_1'$ disagree in the value of at least one variable. (Case 2) Let
$A^- = A_1^- \setminus A_2^-$. By induction hypothesis, each two mappings in $A_1^-$
are pairwise incompatible. By semantics, $A^-$ is a subset of
$A_1^-$, so the incompatibility property still holds for $A^-$. (Case 3) Let
$A^- = A_1^- \leftouterjoin A_2^-$. We rewrite the left outer join according to its
semantics: $A^- = A_1^- \leftouterjoin A_2^- = (A_1^- \Join A_2^-) \cup (A_1^- \setminus A_2^-)$.
As argued in cases (1) and (2), the incompatibility property holds for both subexpressions
$A^\Join = A_1^- \Join A_2^-$ and $A^\setminus = A_1^- \setminus A_2^-$, so it suffices
to show that the mappings in $A^\Join$ are pairwise incompatible to those in $A^\setminus$.
We observe that $A^\setminus$ is a subset of $A_1^-$. Further, each mapping $\mu \in A^\Join$
is of the form $\mu = \mu_1 \cup \mu_2$, where $\mu_1 \in A_1^-$, $\mu_2 \in A_2^-$, and $\mu_1\sim\mu_2$.
By assumption, each mapping in $A_1^-$, and hence each mapping $\mu_1' \in A^\setminus$ is either
identical to or incompatible with $\mu_1$. (a) If $\mu_1 \not = \mu_1'$,
then $\mu_1'$ is incompatible with $\mu_1$, and consequently incompatible with
$\mu_1 \cup \mu_2 = \mu$, so we are done. (b) Let $\mu_1 = \mu_1'$. We observe that, by assumption,
there is a compatible mapping (namely $\mu_2$ in $A_2^-$). This means that $A_1^- \setminus A_2^-$
does not generate $\mu_1'$, so we have a contradiction (i.e., the assumption $\mu_1 = \mu_1'$
was invalid). (Case 4) Let
$A = \select_C(A_1^-)$. Analogously to case 2, $A^-$ is a subset of $A_1^-$, for which the
property holds by induction hypothesis. $\qed$

\subsubsection*{Proof of Lemma~\ref{lemma:completeness}}

We provide an exhaustive set of counterexamples.

\noindent
{\it Proof of Claim 1}. In this part, we give counterexamples for two fragments
(a)~$\cal{A}$$^{\{\Join,\setminus,\leftouterjoin,\sigma,\cup\}}$
and (b)~$\cal{A}$$^{\{\Join,\setminus,\leftouterjoin,\sigma,\pi\}}$.
The result for the full algebra
(i.e., $\cal{A}$$^{\{\Join,\setminus,\leftouterjoin,\sigma,\cup,\pi\}}$)
follows.

(1a) Fragment $\cal{A}$$^{\{\Join,\setminus,\leftouterjoin,\sigma,\cup\}}$.
We use the fixed database $D = \{ (0,c,1) \}$.
Consider the algebra expression $A = \evalcd{(?x,c,1)} \cup \evalcd{(0,c,?y)}$.
It is easy to see that both
$A \Join A = \{ \{ ?x \mapsto 0 \}, \{ ?y \mapsto 1 \}, \{ ?x \mapsto 0, ?y \mapsto 1 \}\}$
and
$A \leftouterjoin A = \{ \{ ?x \mapsto 0 \}, \{ ?y \mapsto 1 \}, \{ ?x \mapsto 0, ?y \mapsto 1 \}\}$
differ from $A = \{ \{?x \mapsto 0 \}, \{ ?y \mapsto 1 \} \}$, which shows that neither
{\it (JIdem)} nor {\it (LIdem)} holds for this fragment.

(1b) Fragment $\cal{A}$$^{\{\Join,\setminus,\leftouterjoin,\sigma,\pi\}}$.
We use the fixed database $D = \{ (0,f,0), (1,t,1), (a,tv,0), (a,tv,1) \}$.
Consider the algebra expression

\begin{tabbing}
x \= \kill
\>$A = \pi_{\{?x,?y\}}((t_1 \leftouterjoin t_2) \leftouterjoin t_3)$, where\\
\\[-0.2cm]
\>$t_1 \stackrel{def}{:=} \evalcd{(a,tv,?z)}$,\\
\>$t_2 \stackrel{def}{:=} \evalcd{(?z,f,?x)}$, and\\
\>$t_3 \stackrel{def}{:=} \evalcd{(?z,t,?y)}$.
\end{tabbing}

It is easy to verify that $A = \{ \{?x \mapsto 0\}, \{?y \mapsto 1\} \}$.
For $A \Join A$ and $A \leftouterjoin A$ we then get exactly the same results
as in part~(1a), and we conclude that neither {\it (JIdem)} nor {\it (LIdem)}
holds for the fragment under consideration.

\medskip

\noindent
{\it Proof of Claim 2}. Trivial.

\noindent
{\it Proof of Claims 3 + 4}. We provide counterexamples for each possible operator
constellations. All counterexamples are designed for the database $D = \{ (0,c,1) \}$.

%%%%%%%%%%%%%% remaining cases \cup
{\it Distributivity over $\cup$ (Claim 3 in Lemma~\ref{lemma:completeness}):}
\begin{itemize}
\item $A_1 \setminus (A_2 \cup A_3) \equiv (A_1 \setminus A_2) \cup (A_1 \setminus A_3)$
{\bf does not hold}, e.g.~$A_1 = \evalcd{(0,c,?a)}$, $A_2 = \evalcd{(?a,c,1)}$, and $A_3 = \evalcd{(0,c,?b)}$
violates the equation.

\item $A_1 \leftouterjoin (A_2 \cup A_3) \equiv (A_1 \leftouterjoin A_2) \cup (A_1 \leftouterjoin A_3)$
{\bf does not hold}, e.g.~$A_1 = \evalcd{(0,c,?a)}$, $A_2 = \evalcd{(?a,c,1)}$, and $A_3 = \evalcd{(0,c,?b)}$.
violates the equation.
\end{itemize}

%%%%%%%%%%%%%% \Join
{\it Distributivity over $\Join$ (Claim 4 in Lemma~\ref{lemma:completeness}):}
\begin{itemize}
\item $A_1 \cup (A_2 \Join A_3) \equiv (A_1 \cup A_2) \Join (A_1 \cup A_3)$
{\bf does not hold}, e.g.~$A_1 = \evalcd{(?a,c,1)}$, $A_2 = \evalcd{(?b,c,1)}$, and $A_3 \evalcd{(0,c,?b)}$
violates the equation.

\item $(A_1 \Join A_2) \cup A_3 \equiv (A_1 \cup A_3) \Join (A_2 \cup A_3)$
{\bf does not hold} (symmetrical to the previous one).

\item $A_1 \setminus (A_2 \Join A_3) \equiv (A_1 \setminus A_2) \Join (A_1 \setminus A_3)$
{\bf does not hold}, e.g.~$A_1 = \evalcd{(?a,c,1)}$, $A_2 = \evalcd{(?b,c,1)}$, and $A_3 \evalcd{(0,c,?b)}$
violates the equation.

\item $(A_1 \Join A_2) \setminus A_3 \equiv (A_1 \setminus A_3) \Join (A_2 \setminus A_3)$
{\bf does not hold}, e.g.~$A_1 = \evalcd{(0,c,?a)}$, $A_2 = \evalcd{(0,c,?b)}$, and $A_3 \evalcd{(?a,c,1)}$
violates the equation.

\item $A_1 \leftouterjoin (A_2 \Join A_3) \equiv (A_1 \leftouterjoin A_2) \Join (A_1 \leftouterjoin A_3)$
{\bf does not hold}, e.g.~$A_1 = \evalcd{(?a,c,1)}$, $A_2 = \evalcd{(?b,c,1)}$, and $A_3 \evalcd{(0,c,?a)}$
violates the equation.

\item $(A_1 \Join A_2) \leftouterjoin A_3 \equiv (A_1 \leftouterjoin A_3) \Join (A_2 \leftouterjoin A_3)$
{\bf does not hold}, e.g.~$A_1 = \evalcd{(0,c,?a)}$, $A_2 = \evalcd{(0,c,?b)}$, and $A_3 \evalcd{(?a,c,1)}$
violates the equation.
\end{itemize}

%%%%%%%%%%%%%% \setminus
{\it Distributivity over $\setminus$ (Claim 4 in Lemma~\ref{lemma:completeness}):}
\begin{itemize}
\item $A_1 \cup (A_2 \setminus A_3) \equiv (A_1 \cup A_2) \setminus (A_1 \cup A_3)$
{\bf does not hold}, e.g.~$A_1 = \evalcd{(?a,c,1)}$, $A_2 = \evalcd{(0,c,?a)}$, and $A_3 \evalcd{(?a,c,1)}$
violates the equation.

\item $(A_1 \setminus A_2) \cup A_3 \equiv (A_1 \cup A_3) \setminus (A_2 \cup A_3)$
{\bf does not hold} (symmetrical to the previous one).

\item $A_1 \Join (A_2 \setminus A_3) \equiv (A_1 \Join A_2) \setminus (A_1 \Join A_3)$
{\bf does not hold}, e.g.~$A_1 = \evalcd{(?a,c,1)}$, $A_2 = \evalcd{(?b,c,1)}$, and $A_3 \evalcd{(0,c,?a)}$
violates the equation.

\item $(A_1 \setminus A_2) \Join A_3 \equiv (A_1 \Join A_3) \setminus (A_2 \Join A_3)$
{\bf does not hold} (symmetrical to the previous one).

\item $A_1 \leftouterjoin (A_2 \setminus A_3) \equiv (A_1 \leftouterjoin A_2) \setminus (A_1 \leftouterjoin A_3)$
{\bf does not hold}, e.g.~$A_1 = \evalcd{(?a,c,1)}$, $A_2 = \evalcd{(?b,c,1)}$, and $A_3 \evalcd{(?b,c,1)}$
violates the equation.

\item ($A_1 \setminus A_2) \leftouterjoin A_3 \equiv (A_1 \leftouterjoin A_3) \setminus (A_2 \leftouterjoin A_3)$
{\bf does not hold}, e.g.~$A_1 = \evalcd{(?a,c,1)}$, $A_2 = \evalcd{(?b,c,1)}$, and $A_3 \evalcd{(0,c,?b)}$
violates the equation. 
\end{itemize}

%%%%%%%%%%%%%% \leftouterjoin
{\it Distributivity over $\leftouterjoin$ (Claim 4 in Lemma~\ref{lemma:completeness}):}
\begin{itemize}
\item $A_1 \cup (A_2 \leftouterjoin A_3) \equiv (A_1 \cup A_2) \leftouterjoin (A_1 \cup A_3)$
{\bf does not hold}, e.g.~$A_1 = \evalcd{(?a,c,1)}$, $A_2 = \evalcd{(c,c,c)}$, and $A_3 \evalcd{(?b,c,1)}$
violates the equation.

\item $(A_1 \leftouterjoin A_2) \cup A_3 \equiv (A_1 \cup A_3) \leftouterjoin (A_2 \cup A_3)$
{\bf does not hold} (symmetrical to the previous one).

\item $A_1 \Join (A_2 \leftouterjoin A_3) \equiv (A_1 \Join A_2) \leftouterjoin (A_1 \Join A_3)$
{\bf does not hold}, e.g.~$A_1 = \evalcd{(?a,c,1)}$, $A_2 = \evalcd{(?b,c,1)}$, and $A_3 \evalcd{(0,c,?a)}$
violates the equation.

\item $(A_1 \leftouterjoin A_2) \Join A_3 \equiv (A_1 \Join A_3) \leftouterjoin (A_2 \Join A_3)$
{\bf does not hold} (symmetrical to the previous one).

\item $A_1 \setminus (A_2 \leftouterjoin A_3) \equiv (A_1 \setminus A_2) \leftouterjoin (A_1 \setminus A_3)$
{\bf does not hold}, e.g.~$A_1 = \evalcd{(?a,c,1)}$, $A_2 = \evalcd{(?b,c,1)}$, and $A_3 \evalcd{(0,c,?a)}$
violates the equation.

\item $(A_1 \leftouterjoin A_2) \setminus A_3 \equiv (A_1 \setminus A_3) \leftouterjoin (A_2 \setminus A_3)$
{\bf does not hold}, e.g.~$A_1 = \evalcd{(?a,c,1)}$, $A_2 = \evalcd{(?b,c,1)}$, and $A_3 \evalcd{(0,c,?b)}$
violates the equation.
\end{itemize}

The list of counterexamples is exhaustive.$\qed$

\subsubsection*{Proof of Proposition~\ref{prop:minus}}

\noindent
{\it (MReord)}. We consider all possible mappings $\mu$. Clearly, if $\mu$ is not
contained in $A_1$, it will be neither contained in the right side nor in the left side
of the expressions (both are subsets of $A_1$). So we can restrict our discussion to mappings
$\mu \in A_1$. We distinguish three cases. Case (1): consider a mapping $\mu \in A_1$ and
assume there is a compatible mapping in $A_2$. Then $\mu$ is not contained in
$A_1 \setminus A_2$, and also not in
$(A_1 \setminus A_2) \setminus A_3$, which by definition
is a subset of the former. Now consider the right-hand side of the equation and let us assume
that $\mu \in A_1 \setminus A_3$ (otherwise we are done). Then, as there is a
compatible mapping to $\mu$ in  $A_2$, the expression
$\mu \in (A_1 \setminus A_3) \setminus A_2$ will not contain $\mu$. Case (2):
The case of $\mu \in A_1$ being compatible with any mapping from $A_3$ is
symmetrical to (2). Case (3): Let $\mu \in A_1$ be a mapping that is not compatible with
any mapping in $A_2$ and $A_3$. Then both $(A_1 \setminus A_2) \setminus A_3$
on the left side and $(A_1 \setminus A_3) \setminus A_2$ on the right side
contain $\mu$. In all cases, $\mu$ is contained
in the right side exactly if it is contained in the left side.$\qed$

\medskip

\noindent
{\it (MMUCorr)}. We show both directions of the equivalence.
$\Rightarrow$: Let $\mu \in (A_1 \setminus A_2) \setminus A_3$. Then
$\mu \in A_1$ and there is neither a compatible mapping $\mu_2 \in A_2$
nor a compatible mapping $\mu_3 \in A_3$. Then both $A_2$ and $A_3$
contain only incompatible mappings, and clearly
$A_2 \cup A_3$ contains only incompatible mappings. Hence, the right
side $A_1 \setminus (A_2 \cup A_3)$ produces $\mu$.
$\Leftarrow$: Let $\mu \in A_1 \setminus (A_2 \cup A_3)$. Then
$\mu \in A_1$ and there is no compatible mapping in $A_2 \cup A_2$,
which means that there is neither a compatible mapping in $A_2$ nor in
$A_3$. It follows that $A_1 \setminus A_2$ contains $\mu$ (as there
is no compatible mapping in $A_2$ and $\mu \in A_1$). From the fact
that there is no compatible mapping in $A_3$, we deduce
$\mu \in (A_1 \setminus A_2) \setminus A_3$.
$\qed$

\medskip

\noindent
{\it (MJ)}. See Lemma~3(2) in~\cite{pag2006}.

\noindent
{\it (LJ)}. Let $A_1^-$, $A_2^-$ be $\mathds{A}$$^-$-expressions.
The following sequence of rewriting steps proves the
equivalence.

\begin{tabbing}
x \= xxxxxxxxxxxxxxxxxxxxxxxxxxxxxxxxxxxxxxx \= \kill
\>$A_1^- \leftouterjoin A_2^-$\\
\>$= (A_1^-\Join A_2^-) \cup (A_1^- \setminus A_2^-)$\>[sem.]\\
\>$= (A_1^- \Join (A_1^- \Join A_2^-)) \cup (A_1^- \setminus (A_1^- \Join A_2^-))$\>[*]\\
\>$= (A_1^- \leftouterjoin (A_1^- \Join A_2^-))$\>[sem.]
\end{tabbing}

* denotes application of {\it (JIdem)}, {\it (JAss)}, and {\it (MJ)}.$\qed$

\subsubsection*{Proof of Lemma~\ref{lemma:minus}}

Let $A_1^-$, $A_2^-$ be $\mathds{A}^-$-expressions, $R$ a filter
condition, and $?x \in {\it safeVars(A_2)} \setminus {\it vars}(A_1)$ a variable
that is contained in the set of safe variables of $A_2$, but not in $A_1$.
We transform the left side expression into the right side expression as follows.

\begin{tabbing}
x \= xx \= xxxxxxxxxxxxxxxxxxxxxxxxxxxxxxx \= \kill
\>$\sigma_{\neg {\it bnd(?x)}}(A_1^- \leftouterjoin A_2^-)$\\
\>$=\sigma_{\neg {\it bnd(?x)}}((A_1^- \Join A_2^-) \cup (A_1^- \setminus A_2^-))$\>\>[semantics]\\
\>$=\sigma_{\neg {\it bnd(?x)}}(A_1^- \Join A_2^-)\ \cup$\\
\>\>$\sigma_{\neg {\it bnd(?x)}}(A_1^- \setminus A_2^-)$\>[{\it (SUPush)}]\\
\>$=\sigma_{\neg {\it bnd(?x)}}(A_1^- \setminus A_2^-)$\>\>[$*_1$]\\
\>$=A_1^- \setminus A_2^-$\>\>[$*_2$]
\end{tabbing}

We first show that rewriting step $*_1$ holds. Observe that $?x \in {\it safeVars}(A^-_2$),
which implies that (following Proposition~\ref{prop:safety}) variable $?x$ is bound
in each mapping generated by $A_2$. Consequently, $?x$ is also bound in each mapping generated by
$A_1 \Join A_2$ and the condition $\neg {\it bnd(?x)}$ is never satisfied for the join part,
so it can be eliminated.
Concerning step $*_2$ we observe that $?x \in {\it safeVars}(A_2) \setminus {\it vars}(A_1)$
implies $?x \not \in {\it vars}(A_1)$. It follows immediately that $?x$ is unbound in any
mapping generated by $A_1^- \setminus A_2^-$, so the surrounding filter condition always holds
and can be dropped.
$\qed$
     % ms
\section{Proofs of the SQO Results}
\label{app:sqo}

\subsection*{Proof of Lemma \ref{direct}}

\begin{itemize}
	\item Let $Q' \in C_1^{-1}(cb_{\Sigma}(C_1(Q))) \cap \mathcal{A}_+$. Then $C_1(Q') \in cb_{\Sigma}(C_1(Q))$. This implies $C_1(Q') \equiv_{\Sigma} C_1(Q)$. It follows that $Q' \equiv_{\Sigma} Q$. 
	
	\item Follows directly from the definition of the second translation scheme.
	
	\item Follows from the last two points.$\qed$
\end{itemize}

\subsection*{Proof of Lemma \ref{direct2}}

\begin{itemize}
	\item Let $Q' \equiv_{\Sigma} Q$. We have that $C_1^{-1}(U(C_1(Q'))),$\\ $C_1^{-1}(U(C_1(Q))) \in \mathcal{A}_+$, therefore $C_1(Q') \equiv_{\Sigma} C_1(Q)$. Then, it follows that $C_1(Q') \in cb_{\Sigma}(C_1(Q))$ and $Q' \in C_1^{-1}(cb_{\Sigma}(C_1(Q)))$.

	\item Follows from the last point and bullet two in lemma \ref{direct}.$\qed$
\end{itemize}

\subsection*{Proof of Lemma \ref{elim-opt}}

\squishlist
\item We transform $Q$ systematically. Let $D$ be an RDF database such that $D \models \Sigma$.\\

\begin{tabbing}
x \= xxx \= \kill
\>$\evalcd{Q}$\\
\>$= \evalcd{(Q_1 \oopt Q_2)}$\\
\>$= \evalcd{(Q_1 \oand Q_2)} \cup (\evalcd{Q_1} \setminus \evalcd{Q_2})$\\ 
\>$= \evalcd{(Q_1 \oand Q_2)}\ \cup$\\
\>\>$(\pi_{{\it vars}(Q_1)}\evalcd{(Q_1 \oand Q_2)} \setminus \evalcd{Q_2})$\\ 
\end{tabbing}

It is easy to verify that each mapping in $\evalcd{Q_1 \oand Q_2}$ is compatible with at least one mapping in $Q_2$, and the same still
holds for the projection $\pi_{{\it vars}(Q_1)}\evalcd{(Q_1 \oand Q_2)}$. Hence, the right side of the union can be dropped and the elimination simplifies to $Q \equiv_\Sigma (Q_1 \oand Q_2)$.

\item Let $D$ be an RDF database such that $D \models \Sigma$. Then we have that

\begin{tabbing}
x \= xxx \= \kill
\>$\evalcd{Q} = \evalcd{(Q_1 \oopt (Q_2 \oand Q_3))}$\\
\>$= \evalcd{(Q_1 \oand Q_2 \oand Q_3)}\ \cup$\\
\>\>$(\evalcd{Q_1} \setminus \evalcd{Q_2 \oand Q_3})$\\ 
\>$= \evalcd{(Q_1 \oand Q_3)}\ \cup$\\
\>\>$(\evalcd{Q_1 \oand Q_2} \setminus \evalcd{Q_2 \oand Q_3})$.
\end{tabbing}

We now show that $(\evalcd{Q_1 \oand Q_2} \setminus \evalcd{Q_2 \oand Q_3}) = (\evalcd{Q_1 \oand Q_2} \setminus \evalcd{Q_3})$. Assume that there is some $\mu \in (\evalcd{Q_1 \oand Q_2} \setminus \evalcd{Q_2 \oand Q_3})$. Then, for all $\mu' \in \evalcd{Q_2 \oand Q_3}$ it holds that $\mu'$ is incompatible to $\mu$. As $\mu$ is, by choice, compatible to some element in $\evalcd{Q_2}$, it must be in compatible to all elements in $\evalcd{Q_3}$. This implies $\mu \in (\evalcd{Q_1 \oand Q_2} \setminus \evalcd{Q_3})$. Assume we have $\nu \in (\evalcd{Q_1 \oand Q_2} \setminus \evalcd{Q_3})$. Choose $\nu' \in \evalcd{Q_2 \oand Q_3}$. It follows that the projection of $\nu'$ in the variables in $Q_3$ is not compatible to $\nu$, therefore $\nu'$ is not compatible to $\nu$. This implies $\nu \in (\evalcd{Q_1 \oand Q_2} \setminus \evalcd{Q_2 \oand Q_3})$. Consequently

\begin{tabbing}
x \= \kill
\>$\evalcd{Q}$\\
\>$= \pi_S( \evalcd{(Q_1  \oand Q_3)} \cup (\evalcd{Q_1 \oand Q_2} \setminus \evalcd{Q_3}))$\\
\>$= \pi_S( \evalcd{(Q_1 \oand Q_3)} \cup (\evalcd{Q_1} \setminus \evalcd{Q_3}))$\\
\>$ = \evalcd{Q_1 \oopt Q_3}$. $\punto$
\end{tabbing}
\squishend

\subsection*{Proof of Lemma \ref{elim-filter}}

\squishlist
	\item Let $\mu \in \evalcd{Q_1 \oopt Q_2}$. Then, $\mu(?x)$ is defined because of $Q_1 \equiv_\Sigma \textsc{Select}_{{\it vars}(Q_1)}(Q_1 \oand Q_2)$. So, $\evalcd{\textsc{Filter}_{\neg {\it bnd}(?x)}(Q_1 \oopt Q_2)} = \emptyset$.
	
	\item The proof of this claim is straightforward.

	\item Assume that there is some $\mu \in \evalcd{\textsc{Filter}_{\neg ?x=?y}(Q_2)}$. So, $\mu |_{S} \in \evalcd{\textsc{Select}_{S}(Q_2)}$. It holds that	
	$\textsc{Select}_{S}(Q_2)$ $\equiv_\Sigma$ $\textsc{Select}_{S}(Q_2 \frac{?x}{?y})$	$\equiv_\Sigma$ $\textsc{Select}_{S}(\textsc{Filter}_{?x=?y}(Q_2))$. It follows that $\mu \in \evalcd{\textsc{Select}_{S}(\textsc{Filter}_{?x=?y}(Q_2))}$, which is a contradiction.$\qed$
\squishend
         % mm
\section{Proofs of the Chase Termination Results}
\label{app:chasetermination}

\subsection{Additional Definitions} \label{app:defs-chase}

\textbf{Databases.} We choose three pairwise disjoint infinite sets $\Delta, \Delta_{null}$ and $V$. We will refer to $\Delta$ as the set of \textit{constants}, to $\Delta_{null}$ as the set of \textit{labelled nulls} and to $V$ as the set of variables.  A \textit{database schema} $\mathcal{R}$ is a finite set of relational symbols $\setone{R_1,...,R_n}$. To every $R_i \in \mathcal{R}$ we assign a natural number $ar(R_i) \in \mathbb{N}$, which we call the \textit{arity} of $R_i$. The arity of $\mathcal{R}$, denoted by $ar(\mathcal{R})$, is defined as $max \set{ar(R_i)}{i \in [n]}$.  Throughout the rest of the paper, we assume the database schema, the set of constants and the set of labelled nulls to be fixed. This is why we will suppress these sets in our notation.

A \textit{database instance} $I$ is an $n$-tuple $(I_1,...,I_n)$, where $I_i \subseteq (\Delta \cup \Delta_{null})^{ar(R_i)}$ for every $i \in [n]$. We will denote $(c_1,...,c_{ar(R_i)}) \in I_i$ by the \textit{fact} $R_i(c_1,...,c_{ar(R_i)})$ and therefore represent the instance $I$ as the set if its facts. Abusing notation, we write $I = \set{R_i(t)}{t \in I_i, i \in [n]}$.

A position is a position in a predicate, e.g.~a three-ary predicate $R$ has three positions $R^1, R^2, R^3$. We say  that a variable, labelled null or constant $c$ appears e.g. in a position $R^1$ if there exists a fact $R(c,...)$.\\

\textbf{Constraints.} Let $\overline{x}, \overline{y}$ be tuples of variables. We consider two types of database constraints, i.e.~\textit{tuple-generating} and \textit{equality generating dependencies}. A tuple-generating dependency (TGD) is a first-order sentence 
\begin{center}
$\varphi := \forall \overline{x} (\phi(\overline{x}) \rightarrow \exists \overline{y} \psi(\overline{x},\overline{y}))$,
\end{center}
 such that (a) both $\phi$ and $\psi$ are conjunctions of atomic formulas (possibly with parameters from $\Delta$), (b) $\psi$ is not empty, (c) $\phi$ is possibly empty, (d) both $\phi$ and $\psi$ do not contain equality atoms and (e) all variables from $\overline{x}$ that occur in $\psi$ must also occur in $\phi$. We denote by $body(\varphi)$ the set of atoms in $\phi$ and by $head(\varphi)$ the set of atoms in $\psi$. 

An equality generating dependency (EGD) is a first-order sentence 
\begin{center}
$\varphi := \forall \overline{x} (\phi(\overline{x}) \rightarrow x_i = x_j)$,
\end{center}
 where $x_i, x_j$ occur in $\phi$ and $\phi$ is a non-empty conjunction of equality-free $\mathcal{R}$-atoms (possibly with parameters from $\Delta$). We denote by $body(\varphi)$ the set of atoms in $\phi$ and by $head(\varphi)$ the set $\setone{x_i = x_j}$.\\
 
For brevity, we will often omit the $\forall$-quantifier and the respective list of universally quantified variables.

\textbf{Constraint satisfaction.}
Let $\models$ be the standard first-order model relationship and $\Sigma$ be a set of TGDs and EGDs. We say that a database instance $I=(I_1,...,I_n)$ satisfies $\Sigma$, denoted by $I \models \Sigma$, if and only if $(\Delta \cup \Delta_{null},I_1,...,I_n) \models \Sigma$ in the sense of an $\mathcal{R}$-structure.

It is folklore that TGDs and EGDs together are expressive enough to express foreign key constraints, inclusion, functional, join, multivalued and embedded dependencies. Thus, we can capture all important semantic constraints used in databases. Therefore, in the rest of the paper, all sets of constraints are a union of TGDs and EGDs only.\\

\textbf{Homomorphisms.}
A homomorphism from a set of atoms $A_1$ to a set of atoms $A_2$ is a mapping

\begin{tabbing}
x \= \kill
\>$\mu : \Delta \cup \Delta_{null} \cup V \rightarrow \Delta \cup \Delta_{null} \cup V$
\end{tabbing}

such that the following conditions hold: (a) if $c \in \Delta$, then $\mu(c)=c$, (b) if $c \in \Delta_{null}$, then $\mu(c) \in \Delta \cup \Delta_{null}$ and (c) if $R(c_1,...,c_n) \in A_1$, then $R(\mu(c_1),...,\mu(c_n)) \in A_2$.\\

\textbf{Chase.}
Let $\Sigma$ be a set of TGDs and EGDs and $I$ an instance, represented as a set of atoms. We say that a TGD $\forall \overline{x} \varphi \in \Sigma$ is applicable to $I$ if there is a homomorphism $\mu$ from $body(\forall \overline{x} \varphi)$ to $I$ and $\mu$ cannot be extended to a homomorphism $\mu' \supseteq \mu$ from $head(\forall \overline{x}\varphi)$ to $I$. In such a case the chase step $I \stackrel{\forall \overline{x}\varphi, \mu(\overline{x})}{\longrightarrow} J$ is defined as follows. We define a homomorphism $\nu$ as follows: (a) $\nu$ agrees with $\mu$ on all universally quantified variables in $\varphi$, (b) for every existentially quantified variable $y$ in $\forall \overline{x} \varphi$ we choose a "fresh" labelled null $n_y \in \Delta_{null}$ and define $\nu(y):=n_y$. We set $J$ to be $I \cup \nu(head(\forall \overline{x}\varphi))$. We say that an EGD $\forall \overline{x}\varphi \in \Sigma$ is applicable to $I$ if there is a homomorphism $\mu$ from $body(\forall \overline{x}\varphi)$ to $I$ and $\mu(x_i) \neq \mu(x_j)$. In such a case the chase step $I \stackrel{\forall \overline{x}\varphi, a}{\longrightarrow} J$ is defined as follows. We set $J$ to be 
\squishlist
	\item $I$ except that all occurrences of $\mu(x_j)$ are substituted by $\mu(x_i) =: a$, if $\mu(x_j)$ is a labelled null,
	\item $I$ except that all occurrences of $\mu(x_i)$ are substituted by $\mu(x_j) =: a$, if $\mu(x_i)$ is a labelled null,
	\item undefined, if both $\mu(x_j)$ and $\mu(x_i)$ are constants. In this case we say that the chase fails.
\squishend

A chase sequence is an exhaustive application of applicable constraints 
\begin{center}
$I_0 \stackrel{\varphi_0, \overline{a}_0}{\longrightarrow} I_1 \stackrel{\varphi_1, \overline{a}_1}{\longrightarrow} \ldots$,
\end{center}
 where we impose no strict order what constraint must be applied in case several constraints apply. If this sequence is finite, say $I_r$ being its final element, the chase terminates and its result $I_0^{\Sigma}$ is defined as $I_r$. The length of this chase sequence is $r$. Note that different orders of application of applicable constraints may lead to a different chase result. However, as proven in \cite{fkmp2003}, two different chase orders lead to homomorphically equivalent results, if these exist. Therefore, we write $I^{\Sigma}$ for the result of the chase on an instance $I$ under constraints $\Sigma$. It has been shown in \cite{mms1979,bv1984,jk1982} that $I^{\Sigma} \models \Sigma$. In case that a chase step cannot be performed (e.g., because a homomorphism would have to equate two constants) the chase result is undefined. In case of an infinite chase sequence, we also say that the result is undefined.\\
 
\textit{Provisio.} We will make a simplifying assumption. Let $I$ be a database instance and $\Sigma$ some constraint set. Without loss of generality we can assume that whenever two labelled nulls, say $y_1, y_2$, are equated by the chase and $y_1 \in dom(I)$, then all occurrences of $y_2$ are mapped to $y_1$ in the chase step. This does not affect chase termination as substituting $y_1$ with $y_2$ would lead to an isomorphic instance.

\subsection{Previous Results}
In the following we are only interested in constraints for which any chase sequence
is finite. In \cite{fkmp2003} weak acyclicity was introduced, which is the starting
point for our work.

\begin{definition} (see \cite{fkmp2003}) \em
Given a set of constraints $\Sigma$, its dependency graph $\mbox{dep}(\Sigma):=(V,E)$
is the directed graph defined as follows. $V$ is the set of positions that occur in the TGDs in $\Sigma$. There are two kind of edges in $E$. Add them as follows: for every TGD 
\begin{center}
$\forall \overline{x} (\phi(\overline{x}) \rightarrow \exists \overline{y} \psi(\overline{x},\overline{y})) \in \Sigma$
\end{center}
 and for every $x$ in $\overline{x}$ that occurs in $\psi$ and every occurrence of $x$ in $\phi$ in position $\pi_1$
  
\squishlist
	\item for every occurrence of $x$ in $\psi$ in position $\pi_2$, add an edge $\pi_1 \rightarrow \pi_2$ (if it does not already exist).
	\item for every existentially quantified variable $y$ and for every occurrence of $y$ in a position $\pi_2$, add a special edge $\pi_1 \stackrel{*}{\rightarrow} \pi_2$ (if it does not already exist).
\squishend
A set $\Sigma$ of TGDs and EGDs is called \textit{weakly acyclic} iff $\mbox{dep}(\Sigma)$ has no cycles  through a special edge. $\punto$
\end{definition}

Then, in \cite{dnr2008} stratification was set on top of the definition of weak acyclicity. The main idea is that we can test if a constraint can cause another constraint to fire, which is the intuition of the following definition.

\begin{definition} (see \cite{dnr2008}) \em
Given two TGDs or EGDs $\alpha = \forall \overline{x_1}\varphi, \beta = \forall \overline{x_2}\psi$, we define $\alpha \prec \beta$ iff there exist database instances $I, J$ and $\overline{a} \in dom(I)$, $\overline{b} \in dom(J)$ such that
\squishlist
	\item $I \nvDash \varphi(\overline{b})$, possibly $\overline{b}$ is not in $dom(I)$,
	\item $I \stackrel{\alpha, \overline{a}}{\longrightarrow} J$ and
	\item $J \nvDash \psi(\overline{b})$. $\punto$
\squishend
\end{definition}

The actual definition of stratification then relies on weak acyclicity.

\begin{definition} (see \cite{dnr2008}) \em
The chase graph $G(\Sigma)=(\Sigma,E)$ of a set of TGDs $\Sigma$ contains a directed edge $(\alpha,\beta)$ between two constraints iff $\alpha \prec \beta$. We call $\Sigma$ stratified iff the set of constraints in every cycle of $G(\Sigma)$ are weakly acyclic. $\punto$
\end{definition}

\begin{theorem} (see \cite{dnr2008}) \em
If a set of constraints of weakly acyclic, then it is also stratified. It can be decided by a $\mbox{coNP}$-algorithm whether a set of constraints is stratified. $\punto$
\end{theorem}

The crucial property of stratification is that it guarantees the termination of the chase in polynomially many chase steps.

\begin{theorem} (see \cite{dnr2008}) \em
Let $\Sigma$ be a fixed and stratified set of constraints. Then, there exists a polynomial $Q \in \mathbb{N}[X]$ such that for any database instance $I$, the length of every chase sequence is bounded by $Q(||I||)$, where $||I||$ is the number of distinct values in $I$. Thus, the chase terminates in polynomial time data complexity. $\punto$
\end{theorem} 

\subsection{Proofs of the Technical Results}

\subsection*{Proof of Theorem \ref{rel-safety}}
% \begin{example}
%Consider the TGD $R(X_1,X_2,X_3), S(X_2) \rightarrow \exists Y R(X_2,Y,X_1)$ from earlier. The dependency graph and the propagation graph are depicted in figure \ref{tab:notweak--safe}. The only affected position is $R^2$. From the respective definitions it follows that this constraint is safe, but not weakly acyclic.
%\end{example}
%
%\begin{figure}[t]
%\begin{center}
%	\begin{tabular}[t]{ccc} 
%\psfig{figure=notweak.eps,width=5cm}	& \hspace{0.5cm} & \raisebox{2cm}{\psfig{figure=safe.eps,width=1cm}}\\
%	\end{tabular}
%\end{center}
%\caption{On the left hand side is the dependency graph and on the right hand side the propagation graph (it has no edges).}
%\label{tab:notweak--safe}
%\end{figure}
% 

\squishlist
	\item Follows directly from the definition of the propagation graph. In the propagation graph stronger conditions have to be satisfied than in the dependency graph in order to add special or non-special edges.
	
	\item Let $\alpha := S(X_2,X_3), R(X_1,X_2,X_3) \rightarrow \exists Y R(X_2,Y,X_1)$ and $\beta := R(X_1,X_2,X_3) \rightarrow S(X_1,X_3)$. It can be seen that $\alpha \prec \beta$ and $\beta \prec \alpha$. Together with the fact that $\setone{\alpha, \beta}$ is not weakly acyclic it follows that $\setone{\alpha, \beta}$ is not stratified. However, $\setone{\alpha, \beta}$ is safe.
	
	\item (see \cite{dnr2008}) \label{zwei}
Let $\gamma$ $:=$ $T(X_1,X_2),$ $T(X_2,X_1)$ $\rightarrow$ $\exists$ $Y_1,Y_2$ $T(X_1,Y_1),$ $T(Y_1,Y_2),$ $T(Y_2,X_1)$. It was argued in \cite{dnr2008} that $\setone{\gamma}$ is stratified. However, it is not safe because both $T^1$ and $T^2$ are affected and therefore $\mbox{dep}(\setone{\gamma})=\mbox{prop}(\setone{\gamma})$ and it was argued in \cite{dnr2008} that it is not weakly acyclic. $\punto$
\squishend

\subsection*{Proof of Theorem \ref{safeterm}}

First we introduce some additional notation. We denote constraints in the form 
$\phi(\overline{x_1},\overline{x_2},\overline{u}) \rightarrow \exists \overline{y} \psi(\overline{x_1},\overline{x_2},\overline{y})$, where $\overline{x_1},\overline{x_2},\overline{u}$ are all the universally quantified variables and 
\squishlist
	\item $\overline{u}$ are those variables that do not occur in the head,
	\item every element in $\overline{x_1}$ occurs in a non-affected position in the body, and
	\item every element in $\overline{x_2}$ occurs only in affected positions in the body.
\squishend

The proof is inspired by the proof of Theorem 3.8 in \cite{fkmp2003}, especially the notation and some introductory definitions are taken from there. In a first step we will give the proof for TGDs only, i.e.~we do not consider EGDs. Later, we will see what changes when we add EGDs again.

Note that $\Sigma$ is fixed. Let $(V,E)$ be the propagation graph $\mbox{prop}(\Sigma)$. For every position $\pi \in V$ an incoming path is a, possibly infinite, path ending in $\pi$. We denote by $rank(\pi)$ the maximum number of special edges over all incoming paths. It holds that $rank(\pi) < \infty$
 because $\mbox{prop}(\Sigma)$ contains no cycles through a special edge. Define $r := max\set{rank(\pi)}{\pi \in V}$ and $p := |V|$. It is easily verified that $r \leq p$, thus $r$ is bounded by a constant. This allows us to partition the positions into sets $N_0,...,N_p$ such that $N_i$ contains exactly those positions $\pi$ with $rank(\pi) = i$. Let $n$ be the number of values in $I$. We define $dom(\Sigma)$ as the set of constants in $\Sigma$.

Choose some $\alpha := \phi(\overline{x_1},\overline{x_2},\overline{u}) \rightarrow \exists \overline{y} \psi(\overline{x_1},\overline{x_2},\overline{y}) \in \Sigma$. Let $I \rightarrow \ldots \rightarrow \overline{G} \stackrel{\alpha,\overline{a_1}\overline{a_2}\overline{b}}{\longrightarrow} G'$ and let $\overline{c}$ be the newly created null values in the step from $\overline{G}$ to $G'$. Then

\begin{enumerate}
  \item newly introduced labelled nulls occur only in affected positions,
	\item $\overline{a_1} \subseteq dom(I) \cup dom(\Sigma)$ and
	\item for every labelled null $Y \in \overline{a_2}$ that occurs in $\pi$ in $\phi$ and every $c \in \overline{c}$ that occurs in $\rho$ in $\psi$ it holds that $rank(\pi) < rank(\rho)$.
\end{enumerate}

This intermediate claim is easily proved by induction on the length of the chase sequence. Now we show by induction on $i$ that the number of values that can occur in any position in $N_i$ in $G'$ is bounded by some polynomial $Q_i$ in $n$ that depends only on $i$ (and, of course, $\Sigma$). As $i \leq r \leq p$, this implies the theorem's statement because the maximal arity $ar(\mathcal{R})$ of a relation is fixed. We denote by ${\it body}(\Sigma)$ the number of characters of
the largest body of all constraints in $\Sigma$.\\

\underline{Case 1: $i=0$.}  We claim that $Q_0(n)$:=$n + |\Sigma| \cdot n^{ar(\mathcal{R}) \cdot body(\Sigma)}$
 is sufficient for our needs. We consider a position $\pi \in N_0$ and an arbitrary TGD from $\Sigma$ such that $\pi$ occurs in the head of $\alpha$. For simplicity we assume that it has the syntactic form of $\alpha$. In case that there is a universally quantified variable in $\pi$, there can occur at most $n$ distinct elements in $\pi$. Therefore, we assume that some existentially quantified variable occurs in $\pi$ in $\psi$. Note that as $i=0$ it must hold that $|x_2|=0$. Every value in $I$ can occur in $\pi$. But how many labelled nulls can be newly created in $\pi$? For every choice of $\overline{a_1} \subseteq dom(\overline{G})$ such that 
$\overline{G} \models \phi(\overline{a_1},\lambda,\overline{b})$ and $\overline{G} \nvDash \exists \overline{y} \psi(\overline{a_1},\lambda,\overline{y})$ at most one labelled null can be added to $\pi$ by $\alpha$. Note that in this case it holds that $\overline{a_1} \subseteq dom(I)$ due to (1). So, there are at most 
$n^{ar(\mathcal{R}) \cdot body(\Sigma)}$ such choices. Over all TGDs at most $|\Sigma| \cdot n^{ar(\mathcal{R}) \cdot body(\Sigma)}$ labelled nulls are created in $\pi$.\\

\underline{Case 2: $i \rightarrow i+1$.}  We claim that $Q_{i+1}(n) := \sum_{j=0}^{i} Q_i(n) + |\Sigma| \cdot (\sum_{j=0}^{i} Q_i(n))^{ar(\mathcal{R}) \cdot body(\Sigma)}$ is such a polynomial. Consider the fixed TGD $\alpha$. Let $\pi \in N_{i+1}$. Values in $\pi$ may be either copied from a position in $N_0 \cup ... \cup N_i$ or may be a new labelled null. Therefore w.l.o.g. we assume that some existentially quantified variable occurs in $\pi$ in $\psi$. In case a TGD, say $\alpha$, is violated in $G'$ there must exist $\overline{a_1}, \overline{a_2} \subseteq dom_{G'}(N_0,...,N_i)$ and $\overline{b} \subseteq dom(G')$ such that $G' \models \phi(\overline{a_1},\overline{a_2},\overline{b})$, but $G' \nvDash \exists \overline{y} \psi(\overline{a_1},\overline{a_2},\overline{y})$.  If newly introduced labelled null occurs in $\overline{a_2}$, say in some position $\rho$, then $\rho \in \bigcup_{j=0}^{i} N_j$. As there are at most 
$(\sum_{j=0}^{i} Q_i(n))^{ar(\mathcal{R}) \cdot body(\Sigma)}$ many such choices for $\overline{a_1}, \overline{a_2}$, at most $(\sum_{j=0}^{i} Q_i(n))^{ar(\mathcal{R}) \cdot body(\Sigma)}$ many labelled nulls can be newly created in $\pi$.\\

When we allow EGDs among our constraints, we have that the number of values that can occur in any position in $N_i$ in $G'$ can be bounded by the same polynomial $Q_i$ because equating labelled nulls does not increase the number of labelled nulls and the fact that EGDs preserve valid existential conclusions of TGDs. $\qed$

\subsection*{Proof of Theorems \ref{rest-safe-weak} and~\ref{rest-comp}}

Theorem~\ref{rest-safe-weak} Follows from Theorem \ref{safestrat}.
Before we prove Theorem~\ref{rest-comp}, we introduce some additional tool.

In general,
a set of constraints may have several restriction systems. 
A restriction system is {\it minimal} if it is obtained from
($(\Sigma,\emptyset)$,\{$(\alpha,\emptyset)$ $\mid$ $\alpha \in \Sigma$\}) by
a repeated application of the constraints from bullets one to three in Definition~\ref{rest}
(until all constraints hold) s.t., in case of the first and second bullet, the image of
$f(\beta)$ is extended only by those positions that are required to satisfy
the condition. Thus, a minimal restriction system can be computed by a fixedpoint iteration.

\begin{lemma} \em Let $\Sigma$ be a set of constraints, $(G'(\Sigma),f)$ a restriction system for $\Sigma$ and $(G_{min}'(\Sigma),f_{min})$ its minimal one.

\squishlist
	\item Let $P$ be a set of positions and $\alpha, \beta$ constraints. Then, the mapping $(P,\alpha,\beta) \mapsto \alpha \prec_P \beta ?$ can be computed by an $\mbox{NP}$-algorithm.

	\item The minimal restriction system for $\Sigma$ is unique. It can be computed from $\Sigma$ in non-deterministic polynomial time. 
	
	\item It holds that $\Sigma$ is safely restricted if and only if every strongly connected component in $G_{min}'(\Sigma)$ is safe. $\punto$
\squishend
\end{lemma}

{\it Proof.} The proof of part one of the lemma proceeds like the proof of Theorem 3 in \cite{dnr2008}. It is enough to consider candidate databases for $A$ of size at most $|\alpha| + |\beta|$, i.e.~unions of homomorphic images of the premises of $\alpha$ and $\beta$ s.t.~null values occur only in positions from $P$. This concludes part one.

Uniqueness holds by definition. It can be computed via successive application of the constraints (note that $f$ and $E$ are changed in each step) in definition \ref{rest} by a Turing machine that guesses answers to the question $\alpha \prec_P \beta ?$. As the mapping $(P,\alpha,\beta) \mapsto \alpha \prec_P \beta ?$ can be computed by an $\mbox{NP}$-algorithm and the fixedpoint is reached after polynomially many applications of the constraints from definition \ref{rest}, this implies the second claim.

Concerning the second claim, observe that every strongly connected component in $G_{min}'(\Sigma)$ is contained in a single strongly connected component of any other restriction system. This implies the third claim.$\qed$

Now we turn towards the proof of Theorem \ref{rest-comp}. By the previous lemma it suffices to check the conditions from definition \ref{def-rest} only for the minimal restriction system. 
To decide whether $\Sigma$ is not safely restricted, compute the minimal restriction system, guess a strongly connected component and check if it is not safe. Clearly, this can be done in non-deterministic polynomial time. $\qed$

\subsection*{Proof of Theorem \ref{chase-main} (Sketch)}

Before proving this theorem, we need a technical lemma. It states the most important property of restriction systems.

\begin{lemma} 
\em
Let $\alpha \in \Sigma$ and $(G(\Sigma),f)$ a restriction system for $\Sigma$ and $I$ be a database instance. If during the chase it occurs  that $J_1 \stackrel{\alpha, \overline{a}}{\rightarrow} J_2$, then the set of positions in which null values from $\overline{a}$ that are not in $dom(I)$ occur in the body of $\alpha$ is contained in $f(\alpha)$. $\punto$
\end{lemma}

The proof of this lemma is by induction on the length of the chase sequence with which $J_1$ was obtained from $I$ and is straightforward. Note that it uses the simplifying assumption that we introduced at the end of appendix \ref{app:defs-chase}.

Let us now turn to the proof of Theorem \ref{chase-main}. Let $(G'(\Sigma),f)=((G,E),f)$ be the minimal restriction system of $\Sigma$. Let $C_1, ... , C_m$ be all the pairwise different strongly connected components of the reflexive closure of $G'(\Sigma)$. The graph $H$ is defined as the quotient graph with respect to $\set{(\alpha,\beta) \in E}{\exists i \in [m]: \alpha, \beta \in C_i}$, i.e. $ H:= G'(\Sigma) / \set{(\alpha,\beta) \in E}{\exists i \in [m]: \alpha, \beta \in C_i}$. $H$ is acyclic and depends only on $\Sigma$. We show the claim by induction on the number of nodes $n$ in $H$.

\textbf{Case: $n=1$:} $G'(\Sigma)$ has a single strongly connected component. This component is safe by prerequisite. It follows from Theorem \ref{safeterm} that the chase terminates in polynomial time data complexity in this case.

\textbf{Case: $n \mapsto n+1$:} Let $h$ be a node in $H$ that has no successors and $H_{-}$ the union of constraints from all other nodes in $H$. The chase with $H_{-}$ terminates by induction hypothesis, say that the number of distinct value in this result is bounded by some polynomial $Q_{-}$. Chasing the constraints in $h$ alone terminates, too, say that the number of distinct value in this result is bounded by the polynomial $q$. The firing of constraints from $H_{-}$ can cause some constraints from $h$ to copy null values in their heads. Yet, the firing of constraints in $h$ cannot enforce constraints from $H_{-}$ to copy null values to their head (by construction of the minimal restriction system). If $I$ is the database instance to be chased, then the number of distinct value in this result is bounded by $Q_{-}(||I||) + q(||I|| + Q_{-}(||I||) )$. As $\Sigma$ is fixed we can conclude that the chase terminates in polynomial time data complexity. $\qed$

\subsection*{Proof of Theorem \ref{safestrat}}

\squishlist
	\item Let $\Sigma$ be weakly acyclic. Every cycle in $G(\Sigma)$ is safe, because $\Sigma$ is safe and weak acyclicity implies safety. 
	Let $\Sigma$ be safe. Every cycle in $G(\Sigma)$ is safe, because $\Sigma$ is.
	
	\item Follows from Example \ref{alle} and the following proposition.

 \begin{proposition} \label{verf} \em
 Let $P \subseteq P' \subseteq pos(\Sigma)$. If $\alpha \prec_{P} \beta$, then $\alpha \prec_{P'} \beta$. It holds that if $\alpha \prec_{P} \beta$, then $\alpha \prec \beta$. $\punto$
 \end{proposition}
	
	The proof follows from the definition of $\prec_{P}$ and $\prec$.
	
	\item Consider the following TGDs. $\Sigma := \setone{\alpha, \beta, \chi, \delta}$.\\
	$\alpha := R_1(x_1,x_2) \rightarrow \exists y S(x_1,x_2,y)$,\\
	$\beta := R_1(x_1,x_2) \rightarrow \exists y T(x_1,x_2,y)$,\\
	$\chi := S(x_1,x_2,x_3), T(x_4,x_5,x_6) \rightarrow T(x_5,x_1,x_4)$ and\\
	$\delta := S(x_1,x_2,x_3), T(x_4,x_5,x_3) \rightarrow T(x_1,x_3,x_3),$\\ $R_1(x_3,x_1), R_2(x_3,x_1)$.\\
	It can be seen that $\alpha \prec \chi$, $\beta \prec \chi$, $\chi \prec \delta$, $\delta \prec \alpha$ and $\delta \prec \beta$ holds. Thus , there is a cycle in the chase graph that involves all constraints. Unfortunately, the constraint set is not safe. Therefore, it is also not safely stratified.
	
	The minimal restriction system is $((\Sigma,E),f)$, where $E = \emptyset$ and $f = \set{(\gamma,\emptyset)}{\gamma \in \Sigma}$. Obviously, every cycle in $(\Sigma,E)$ is safe. Hence, $\Sigma$ is safely restricted. $\qed$
\squishend

\subsection*{Proof of Proposition \ref{falsch}}
Let $(G'(\Sigma),f)$ be a restriction system  for $\Sigma$ such that every strongly connected component in $G'(\Sigma)$ is safely stratified. Choose some strongly connected component $C$ and two constraints $\alpha, \beta \in C$ such that $\alpha \prec_P \beta$ for some set of positions $P$. By Proposition \ref{verf}, $\alpha \prec \beta$ holds. As $C$ is safely stratified, this means that $C$ must also be safe. So, every cycle in $G'(\Sigma)$ is also safe. $\punto$

\subsection*{Proof of Proposition \ref{schemes-vgl}}

\squishlist
	\item Let $\Sigma := \setone{\forall x_1, x_2 (T(e,x_1,x_2),T(x_2,d,d) \rightarrow T(x_1,x_2,x_1))}$, where $d$ is a constant and $x_i$ are variables. Then, $\Sigma' = \emptyset$.
	
	\item An example for such a set of constraints $\mathcal{C}$ is constituted as follows.\\
	$T(x_1,d,x_2) \rightarrow \exists y T(g,e,y), T(f,d,y)$\\
	$T(x_1,e,x_2) \rightarrow \exists y T(g,e,y), T(f,d,y)$\\
	$T(x_1,d,x_2) \rightarrow T(x_2,e,x_1)$\\
	$T(x_1,e,x_2) \rightarrow \exists y T(x_2,d,y)$\\
	Note that $d,e,f,g$ are constants. $\punto$
\squishend
 % mm

\end{document}